\PassOptionsToClass{12pt}{revtex4-1}
\documentclass[ApJ]{aastex631}
\usepackage{graphics,amsmath,bm}
\usepackage{amsbsy}
\graphicspath{{./figures/}}

\newcommand\eps{\epsilon}
\newcommand{\lapprox} {\, \lower3pt\hbox{$\sim$}\llap{\raise2pt\hbox{$<$}}\,}
\newcommand{\gapprox} {\, \lower3pt\hbox{$\sim$}\llap{\raise2pt\hbox{$>$}}\,}

\newcommand\mmatrix[1]{\textsf{\textbf{#1}}}
\renewcommand{\vec}[1]{ \protect {{\mathbf{\boldsymbol{#1}}}}}

\shorttitle{Anisotropic Density Turbulence Model from the Sun to 1~au}
\shortauthors{Kontar et al.}

\begin{document}

\title{An Anisotropic Density Turbulence Model from the Sun to 1~au Derived From Radio Observations}

\author[0000-0002-8078-0902]{Eduard P. Kontar}
\affiliation{School of Physics \& Astronomy, University of Glasgow, Glasgow, G12 8QQ, UK}

\author[0000-0001-8720-0723]{A. Gordon Emslie}
\affiliation{Department of Physics \& Astronomy, Western Kentucky University, Bowling Green, KY 42101, USA}

\author[0000-0003-1967-5078]{Daniel L. Clarkson}
\affiliation{School of Physics \& Astronomy, University of Glasgow, Glasgow, G12 8QQ, UK}

\author[0000-0002-1810-6706]{Xingyao Chen}
\affiliation{School of Physics \& Astronomy, University of Glasgow, Glasgow, G12 8QQ, UK}

\author[0000-0002-4389-5540]{Nicolina Chrysaphi}
\affiliation{Sorbonne Universit\'{e}, \'{E}cole Polytechnique, Institut Polytechnique de Paris, CNRS, Laboratoire de Physique des Plasmas (LPP), 4 Place Jussieu, 75005 Paris, France}
\affiliation{School of Physics \& Astronomy, University of Glasgow, Glasgow, G12 8QQ, UK}

\author[0009-0001-7368-0938]{Francesco Azzollini}
\affiliation{School of Physics \& Astronomy, University of Glasgow, Glasgow, G12 8QQ, UK}

\author[0000-0001-6583-1989]{Natasha L.S. Jeffrey}
\affiliation{Department  of  Mathematics,  Northumbria  University,  Physics and Electrical  Engineering, Newcastle upon Tyne, NE1 8ST, UK}

\author[0000-0003-2291-4922]{Mykola Gordovskyy}
\affiliation{Department of Physics, Astronomy \& Mathematics. University of Hertfordshire, Hatfield AL10 9AB, UK}

\begin{abstract}
  Solar radio bursts are strongly affected by radio-wave scattering
  on density inhomogeneities, changing their observed time characteristics,
  sizes, and positions. The same turbulence causes angular broadening
  and scintillation of galactic and extra-galactic compact radio sources observed
  through the solar atmosphere. Using large-scale simulations of radio-wave transport,
  the characteristics of anisotropic density turbulence from $0.1 \, R_\odot$ to $1$~au
  are explored. For the first time, a profile of heliospheric density fluctuations
  is deduced that accounts for the properties of extra-solar radio sources, solar radio bursts,
  and in-situ density fluctuation measurements in the solar wind at $1$~au.
  The radial profile of the spectrum-weighted mean wavenumber of density fluctuations
  (a quantity proportional to the scattering rate of radio-waves) is found
  to have a broad maximum at around $(4-7) \, R_\odot$, where the slow solar wind becomes
  supersonic. The level of density fluctuations at the inner scale (which is consistent
  with the proton resonance scale) decreases with heliocentric distance
  as $\langle\delta{n_i}^2 \rangle (r) \simeq 2 \times 10^7 \, (r/R_\odot-1)^{-3.7}$~cm$^{-6}$.
  Due to scattering, the apparent positions of solar burst sources observed at frequencies
  between $0.1$ and $300$~MHz are computed to be essentially cospatial
  and to have comparable sizes, for both fundamental and harmonic emission. Anisotropic scattering
  is found to account for the shortest solar radio burst decay times observed,
  and the required wavenumber anisotropy is  $q_\parallel/q_\perp =0.25-0.4$,
  depending on whether fundamental or harmonic emission is involved.
  The deduced radio-wave scattering rate paves the way to quantify intrinsic solar radio burst characteristics.
\end{abstract}


\section{Introduction}

Radio-wave scattering affects the temporal characteristics, sizes, and positions of both solar radio bursts
and extra-solar sources observed through the turbulent solar atmosphere.
Solar radio bursts (such as Type~I, Type~II, Type~III, etc.) are produced predominantly via plasma mechanisms
at frequencies that are close to either the local plasma frequency or its double (harmonic),
and are thus particularly strongly affected by scattering
\citep[e.g.,][]{1965BAN....18..111F,1972A&A....18..382S,1974SoPh...35..153R,
  1994ApJ...426..774B,1999A&A...351.1165A,2008ApJ...676.1338T,2017NatCo...8.1515K,
  2018ApJ...868...79C,2018ApJ...857...82K,2018SoPh..293..132M,
  2019ApJ...873...48G,2019ApJ...884..122K,2020ApJ...893..115C,
  2020ApJS..246...57K,2021A&A...645A..11M,2021ApJ...909....2M,
  2021A&A...655A..77M,2021A&A...656A..34M,2021ApJ...913..153S}.
Similarly, extra-solar point sources observed through the solar atmosphere and solar wind are noticeably
broadened \citep[e.g.,][]{1952Natur.170..319M,1958MNRAS.118..534H,1972PASAu...2...86D,
  1994JApA...15..387A,2017ApJ...850..129S}
and scintillate \citep{1964Natur.203.1214H,1967ApJ...147..449C,1977Natur.266..514W,
  1978SSRv...21..411C,1987A&A...183..135A,1993SoPh..148..153M,
  1999JGR...104.9847B,2014ApJ...797...51M,2016JGRA..12111605S,
  2018MNRAS.474.4937C,2023MNRAS.tmp.1421T}.
Overall, various observations provide a qualitatively consistent picture of turbulence in the heliosphere.
These observations also strongly suggest that the density fluctuations are anisotropic,
with parallel wavenumbers $q_\parallel$ that are smaller than the perpendicular wavenumbers
$q_\perp$ \citep[e.g.,][]{1972PASAu...2...86D,1989ApJ...337.1023C,1990ApJ...358..685A},
with stronger anisotropy suggested at smaller scales. Observations of solar radio bursts
at higher
frequencies ($\sim$30-50~MHz) originate closer to the Sun and usually require an anisotropy
$q_\parallel/q_\perp \simeq 0.15-0.3$ \citep{2019ApJ...884..122K,2020ApJ...905...43C,2020ApJ...898...94K,2021ApJ...917L..32C},
while simultaneous observations of solar radio bursts in interplanetary space
using multiple spacecraft require a somewhat larger ratio $q_\parallel/q_\perp \simeq 0.3-0.4$ \citep{2021A&A...656A..34M}.
In-situ observations of the wavenumber anisotropy of density fluctuations are very few:
\citet{1987A&A...181..138C} report an anisotropy factor between $1$ and~$2$ with an uncertainty comparable
to the measured values, while \citet{2018AnGeo..36..527R} report an anisotropy factor between $1$ and~$3$.

The scattering regimes for solar radio bursts and extra-solar point sources are very different.
Extra-solar sources are typically observed at frequencies $\omega$ that are much greater than the plasma frequency
of the solar medium in which their emitted radiation propagates and scatters, i.e., $\omega/\omega_{pe}\gg 1$,
where $\omega _{pe}=\sqrt{4\pi e^2n/m_e}$ (s$^{-1}$) is the local plasma frequency, $n$ (cm$^{-3}$)
is the local number density, and $e$ and $m_e$ are the electronic charge (esu) and mass (g), respectively.
The scattering mean free path $\lambda _{sc}$ is proportional to $(\omega ^2-\omega_{pe}^2)^{2}$ and is typically much
larger than the distance $r$ traveled by the radio waves in the turbulent medium. Consequently, such sources experience weak scattering and their broadening is normally computed using a thin-screen approximation to deduce a wave structure function that measures the spatial correlation of the turbulence at different scales \citep[see, e.g.,][]{1975ApJ...196..695L,1977ApJ...218..557W,1989ApJ...337.1023C}. By contrast, solar radio burst emission is at frequencies that are inherently close to the local plasma frequency (or its second harmonic), so that $\omega /\omega_{pe}\simeq 1-2$. For values of $\omega$ so close to $\omega_{pe}$, the mean free path $\lambda_{sc}$ can be much smaller than the distance a wave propagates between the emission location and the observer, so that the emitted radio waves are (at least close to their origin) subject to multiple scatterings that randomize the wave propagation. Waves emitted in solar radio bursts, at least close to the emitting region, thus propagate diffusively \citep{2019ApJ...873...33B,2019ApJ...884..122K,2020ApJ...905...43C}, with the transport modeled using a stochastic random-phase description \citep{1952MNRAS.112..475C,1965BAN....18..111F,1971A&A....10..362S,
  2018ApJ...857...82K,2008ApJ...676.1338T} with inclusion of anisotropic effects \citep{1999A&A...351.1165A,2019ApJ...884..122K}.

Regarding observations of the density fluctuations, solar corona sounding experiments normally work well for heliocentric distances $\sim (2-100)~R_\odot$ \citep[e.g.,][]{1989ApJ...337.1023C}, but become progressively more difficult at smaller heliocentric distances due to the strong radio emission of the quiet solar corona between $(1-2)$~$R_\odot$. On the other hand, solar radio burst emissions originating between $(1-2)$~$R_\odot$ have brightness temperatures far greater than the corona and are easily observed. As such, solar radio bursts, such as Type~III bursts, are better suited to provide constraints on scattering and density fluctuations close
to the Sun, i.e., below $2$~$R_\odot$ \citep{2017NatCo...8.1515K,2018ApJ...868...79C,2020ApJ...905...43C,
  2021ApJ...913..153S,2021A&A...655A..77M}.

Since it is the same density turbulence that affects the properties of (1) extra-solar sources, (2) solar sources, and (3) density fluctuations measured in-situ in the solar wind, a common density turbulence model must self-consistently explain all three sets of observations. Here, using the numerical model of \cite{2019ApJ...884..122K}, we carry out a number of radio-wave propagation simulations between $0.15 - 300$~MHz  and we consider the results in light of the veryu substantial array of extra-solar, solar, and in-situ observations published in the literature. We show that the level of radio-wave scattering necessary to account for the properties of non-solar sources matches the level required to account for the source sizes and apparent positions of solar radio bursts in the corona and in the heliosphere. Further, the burst decay times and time durations predicted by such a turbulence model are consistent both with typical Type~III burst durations below $1$~MHz and with the shortest radio bursts observed (such as spikes and Type~IIIb striae) above $1$~MHz. Intriguingly, the Type~III burst striae that were observed by Parker Solar Probe \citep[PSP;][]{2016SSRv..204....7F} show a transition from a decay time that is proportional to $1/f$ at lower frequencies to a downshifted $1/f$ behavior at higher frequencies, consistent with the trend predicted by the scattering model.

In Section~\ref{sec:methodology} we discuss the model of radio-wave propagation in a turbulent medium with an anisotropic turbulence wavenumber spectrum $S({\vec q})$. In Section~\ref{extra-solar-source-broadening}, we discuss the angular broadening of extra-solar sources with ray paths that pass close to the turbulent atmosphere of the Sun, and we determine the density and turbulence profiles required to account for various source size observations. In Section~\ref{sec:simulations-of-solar-bursts} we discuss simulations of solar radio bursts that use the turbulence profile (both strength and anisotropy) established by the data on extra-solar source angular broadening, and we compare the model predictions with observations of source decay times (Section~\ref{subsec:decay-time}), sizes (Section~\ref{subsec:size}), and apparent positions (Section~\ref{subsec:positions}). In Section~\ref{sec:in-situ} we discuss the relation of the turbulence wavenumber spectrum to in-situ measurements of the frequency spectrum of density fluctuations, leading to further constraints on both the level of turbulence and its degree of anisotropy. In Section~\ref{sec:amplitude} we use observations of the turbulence level at the inner scale (the boundary between the inertial and dissipative ranges of the turbulence spectrum) to further constrain the turbulence amplitude, and we present results on the inferred variation of the level of density fluctuations with solar distance. Finally, in Section~\ref{sec:summary} we summarize the results obtained and use them to present a self-consistent model of the interplanetary turbulence profile between the Sun and the Earth.

\section{Radio-wave transport in an anisotropic turbulent medium}\label{sec:methodology}

\begin{figure*}
  \centering
  \includegraphics[width=0.45\linewidth]{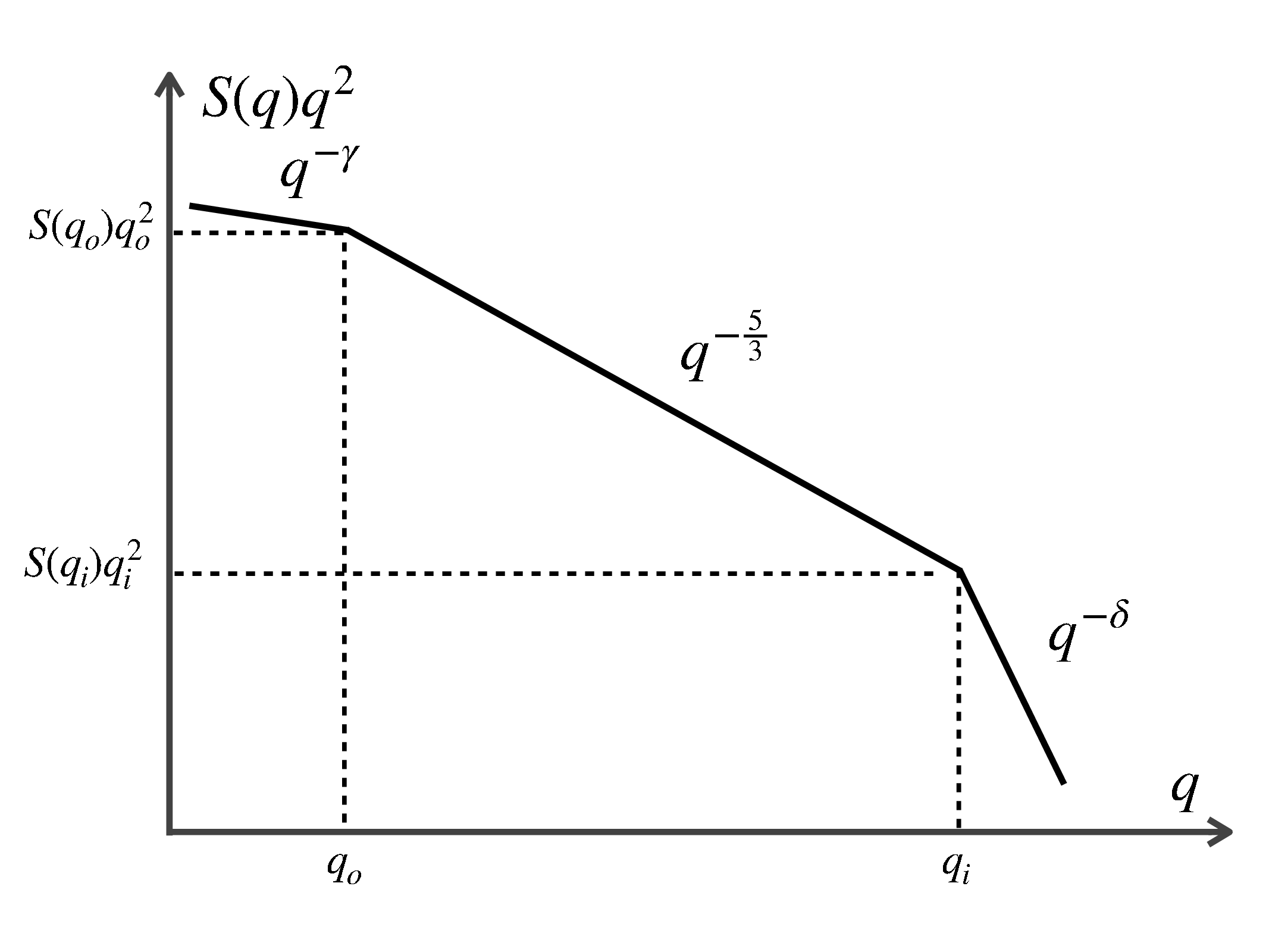}
  \includegraphics[width=0.45\linewidth]{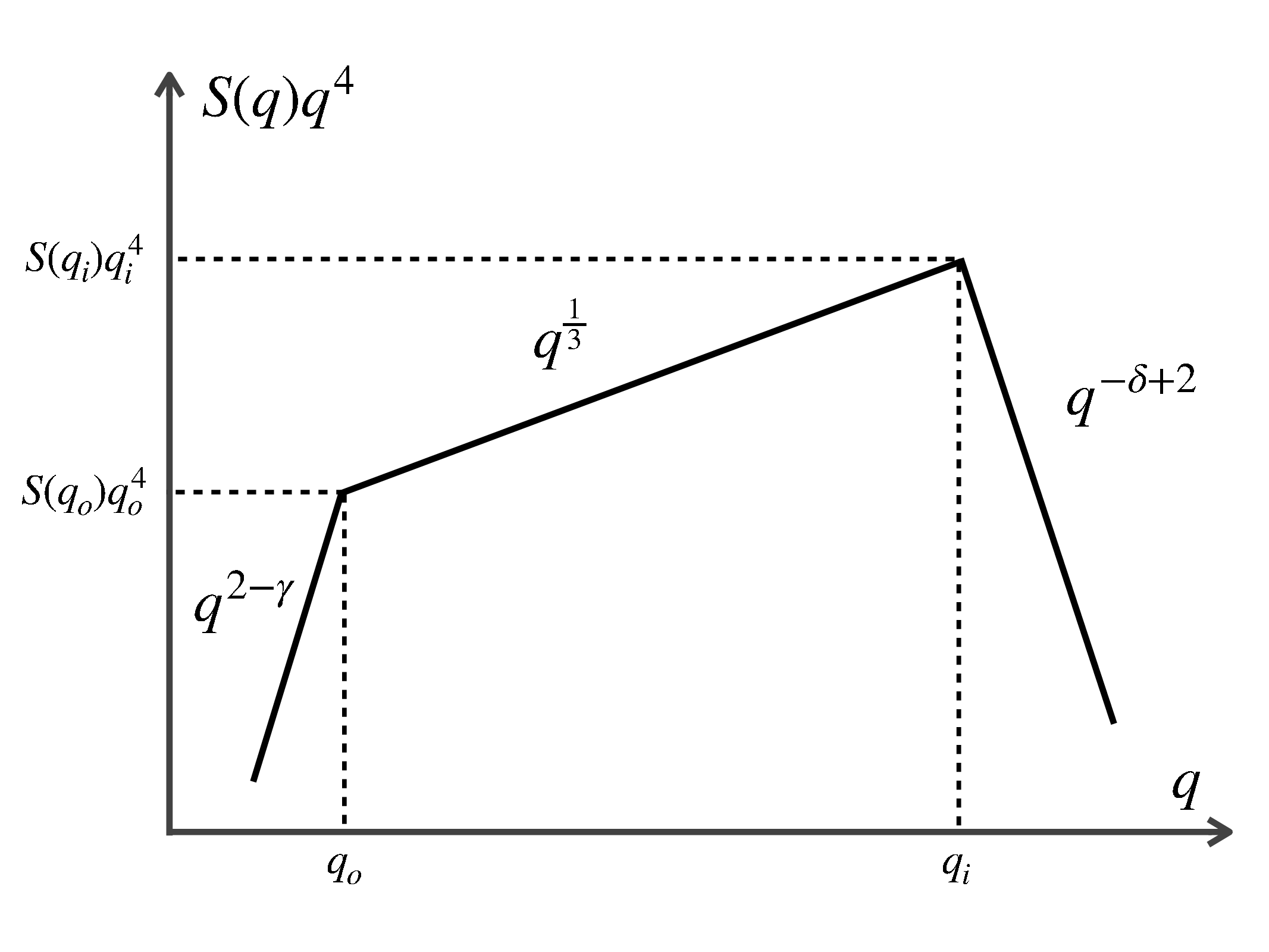}
  \caption{\textit{Left:} Spectrum of density fluctuations showing the $q^2 \, S(q) \sim q^{-\gamma}$ spectrum at large scales $> q_o^{-1}$, the (Kolmogorov) inertial range $q^2 \, S(q) \sim q^{-5/3}$ between the outer scale $q_o^{-1}$ and the inner scale $q_i^{-1}$, and the dissipation range $q^2 \, S(q) \sim  q^{-\delta}$ (with $\delta > 2$) at small scales $< q_i^{-1}$. \textit{Right:} The $\ln q$ space integrand required to compute the mean wavenumber $\overline{q \, \eps ^2} \propto \int q \, S(q) \, q^2 \, dq = \int q^4 \, S(q) \, d\ln q$, highlighting the peak near the inner scale $q_i^{-1}$. All axes are on logarithmic scales.}\label{fig:Sqq2}
\end{figure*}

Density fluctuations in general are characterized by their three-dimensional wavenumber spectrum $S(\vec{q})$, such that the density fluctuation variance $\langle\delta n^2\rangle$ (cm$^{-6}$), normalized by the square of the local density, is

\begin{equation}\label{eq:Sq3d}
  \epsilon^2 \equiv \frac{\langle\delta n^2\rangle}{n^2}=\int S(\vec{q}) \, \frac{d^3q}{(2\pi)^3} \,\,\, ,
\end{equation}
where ${\vec q}$ (cm$^{-1}$) is the wavevector associated with a density fluctuation, $\langle \cdots \rangle$ denotes an ensemble average, and $n(\vec{r})$ is the background plasma density. A typical wavenumber spectrum \citep[e.g.,][]{2013SSRv..178..101A,2022ApJ...928..125T} has three main domains:

\begin{equation}\label{sq-form-general-delta}
  S(q) = S(q_i) \, \times \,
  \left\{
  \begin{array}{lll}
    \left ( \frac{q}{q_o} \right )^{-2-\gamma} \left (\frac{q_o}{q_i} \right )^{-11/3} , & \qquad  q < q_o \,\,          & {\rm (outer)}  \cr
    \left (\frac{q}{q_i} \right )^{-2-5/3} ,                                             & \qquad q_o \le q \le q_i \,\, & {\rm (inertial)} \cr
    \left ( \frac{q}{q_i} \right )^{-2-\delta} ,                                         & \qquad q > q_i \,\,           & {\rm (dissipative)} \,\,\, ,
  \end{array}
  \right.
\end{equation}
where the spectrum is normalized to the value $S(q_i)$ at the inner scale wavenumber $q=q_i$,
the boundary between the inertial and dissipative ranges. We note that the density fluctuation power spectrum shows evidence of flattening between the inertial and dissipation scales \citep[e.g.,][]{1987A&A...181..138C,
  1989ApJ...337.1023C,2015ApJ...803..107S}, that we do not consider here for simplicity.
Qualitatively, since the wavenumber volume element $d^3 q = 4 \pi \, q^2 \, dq$, the wavenumber spectrum
(for an illustrative isotropic density fluctuation spectrum) has three major parts
\citep[left panel of Figure~\ref{fig:Sqq2}; see also Figure~2 of][]{2022ApJ...928..125T}: $q^2 \, S(q) \, q^{-\gamma}$
for large wavelengths ($q < q_o$), with values of $\gamma$ between $0$ and~$1$
\citep[e.g.,][]{1990JGR....9511945M,2002AdSpR..30..447B,2018ApJ...862...18D},
the (Kolmogorov) inertial range $q^2 \, S(q) \sim q^{-5/3}$ between the outer scale $q_o$
and the inner scale $q_i$ \citep[e.g.,][]{1990JGR....9511945M},
and the dissipation range $q^2 \, S(q) \sim q^{-\delta}$ at small wavelengths
\citep[$q > q_i$; e.g.,][]{1983A&A...126..293C}.

As we shall show below \citep[Equation~\eqref{eq:d_ij_aniso_k}; see also][]{2019ApJ...884..122K},
the diffusion coefficient for radio waves is proportional to the spectrum-weighted mean wavenumber

\begin{equation}\label{q-eps2-definition}
  \overline{q \, \eps ^2} = \frac{4 \pi}{(2 \pi)^3} \, \int q \, S(q) \, q^2 \, dq \,\,\, ,
\end{equation}
which is the central quantity of interest in this work. This integral is proportional to $\int q^4 \, S(q) \, d\ln q$; the corresponding integrand is shown schematically in the right panel of Figure~\ref{fig:Sqq2} and is seen to peak near $q_i$, corresponding to the inner scale. Quantitatively, the contribution to $\int_0^\infty q^3 \, S(q) \, dq \equiv [(2 \pi)^3/4\pi] \,\overline{q \, \eps^2}$ from the long-wavelength outer range ($q < q_o$) is $S(q_i) \, q_i^{11/3} \, q_o^{\gamma -5/3} \, \int_0^{q_o} q^{1-\gamma }\, dq = [1/ (2-\gamma)] \, S(q_i) \, q_i^4 \, (q_o/q_i)^{1/3}$, the contribution from the inertial range ($q_o < q < q_i$) is $S(q_i) \, q_i^{11/3} \int_{q_o}^{q_i} q^{-2/3} \, dq = 3 \, S(q_i) \, q_i^4 (1 - [q_o/q_i]^{1/3})$, and the contribution from the dissipative range ($q > q_i$) is $S(q_i) \, q_i^{\delta+2} \int_{q_i}^\infty q^{1-\delta} \, dq = [1/(\delta-2)] \, S(q_i) \, q_i^4$. The inertial range between $q_0$ and $q_i$ is typically about three orders of magnitude \citep[e.g.,][]{2013SSRv..178..101A}, and a typical dissipation range spectral index $\delta$ varies from $2.3$ to $2.9$ \citep[e.g.,][]{1987A&A...181..138C}, so we adopt $\delta =2.5$. The contributions to $\overline{q \, \eps ^2}$ from the three ranges are therefore in the approximate ratio 0.1: 3 : 2. Ignoring the small contribution at large scales ($q<q_o$), the spectrum-weighted mean wavenumber $\overline{q \, \eps^2}$ is
\begin{equation}\label{eq:qeps2_iso}
  \overline{q \, \eps ^2} \simeq \frac{4\pi}{(2\pi)^3} \left ( 3+\frac{1}{\delta-2} \right ) \, S(q_i) \, q_i^4 = \frac{20 \, \pi}{(2\pi)^3} \, S(q_i) \, q_i^4 \,\,\, ,
\end{equation}
where the last equality uses $\delta = 2.5$. Equation~\eqref{eq:qeps2_iso} and Figure~\ref{fig:Sqq2} importantly highlight that $\overline{q \, \eps^2}$ is determined mostly by density fluctuations close to the inner scale $q_i^{-1}$.

Observations often suggest the presence of anisotropy relative to the direction of the magnetic field, suggesting a correspondingly anisotropic wavenumber spectrum. Following previous studies \citep{1972PASAu...2...86D,1977ApJ...218..557W}, we therefore consider a wavenumber spectrum of the form

\begin{equation}\label{eq:Sq_alpha}
  S(\vec{q})=S(\tilde{q})\,,\,\,\, \mbox{where}\;\;\;
  \tilde{q}=\sqrt{q_{\perp_1}^2 + q_{\perp_2}^2 + \frac{q_{\parallel}^2}{\alpha^2}} \,\,\, ,
\end{equation}
which has axial symmetry around the $\parallel$ direction, i.e., along the (assumed radial) magnetic field $\vec{B}$. The wavenumber diffusion tensor describing elastic scattering of radio waves with wavenumber $\vec{k}$
in such an anisotropic medium can be written \citep[e.g.,][]{1999A&A...351.1165A,2019ApJ...873...33B,
  2019ApJ...884..122K}

\begin{equation}\label{eq:D}
  D_{ij} =  \frac{\omega_{pe}^4} {32 \, \pi^2\, \omega \,  c^2 } \int q_i \, q_j \,  S(\vec{q}) \,\, \delta(\vec q \cdot \vec {k}) \, {d^3q} \,\,\, ,
\end{equation}
where $\omega (\vec{k})=\sqrt{\omega_{pe}^2+c^2k^2}$ is the angular frequency and $\vec{k}$ the wavevector of electromagnetic waves in a plasma with local plasma frequency $\omega_{pe} (r)$. For the spectrum given by Equation~\eqref{eq:Sq_alpha}, we can obtain \citep[Equation~(16) of][]{2019ApJ...884..122K} an explicit expression for the components of the wavenumber diffusion tensor (cm$^{-2}$~s$^{-1}$):

\begin{eqnarray}\label{eq:d_ij_aniso_k}
  D_{ij} &=& \left[\frac{{A}^{-2}_{ij}}{(\vec k \mmatrix{A}^{-2} \vec k)^{1/2}}-\frac{(\mmatrix{A}^{-2}\vec k)_i(\mmatrix{A}^{-2}\vec k)_j}{(\vec k \mmatrix{A}^{-2} \vec k)^{3/2}}\right]
  \frac {\omega_{ pe }^4} { 32 \, \pi \, \omega \, c^2 } \,
  \alpha \int {\widetilde q}^3 \, S(\widetilde q) \, d\widetilde q \cr
  &=& \left[\frac{{A}^{-2}_{ij}}{(\vec k \mmatrix{A}^{-2} \vec k)^{1/2}}-\frac{(\mmatrix{A}^{-2}\vec k)_i(\mmatrix{A}^{-2}\vec k)_j}{(\vec k \mmatrix{A}^{-2} \vec k)^{3/2}}\right]
  \frac {\pi \, \omega_{ pe }^4} { 16 \, \omega \, c^2 }  \,  \overline{q \, \eps ^2}\,\,\, ,
\end{eqnarray}
where ${\widetilde {\bf q}} = (q_{\perp,1},  q_{\perp,2}, q_\parallel/\alpha)$ is the wavevector transformed into a space such that $S({\widetilde q})$ is isotropic, ${\bf A}$ is the matrix  ${\rm diag} \, (1,1,\alpha^{-1})$, and we have used $d^3 q = \alpha \, d^3 \widetilde q$. The case $\alpha = 1$ corresponds to isotropic turbulence, while the typically observed case $\alpha < 1$ has power predominantly oriented in the perpendicular direction. In such a situation, the density perturbations are elongated along the field lines, so that the parallel length is larger than the perpendicular length: $q_\parallel^{-1} > q_\perp^{-1}$, i.e., the perpendicular wavenumber is larger than the parallel wavenumber: $q_\parallel < q_\perp$. Equation~\eqref{eq:d_ij_aniso_k}  explicitly shows that the diffusion tensor is proportional to the quantity $\overline{q \, \eps ^2}$.

\section{Angular Broadening of Extra-Solar Radio Sources}\label{extra-solar-source-broadening}

\begin{figure}[htb!]
  \centering
  \includegraphics[width=0.45\textwidth]{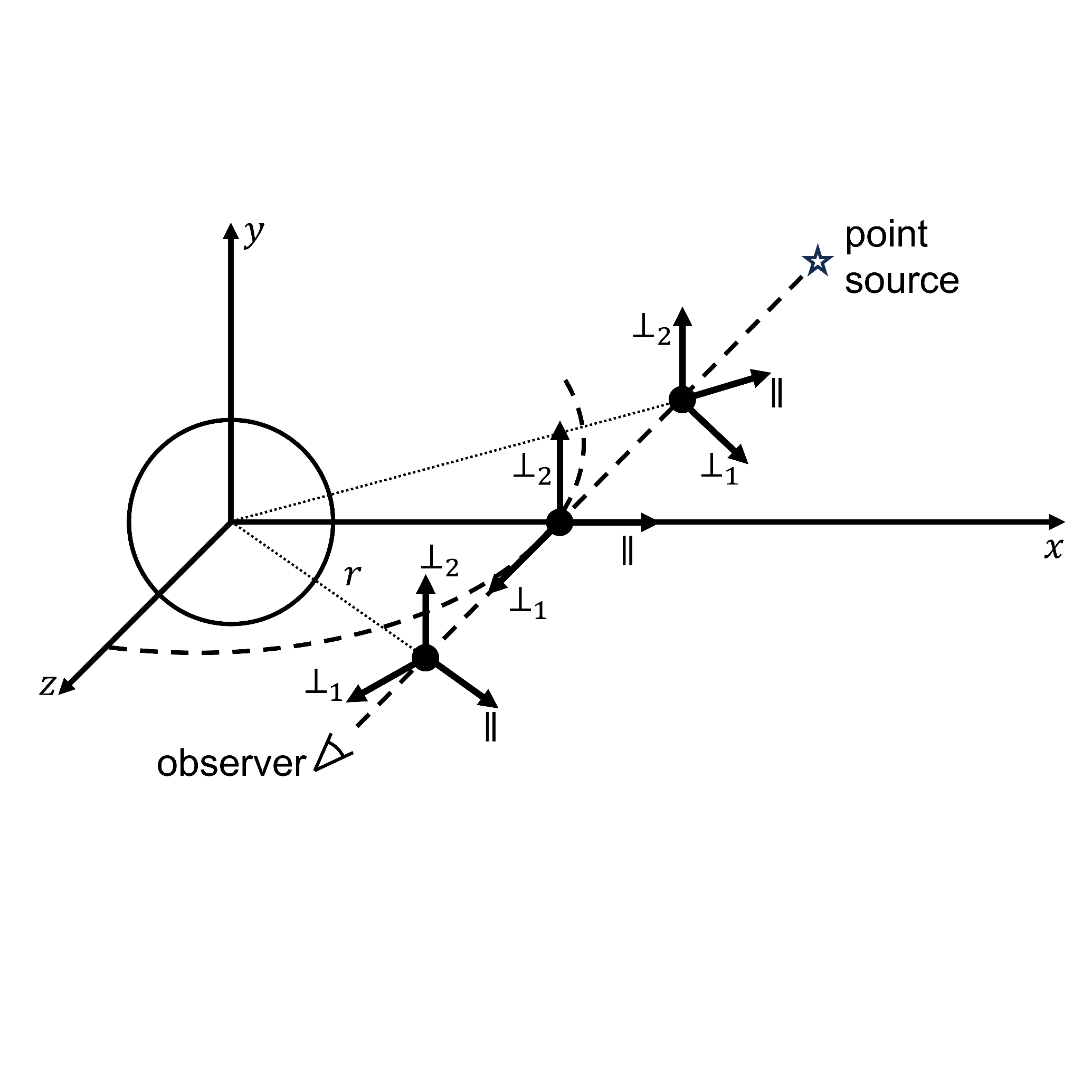}
  \includegraphics[width=0.49\textwidth]{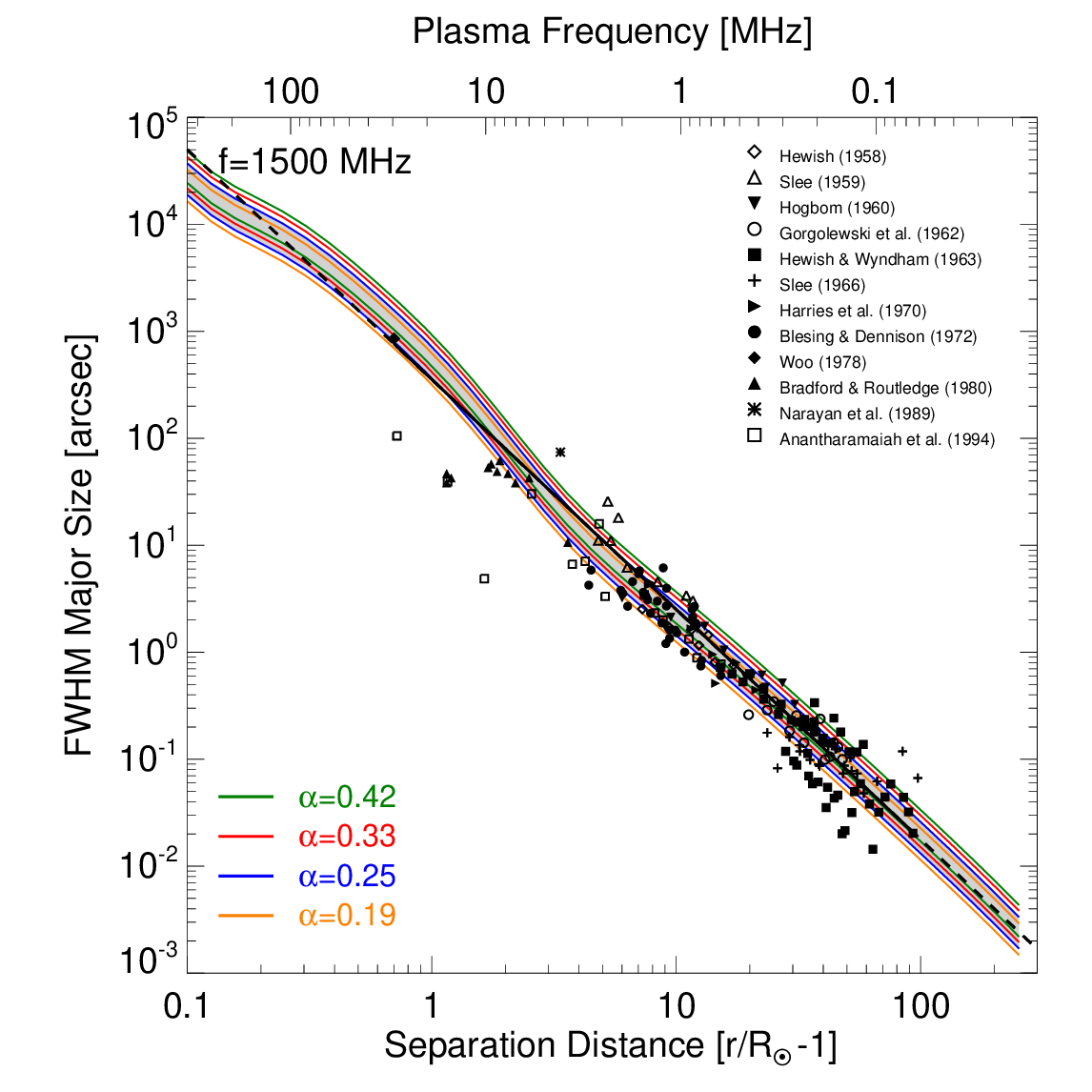}
  \caption{ \textit{Left:} Schematic showing the coordinate system used and its relation to the solar disk and the line of sight to an extra-solar point source with separation distance $r$. The broadening of point sources is calculated as an integral along $z$ in Appendix \ref{appendix:angular}. \textit{Right:} Observed FWHM major-axis source sizes of broadened point sources \citep{1958MNRAS.118..534H,1959AuJPh..12..134S,1960MNRAS.120..530H, 1962AcA....12..251G, 1963MNRAS.126..469H, 1966P&SS...14..255S, 1970PASA....1..319H, 1972PASA....2...84B, 1978ApJ...219..727W, 1980MNRAS.190P..73B, 1989MNRAS.241..403N, 1994JApA...15..387A}. The \cite{1958MNRAS.118..534H} data shows points where the detector baseline orientation was specified, and we take the larger of the measurements as the assumed major axis dimension. The solid black line shows the data fit $\theta=355(r/R_\odot-1)^{-2.15}$~arcsec, and extrapolated by the dashed line. The colored lines show the simulated data for various values of the anisotropy parameter $\alpha$, with the lower and upper bounds of each region denoting values of $\overline{q \, \eps ^2} \, R_\odot$ corresponding to $0.5 \times$ and $2 \times$ the nominal values from Equation~\eqref{eq:qbar}. Different observations are scaled to a frequency of $1.5$~GHz using a frequency dependence $f^{-2}$, as expected from Equation~\eqref{eq:dtheta2_perp}; see also Equation~\eqref{eq:FWHM_perp} in Appendix~\ref{appendix:angular}.}
  \label{fig:point_source_size}
\end{figure}

We consider the propagation of a radio wave from a distant extra-solar source with a heliocentric angular separation of $\simeq 0.27 \, r/R_\odot$~degrees, where  $r$ is the linear distance of closest approach of the ray path to the center of the Sun, and $0^o.27$ is the angular radius of the Sun (see Figure~\ref{fig:point_source_size}). We define the $q_\parallel$ direction as that aligned with the (assumed radial) solar magnetic field $\vec{B}$, $q_{\perp_1}$ as the perpendicular direction that is aligned with the wave propagation direction ${\hat z}$ at the point of closest approach, and $q_{\perp_2}$ as the perpendicular direction ${\hat y}$ on the plane of the sky that is perpendicular to ${\vec B}$ (Figure~\ref{fig:point_source_size}). Evidently, radio waves propagating from such a source will be affected by scattering in the turbulent solar atmosphere and, for emission frequencies $\omega$ that are much larger than the plasma frequencies $\omega_{pe}$ along the ray path though the heliosphere, this can be considered in the weak-scattering limit. In this limit, Equation~\eqref{eq:D} shows that the angular broadening rates per unit travel distance $v_{gr} \, dt = (c^2 \, k /\omega) \, dt\simeq c \, dt$, in the directions $\parallel$ (i.e., radial) and $\perp_2$ (i.e., perpendicular to the radial direction on the plane of the sky) can be written, using Equation~\eqref{eq:d_ij_aniso_k} \citep[see also][]{2019ApJ...884..122K}, as
\begin{equation}\label{eq:dtheta2_par}
  \frac{d \langle \theta ^2_{\parallel\parallel}\rangle}{c \, dt} =\frac{2 \, D_{\parallel\parallel}}{v_{\rm gr} \, k^2}
  =\frac{\pi}{8} \, \frac{\omega _{pe}^4}{\omega^4} \,\, \alpha^2 \, \overline{q \, \eps ^2}
\end{equation}
and
\begin{equation}\label{eq:dtheta2_perp}
  \frac{d \langle \theta ^2_{\perp_2\perp_2}\rangle}{c \, dt} =\frac{2 \, D_{\perp_2\perp_2}}{v_{\rm gr} \, k^2}
  =\frac{\pi}{8} \, \frac{\omega _{pe}^4}{\omega^4} \,\, \overline{q \, \eps ^2} \,\,\, ,
\end{equation}
where  $\vec{v}_{\rm gr} = \partial \omega/\partial {\vec k} = c^2 \, \vec{k}/\omega$ is the group velocity of the radio wave and we
have used the approximate (high-frequency) dispersion relation $\omega = c k$. Since the radio waves propagate along the $z$-direction, the broadening in $x$ and $y$ directions on the plane of the sky become integrals along the line of sight :
\begin{equation}\label{eq:theta2_perp}
  \langle \theta ^2_{yy}\rangle =\int _{los} \frac{d \langle \theta ^2_{\perp_2\perp_2}\rangle}{c \, dt} \, dz ,
\end{equation}
and
\begin{equation}\label{eq:theta2_par}
  \langle \theta ^2_{xx}\rangle = \int _{los}\left( \frac{d \langle \theta ^2_{\parallel\parallel}\rangle}{c \, dt}
  \cos^2 \chi  +\frac{d \langle \theta ^2_{\perp_2\perp_2}\rangle}{c \, dt}\sin^2 \chi \right) \, dz \,\,\, ,
\end{equation}
where $\chi(z)$ is the angle between the radial ($\parallel$) direction of the magnetic field and the $x$-axis (see Figure~\ref{fig:point_source_size}). When the largest contribution to $ \langle \theta ^2_{xx}\rangle $ comes from plasma near $z=0$ (a.k.a., the thin-screen approximation), $\langle \theta ^2_{xx}\rangle \simeq \alpha ^2\langle \theta ^2_{yy}\rangle$, so the source broadening in the $x$-direction is reduced by a factor $\alpha$ relative to that in the $y$-direction. Broadened sources are indeed observed to be elliptical
\citep[e.g.,][]{1972PASAu...2...86D,1994JApA...15..387A}, and this requires that the scattering occurs in both directions in the $x-y$ plane (rather than along $x$ or $y$ only), so that $S(\vec{q})$ should be evaluated with $q_{\perp_1} = 0$, i.e., with $\vec{q}$ lying in the $\perp_2-\parallel$ plane). We note that the observed form of source broadening is not consistent with a simple 2D-plus-slab model, in which $S(\vec{q}) = S_\perp \, \delta(q_\parallel) + S_\parallel \, \delta(\vec{q}_\perp)$ and the scattering tensor (Equation~\eqref{eq:D}) has a non-zero $\perp_2-\parallel$ component; such a model would result instead in ``cross-like,'' rather than the observed elliptical, apparent source structures.

For a spherically symmetric corona, the apparent rms source sizes in the perpendicular direction
(along the $y$-axis in the Figure~\ref{fig:point_source_size}),
observed at frequency $f = \omega/2 \pi$ (Hz), thus satisfy
\begin{equation}\label{eq:perp_diffusion}
  \langle \theta_{yy} ^2 \rangle^{1/2} \, (r) =\int _{los}\frac{d \langle \theta ^2_{\perp_2\perp_2}\rangle}{c \, dt} \, dz \propto \frac{1}{f^2} \, \left [ \int_r^\infty \overline{q \, \eps ^2}(r^\prime) \, n^2 \, (r^\prime) \, dr^\prime \right ]^{1/2} \,\,\, ,
\end{equation}
where the constant of proportionality
depends\footnote{See Appendix~\ref{appendix:angular} for a more formal
expression for Equation~\eqref{eq:perp_diffusion}.}
on the distance of closest approach $r$ of the ray to the Sun, and we have used the relation
$\omega_{pe} \propto n^{1/2}$. At radial distance $r$, the mean (spectrum-averaged)
wavenumber, $\overline{q \, \eps ^2}$,  at the position $r$ is given by substituting Equation~\eqref{eq:Sq_alpha} into Equation~\eqref{eq:Sq3d}:

\begin{equation}\label{eq_q_bar}
  \overline{q \, \eps ^2} \, R_\odot = R_\odot \, \int q \, S(q) \, \frac{d^3q}{(2\pi)^3}
  = \alpha \, R_\odot \, \int {\widetilde q} \, S(\widetilde q) \, \frac{d^3 \widetilde q}{(2\pi)^3} = \alpha \, \frac{4 \pi}{(2 \pi)^3} \, R_\odot \, \int_0^{\infty} {\widetilde q}^3 \, S(\widetilde q) \, d {\widetilde q} \,\,\, ,
\end{equation}
where we have multiplied $\overline{q \, \eps ^2}$ by the solar radius $R_\odot$ to produce a dimensionless quantity $\overline{q \, \eps ^2} \, R_\odot$. The angular broadening rates in both directions on the sky are thus, as expected from Equation~\eqref{eq:d_ij_aniso_k}, proportional to the square root of the key quantity $\overline{q \, \eps ^2} \, R_\odot$.

For a prescribed wave frequency $f$ and distance of closest approach $r$, Equation~\eqref{eq:perp_diffusion} shows that the source rms angular size and shape depend only on the profiles of plasma density $n$
(through the local plasma frequency $\omega _{pe}$) and the spectrum-weighted mean wavenumber of density fluctuations $\overline{q \, \eps ^2} \, R_\odot$,
at distances larger than the closest ray-approach distance. The turbulence profile in turn depends on the spectrum $S({\widetilde q})$ and its degree of anisotropy, characterized by the parameter $\alpha$ (see Equation~\eqref{eq:Sq_alpha}). Further, Equation~\eqref{eq:perp_diffusion} shows that the apparent angular extent of the source in the radial direction is smaller than the apparent extent in the perpendicular direction on the plane of the sky, by a factor $\alpha$. Hereafter we define the source ``size'' as the major axis, i.e., the FWHM source extent in the $\perp_2$ direction, perpendicular to the radial direction on the plane of sky.

If the density and turbulence profiles can be reasonably represented by power-law forms ($n(r) \sim r^{-\eta}$ and $\overline { q \, \eps^2} (r) \sim r^{-\zeta}$, respectively), then Equation~\eqref{eq:perp_diffusion} shows that $\langle \theta^2 \rangle^{1/2} \sim r^{-[\eta + (\zeta - 1)/2]}$. The data on the apparent sizes of extra-solar sources are presented in Figure~\ref{fig:point_source_size}; they show that $\langle \theta^2 \rangle^{1/2}  \propto r^{-2.15}$ for $r\gg R_\odot$. Thus, a radial density profile $n(r) \propto r^{-2.3}$ (valid at $r \gg R_\odot$; see Appendix~\ref{sec:plasma_params}), requires that $(\zeta-1)/2 = 2.15 - 2.3 = -0.15$; i.e., $\zeta = 0.7$.  We therefore adopt the nominal turbulence profile
\begin{equation}\label{eq:qbar}
  \overline{q \, \eps ^2}  \, R_\odot  = 2 \times 10^3 \, \alpha \, \left( 1 - \frac{R_\odot}{r} \right)^{2.7} \left ( \frac{R_\odot}{r} \right )^{0.7}   \,\,\, ,
\end{equation}
which is proportional to $\alpha$ (Equation~\eqref{eq_q_bar}) and also includes an additional factor $( 1 - {R_\odot}/{r})^{2.7} $ so that $\overline{q \, \eps ^2}$ decreases towards the solar surface as required by solar burst observations (see next section). The later is required mostly by $>20$~MHz solar radio burst observations discussed in the next section. The scaling factor $2 \times 10^3$ and the exponent $2.7$ in the additional term are chosen to best match the simulation results with the observational data in Figure~\ref{fig:point_source_size}.

\section{Solar Radio Bursts}\label{sec:simulations-of-solar-bursts}

Radio waves emitted by solar sources (e.g., Type~III bursts) suffer
scattering in the same turbulent environment that leads to the angular broadening of extra-solar sources. Because such bursts are produced by plasma processes at or near the plasma frequency $\omega_{pe}$ or its double \citep[see, e.g.,][]{1958SvA.....2..653G}, the scattering process is much stronger than for extra-solar sources with $\omega \gg \omega_{pe}$ and hence the weak-scattering treatment of the previous section is not applicable. Here, using a code developed by \citet{2019ApJ...884..122K}, we numerically simulate the propagation of radio waves in the presence of a prescribed density fluctuation profile $\overline{q \, \eps ^2} \,$ as function of $r$ and anisotropy parameter $\alpha$,
which we take\footnote{In principle, $\alpha$ could be treated as a function of $r$ in the simulations, but this would make the parameter space too large to study effectively.} to be a constant in our model. We performed simulations with $\overline{q \, \eps ^2} \, R_\odot$ scaled from its nominal value in Equation~\eqref{eq:qbar} by a factor in the range $[1/4, 4]$, and using anisotropy parameters $\alpha$ in the range $[0.19, 0.42]$. The results of these simulations lead to predictions of solar radio source properties (source size, apparent position, time profile) as seen from $1$~au, and are compared with observations to ascertain the veracity of the  $\overline{q \, \eps ^2} \, R_\odot$ profile and anisotropy parameter $\alpha$ used. For the simulations, we assume an initially isotropic distribution of photons emitted near the plasma frequency (fundamental), or twice the plasma frequency (harmonic). Assuming a spherically symmetric corona, we ran numerical simulations of the wave propagation, taking into account anisotropic scattering, large scale refraction, and free-free absorption, using profiles for the ambient density profile $n(r)$ given in
Appendix~\ref{sec:plasma_params}. We performed simulations for radio frequencies covering the range from $\sim$0.15~MHz to $\sim$300~MHz, thus encompassing ground-based and space-based (below the ionospheric cut-off) observations. We simulated burst decay times (Subsection~\ref{subsec:decay-time}), angular source sizes (Subsection~\ref{subsec:size}), and source positions (Subsection~\ref{subsec:positions}), which are then compared with pertinent observations to determine the spectrum-weighted mean wavenumber of density fluctuations $\overline{q \, \eps ^2} \, R_\odot$ as a function of $r$, and anisotropy parameter $\alpha$, that together best account for the observations.

\subsection{Decay Time}\label{subsec:decay-time}

\begin{figure*}
  \includegraphics[width=0.49\textwidth]{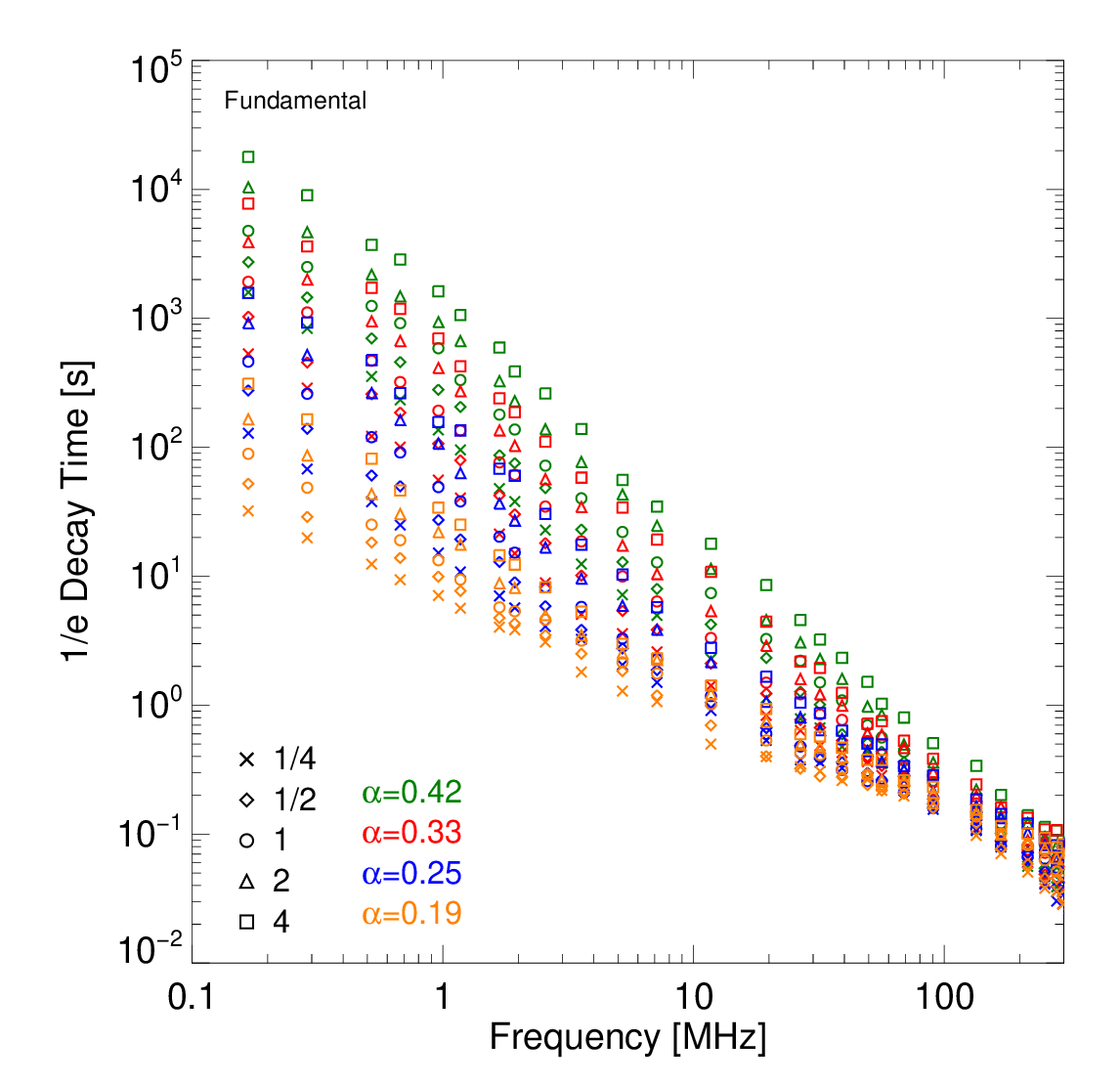}
  \includegraphics[width=0.49\textwidth]{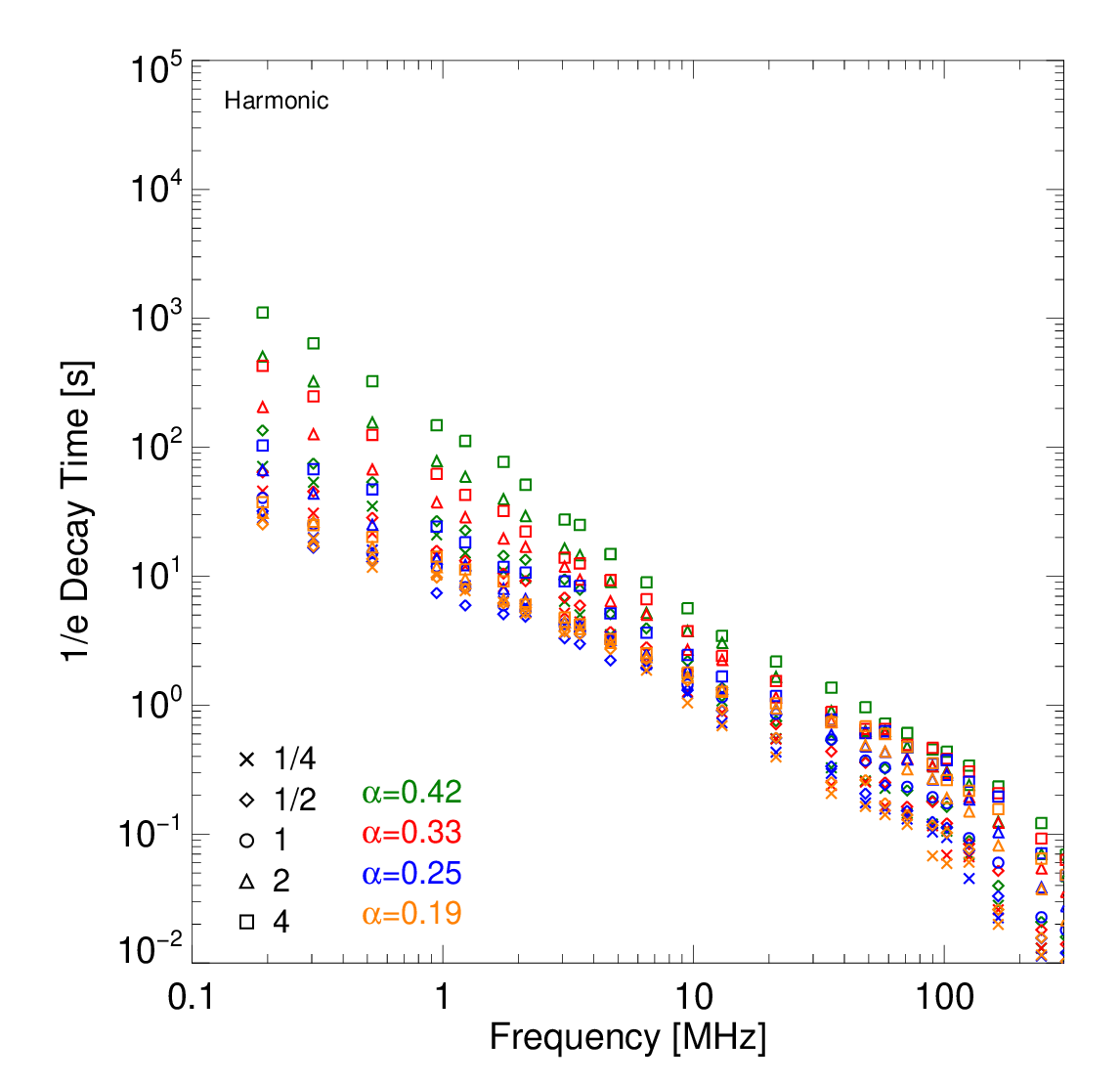}\\
  \includegraphics[width=0.49\textwidth]{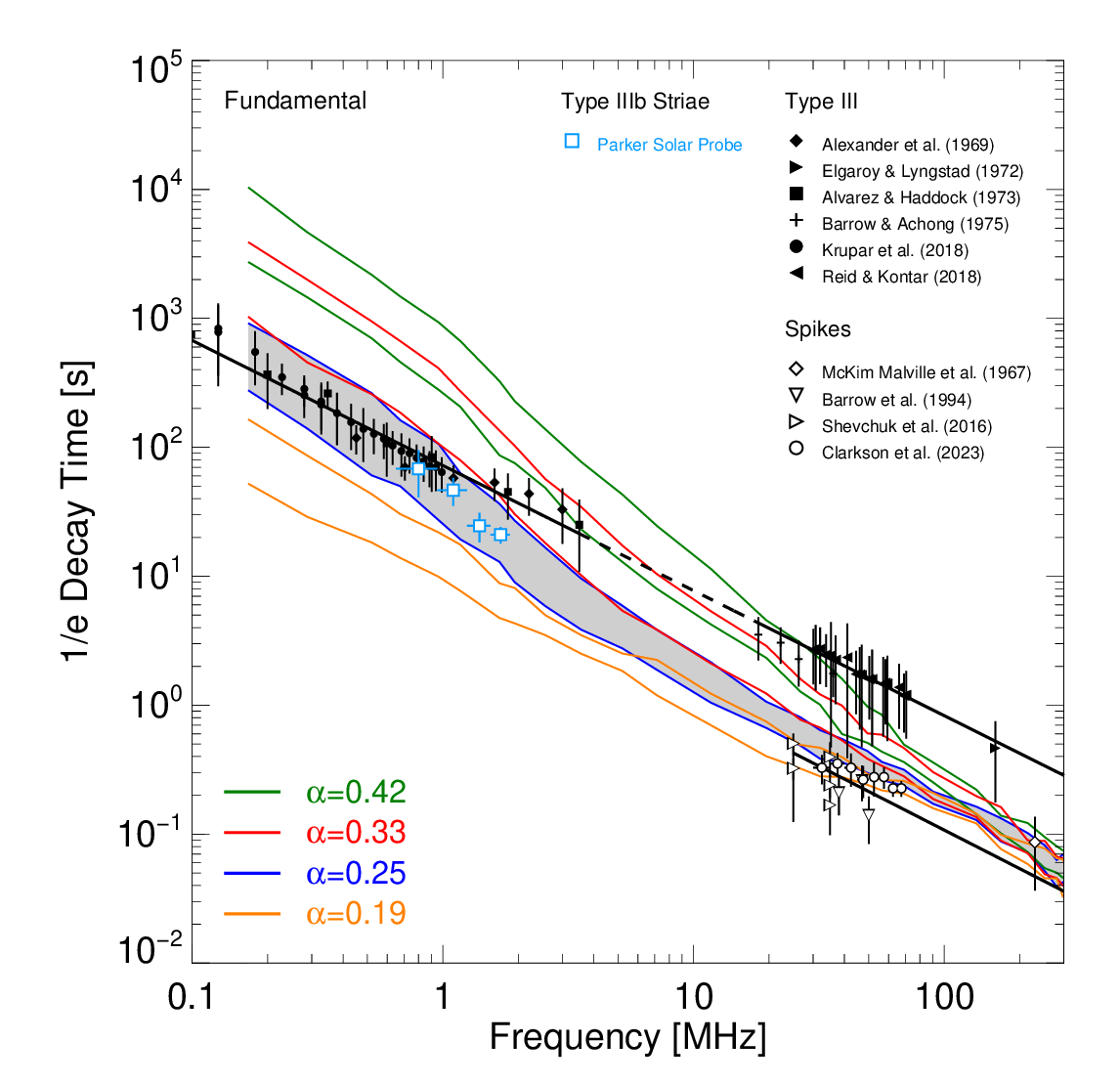}
  \includegraphics[width=0.49\textwidth]{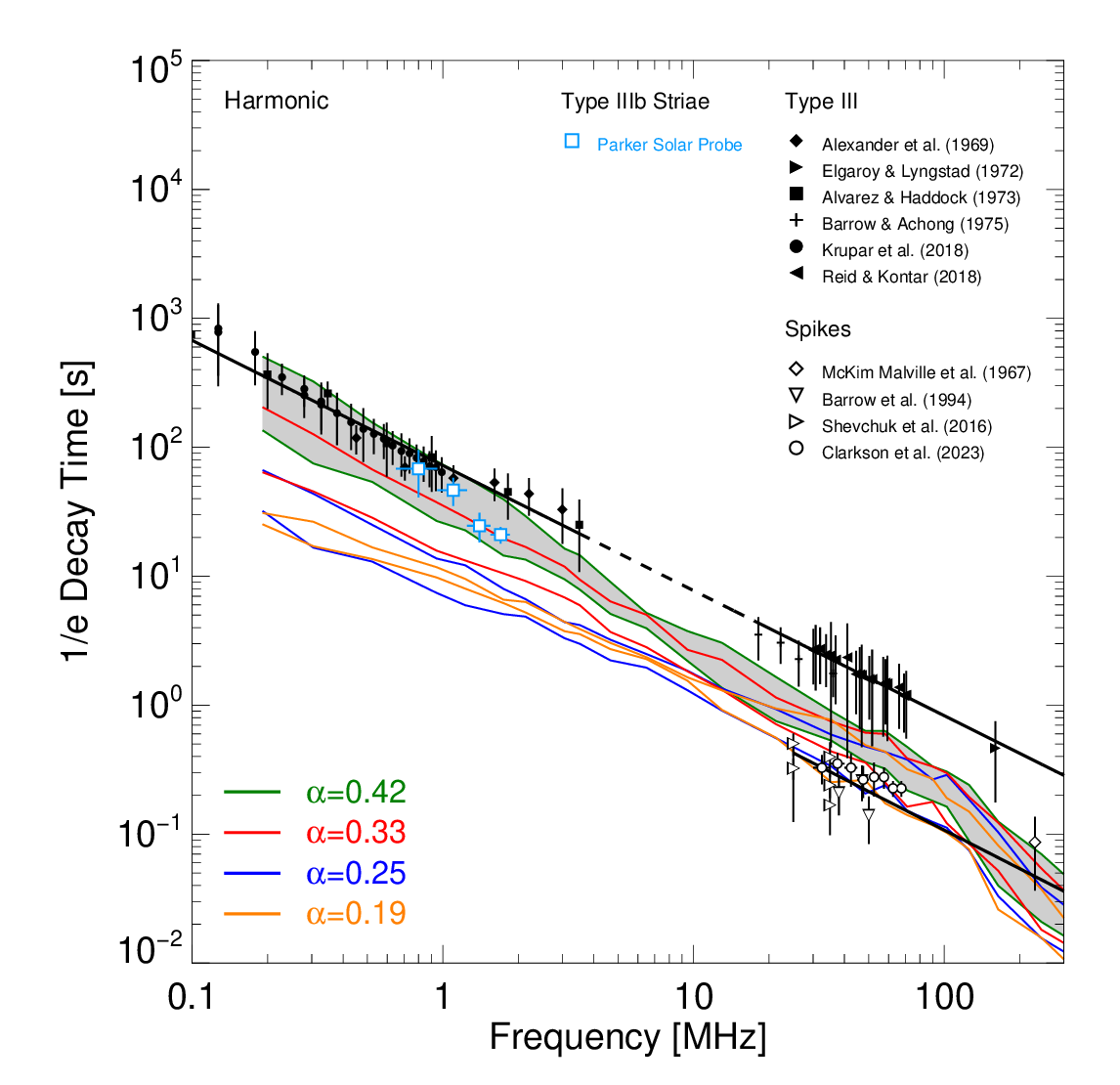}
  \caption{\textit{Top panels}: Simulated solar radio burst $1/e$ decay times versus frequency. The simulations were performed for turbulence profiles $\overline{q \, \eps ^2} \, R_\odot $ given by Equation~\eqref{eq:qbar} multiplied by factors $[1/4,1/2,1,2,4]$, and values of the anisotropy parameter $\alpha=[0.19,0.25,0.33,0.42]$, and for both fundamental (left panels) and harmonic emission (right panels). \textit{Bottom panels}: Observations of source decay times. The solid black points show measurements of Type~III decay times \citep{1969SoPh....8..388A,1972A&A....16....1E,1973SoPh...30..175A,
    1975SoPh...45..459B,2018ApJ...857...82K,2018A&A...614A..69R}, while open black points show spike decay times \citep{1967ApJ...147..711M,1994A&A...286..597B,2016SoPh..291..211S,2023ApJ...946...33C}. The solid black lines represent power-law fits to the spike data from Equation~(3) of \cite{2023ApJ...946...33C} and to the Type~III data from Equation~(51) of \cite{2019ApJ...884..122K}. The light blue points show Type~IIIb striae median decay times observed by Parker Solar Probe between $0.95$ and~$2.15$~MHz. The colored lines represent simulated decay times from anisotropic scattering simulations at the fundamental (left panel) and harmonic (right panel). The lower and upper bound of each region denotes values corresponding to turbulence profiles of $0.5\, \overline{q \, \eps ^2} \, R_\odot $ and $2 \, \overline{q \, \eps ^2} \, R_\odot $, respectively.}
  \label{fig:decay_time}
\end{figure*}

The top panels of Figure~\ref{fig:decay_time} show predicted $1/e$ decay times for the range of simulations discussed above, both for sources that emit at the fundamental (left panel) and for those that emit at the second harmonic (right panel).

The decay time of solar radio bursts has been observed both from space and from the ground, resulting in an extended set of decay-time observations. The bottom panels of Figure~\ref{fig:decay_time} compare observations of characteristic decay times \citep[expanding the collection of decay times presented by][]{2019ApJ...884..122K} with those obtained from our simulations, assuming emission either at the fundamental (left panel) or at the harmonic (right panel). The vertical bars in Figure~\ref{fig:decay_time} show the spread of measurements (where available) rather than measurement uncertainties. The decay profile is normally well-fit by an exponential form, with a generally well-measured characteristic decay time \citep[e.g.,][]{2018ApJ...857...82K,2020ApJS..246...57K} both below $1$~MHz and above $20$~MHz, with a gap in near-Earth data between a few~MHz (due to the limited time-resolution of space-based observations) and $\sim 10-20$~MHz (due to the ionospheric cut-off for ground-based observations). Although there is a noticeable spread in the observations, the grey area, representing a factor of $0.5 \times$ to $2 \times $ the nominal $\overline{q \, \eps^2} \, R_\odot $ profile of Equation~\eqref{eq:qbar}, includes all data below $3$~MHz and is found close to the observed decay times of the high-frequency spikes and Type~IIIb fine structures. This is convincing evidence that scattering effects account for the shortest observed decay times. The decay time of an average Type~III burst at $>10$~MHz is normally quite a bit larger, and is principally determined by the beam width and/or emission processes \citep[e.g.,][]{2019ApJ...885..140Z,2020ApJ...905...43C}. We note that the simulated decay time is dependent on whether the emission is fundamental or harmonic (Figure~\ref{fig:decay_time}). If all observations are due to harmonic emission, the simulations require $\alpha \simeq 0.4$, while the assumption of emission at the fundamental hypothesis requires a smaller anisotropy $\alpha\simeq 0.25$ to explain decay time observations.

\subsection{Angular Size}\label{subsec:size}

\begin{figure*}
  \includegraphics[width=0.49\textwidth]{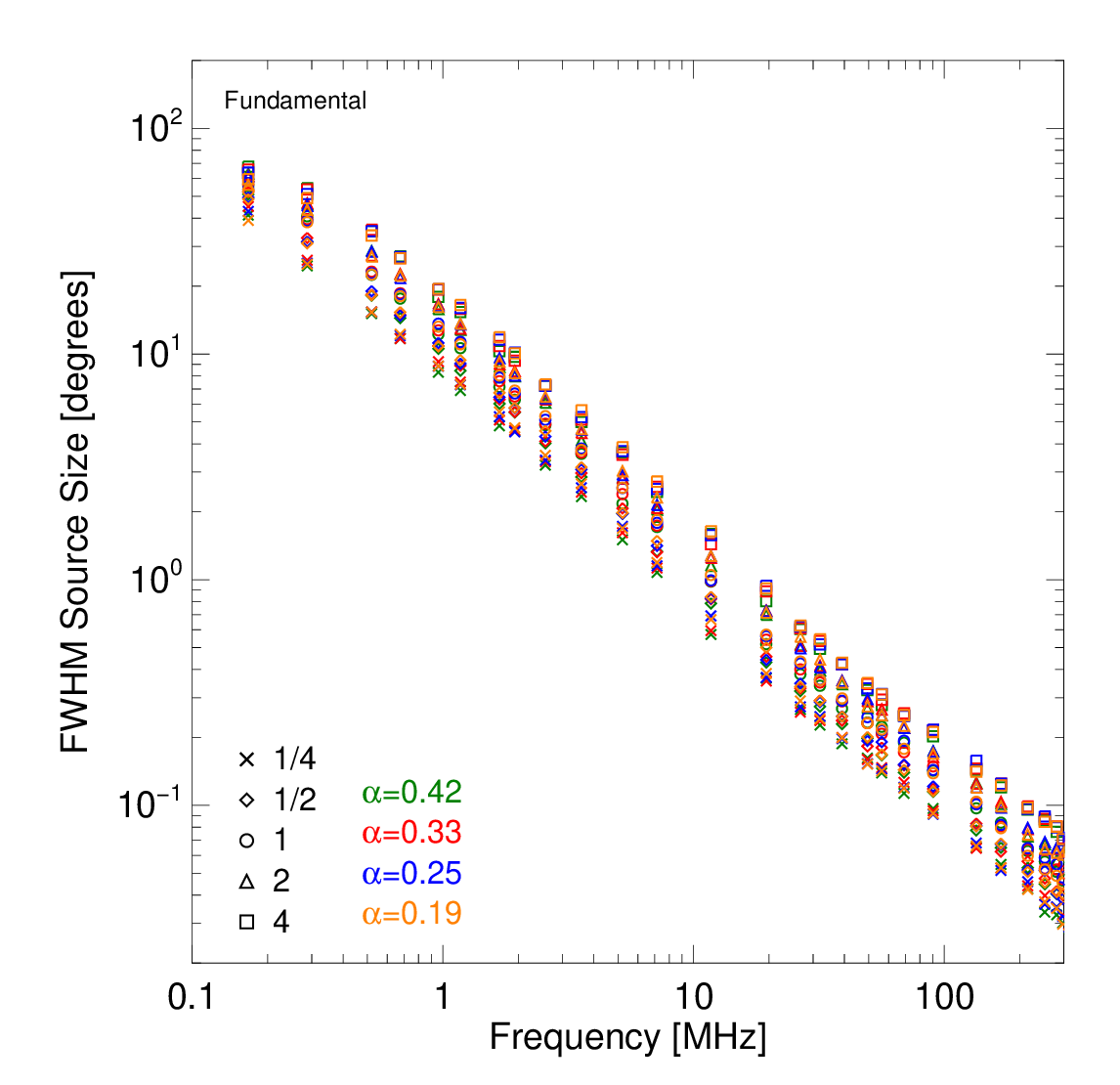}
  \includegraphics[width=0.49\textwidth]{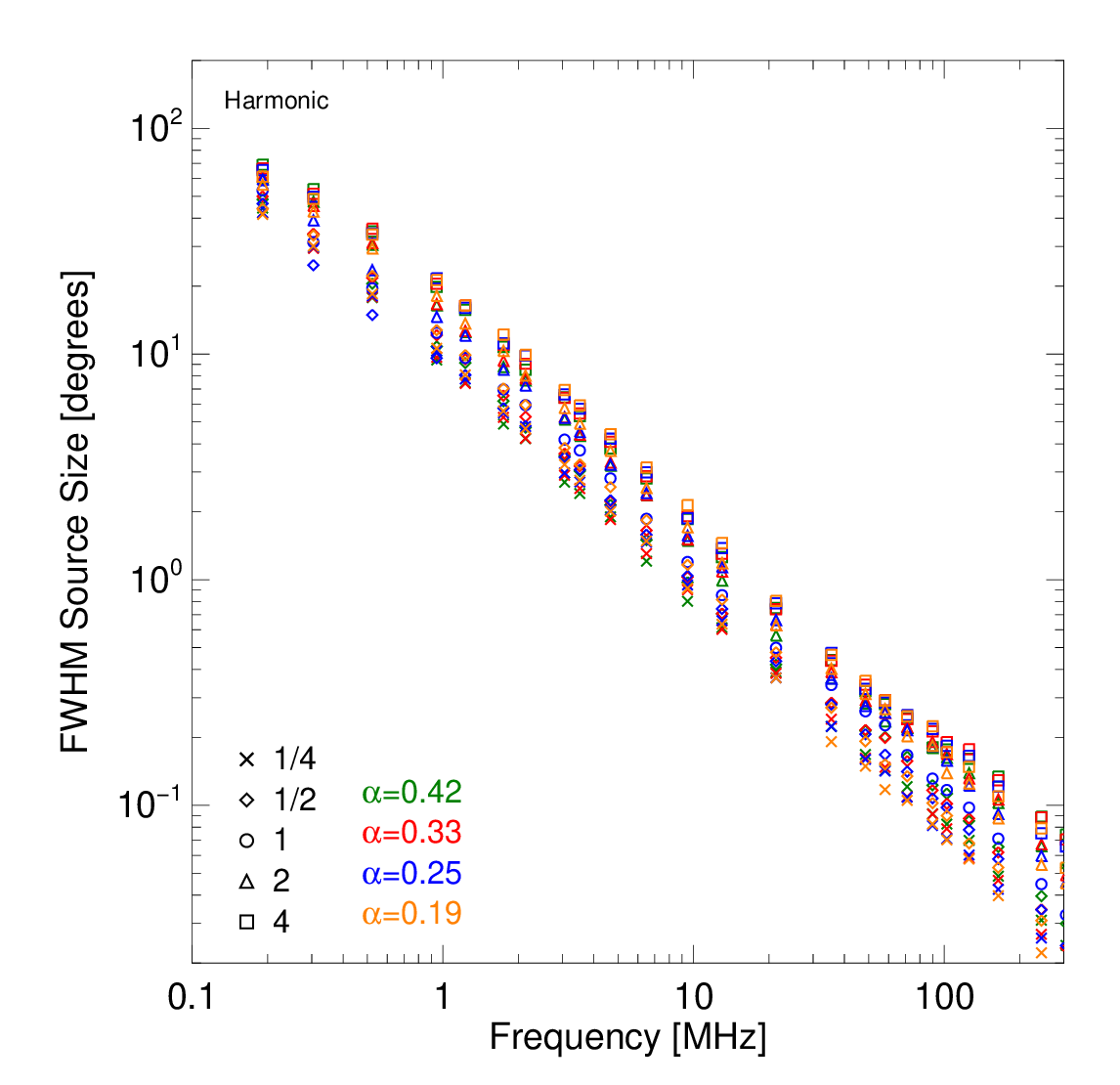}
  \includegraphics[width=0.49\textwidth]{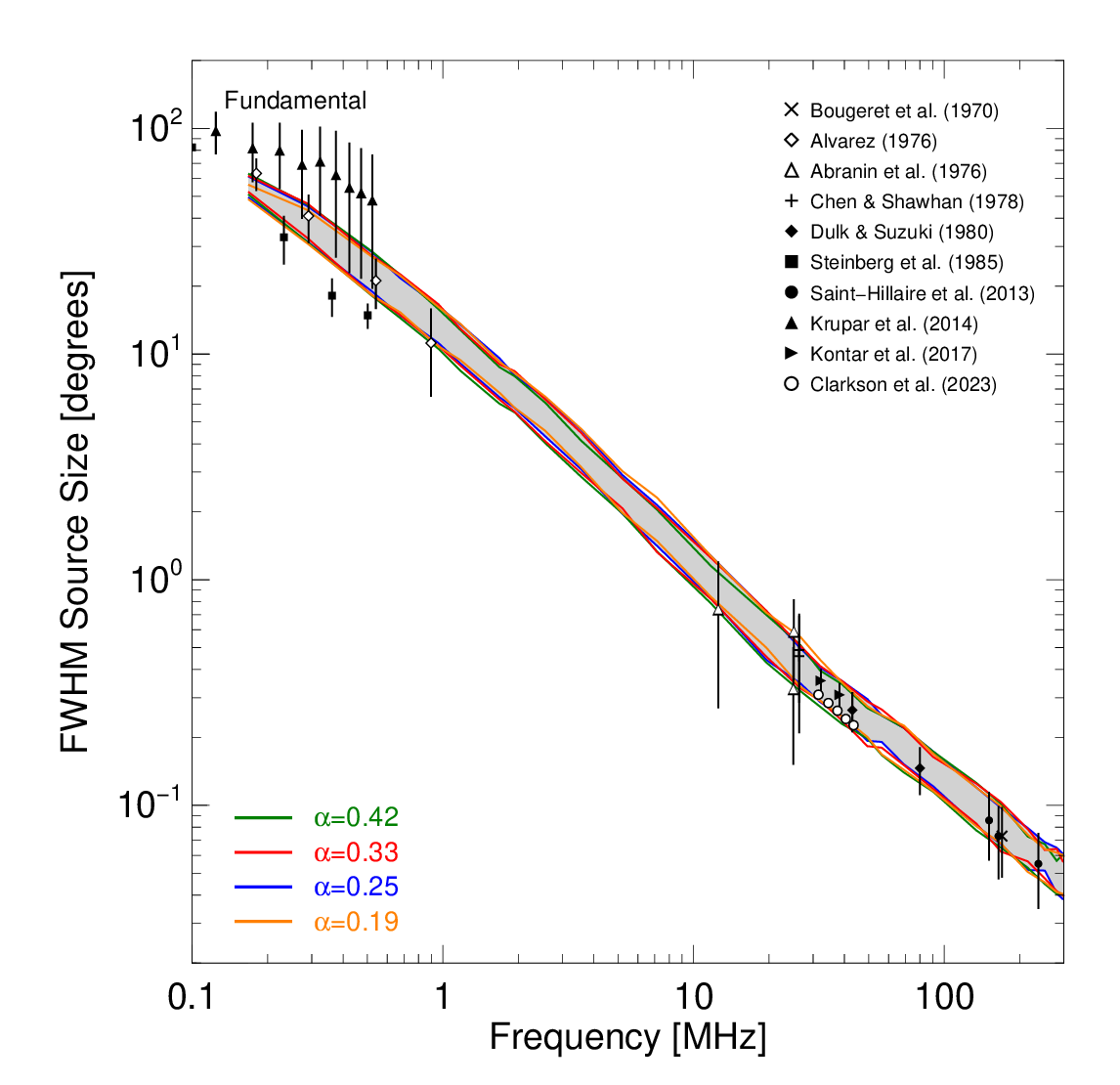}
  \includegraphics[width=0.49\textwidth]{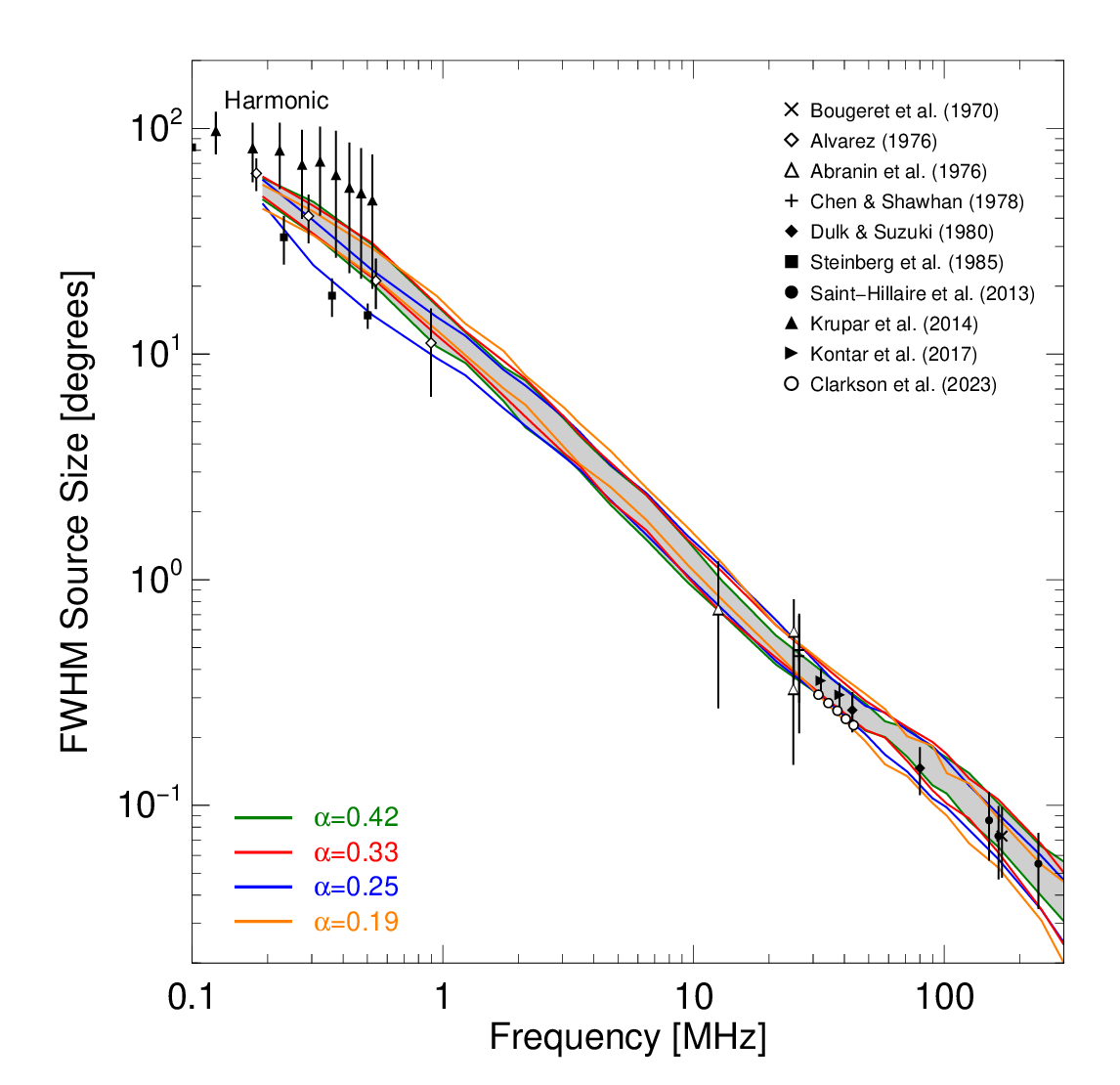}
  \caption{\textit{Top panels}: Simulated FWHM source sizes of Type~III bursts. The simulations were performed for turbulence profiles $\overline{q \, \eps ^2} \, R_\odot$ given by Equation~\eqref{eq:qbar}, multiplied by factors $[1/4,/1/2,1,2,4]$, for values of the anisotropy parameter $\alpha=[0.19,0.25,0.33,0.42]$, and for both fundamental (left panels) and harmonic emission (right panels). \textit{Bottom panels}: average FWHM size observations \citep{1970A&A.....6..406B,1976SvA....19..602A,1976SoPh...46..483A,1978SoPh...57..229A,1978SoPh...57..205C,1980A&A....88..203D,1985A&A...150..205S,2013ApJ...762...60S,2014SoPh..289.4633K,2017NatCo...8.1515K} and spikes \citep{2023ApJ...946...33C}.  Simulated sizes from anisotropic scattering simulations at the fundamental (left) and harmonic (right) are overlaid with colored lines at various values of $\alpha$, with turbulence levels ranging from $0.5 \, \overline{q \, \eps ^2} \, R_\odot$ (lower) to $2 \, \overline{q \, \eps ^2} \, R_\odot$ (upper), where the nominal $\overline{q \, \eps ^2} \, R_\odot $ profile is given by Equation~\eqref{eq:qbar}. The grey region shows the spread for $\alpha=0.25$, that is consistent with the observations.}
  \label{fig:source_size}
\end{figure*}

The top panels of Figure~\ref{fig:source_size} show predicted apparent source sizes for the range of simulations discussed above, both for sources that emit at the fundamental (left panel) and for those that emit at the second harmonic (right panel).

We have expanded the collection of Type~III solar radio burst angular extents initially presented by \cite{2019ApJ...884..122K}. Source sizes at frequencies above 10~MHz are evaluated using ground-based interferometers, while those below $1$~MHz are determined using indirect techniques \citep[e.g., spinning demodulation and/or goniopolarimetric techniques;][]{1976SoPh...46..483A,2008SSRv..136..487B,2014CRPhy..15..441C}. Similarly to the decay time measurements of Figure~\ref{fig:decay_time}, the bars in Figure~\ref{fig:source_size} show the spread of measurements (where available) rather than measurement uncertainties. The bottom panels of Figure~\ref{fig:source_size} show observed FWHM sizes (given as the FWHM semi-major axis as observed at $1$~au) compared with the predictions of the simulations, assuming emission either at the fundamental (left panel) and at the harmonic (right panel). The simulated source sizes are given in terms of their ${\rm FWHM}$ (for sources with emission frequencies below $0.15$~MHz, the source is so large that it extends all the way to the observer at $1$~au). One can see that the majority of observations are within the range $[1/2,2]$ times the nominal $\overline{q \, \eps^2} \, R_\odot $ profile of Equation~\eqref{eq:qbar}.

\bigskip

Publications of observations of Type~III burst decay times and/or source sizes rarely identify whether the observations correspond to emission at the fundamental or at the second harmonic, and we therefore conservatively include the possibility of either in our comparison of observed decay times and sizes with simulation results. With that in mind, it is interesting to note that both fundamental and harmonic sources have similar sizes over a wide range of frequencies. This is because the scattering surfaces that principally determine the source size are located at levels where both the plasma frequency and its double are less than the observed frequency, so that whether the emission is fundamental or harmonic is not a critical factor in determining the apparent source size. This also explains why harmonic and fundamental Type~III burst source sizes measured at the same frequency are nearly identical \citep{1980A&A....88..203D,2017NatCo...8.1515K,2023MNRAS.520.3117C}.

\subsection{Apparent Position}\label{subsec:positions}

\begin{figure*}
  \includegraphics[width=0.49\textwidth]{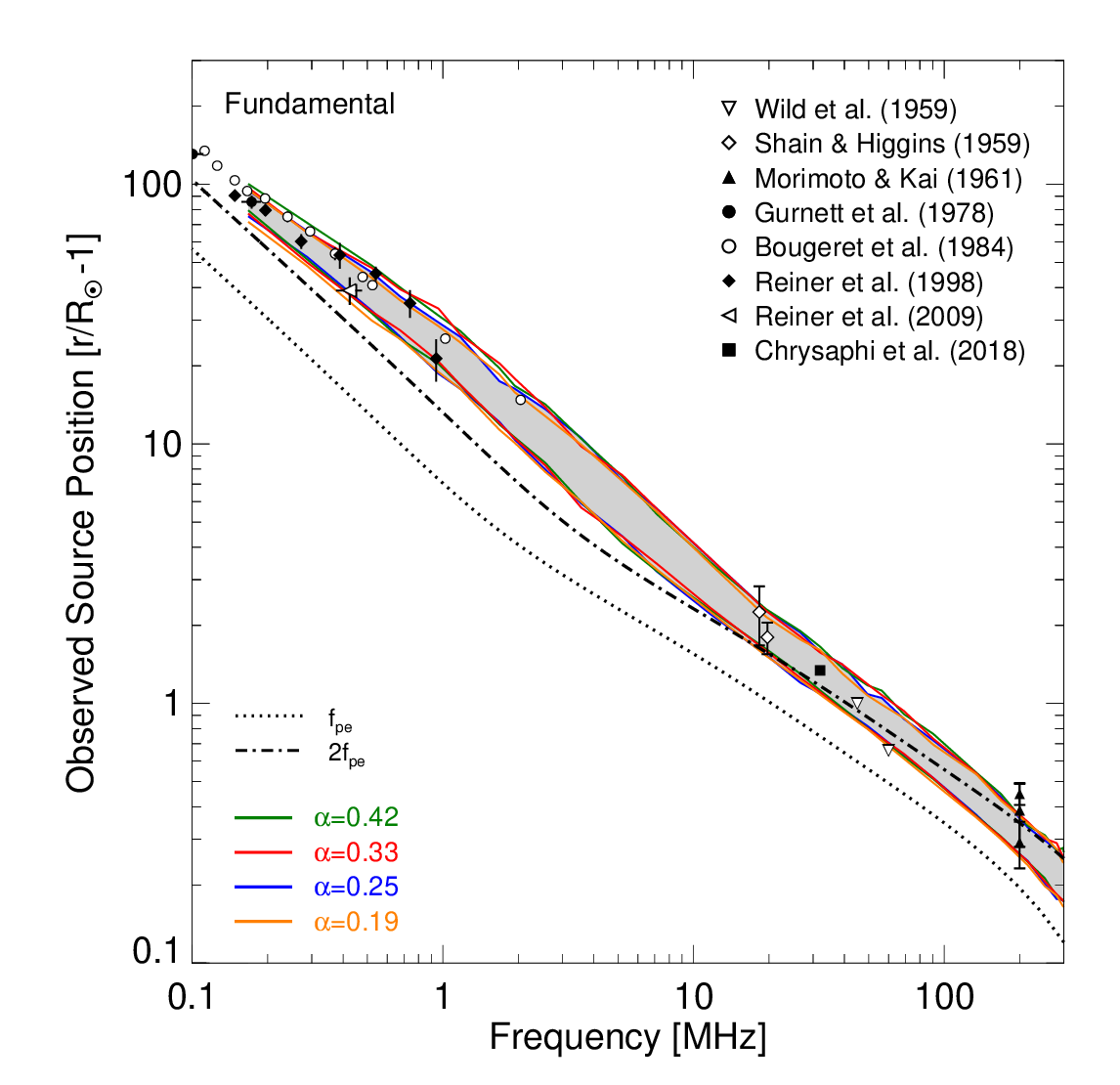}
  \includegraphics[width=0.49\textwidth]{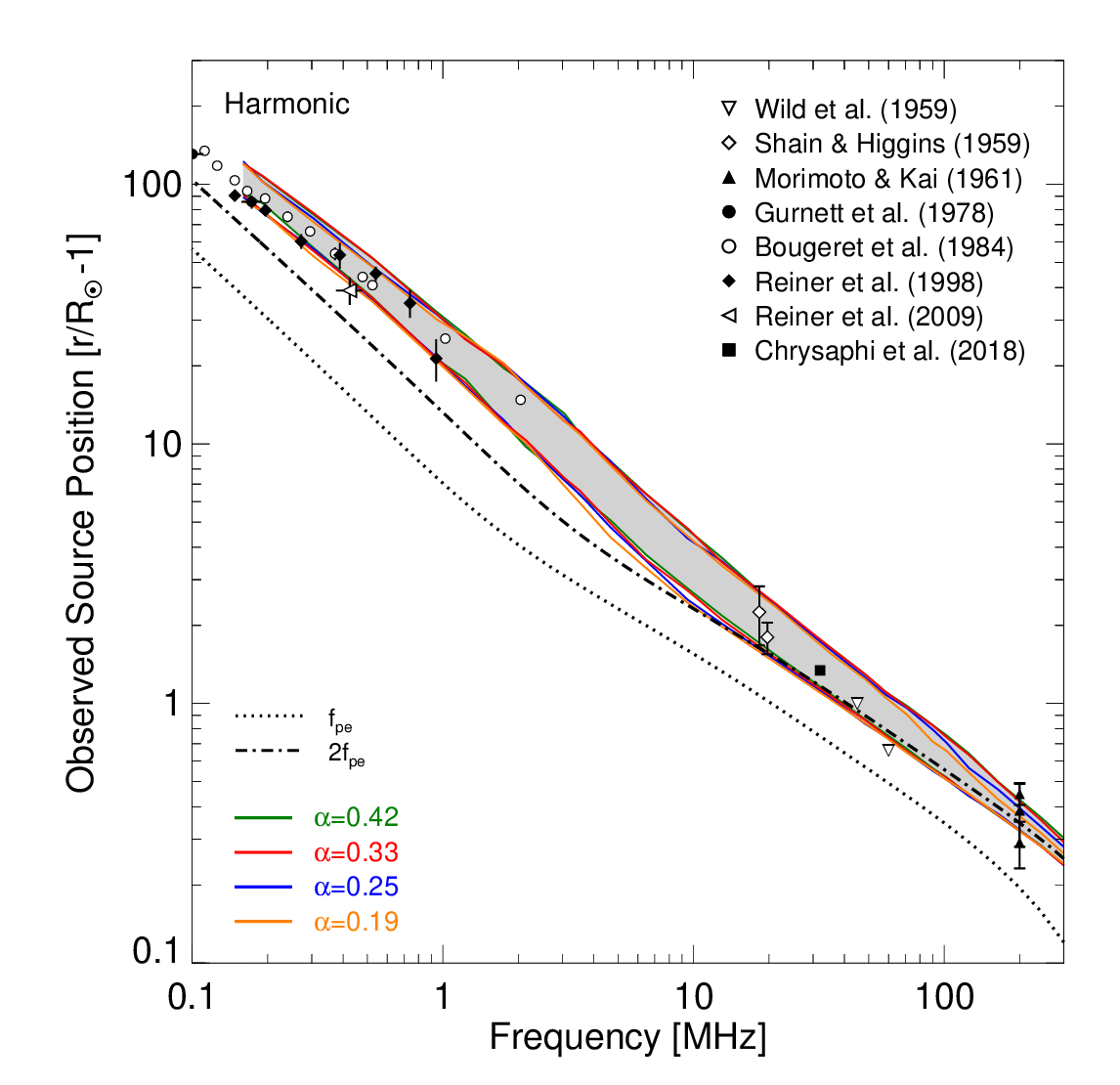}
  \caption{Observed source position versus frequency \citep{1959IAUS....9..176W, 1959AuJPh..12..357S, 1961PASJ...13..294M, 1978JGR....83..616G, 1984A&A...141...17B, 1998JGR...103.1923R, 2009SoPh..259..255R, 2018ApJ...868...79C}. The colored lines represent the observed positions from the simulations for the fundamental (left) and harmonic (right) at various values of the anisotropy parameter $\alpha$ and the turbulence profile $\overline{q \, \eps ^2} \, R_\odot $, multiplied by $1/4$ (lower bound) to $4$ (upper bound). The grey region shows the spread across values for an anisotropy parameter $\alpha=0.25$. The black dashed and dotted lines show the locations corresponding to the fundamental (plasma frequency) and harmonic (double plasma frequency), respectively.}
  \label{fig:obs_source_pos}
\end{figure*}

Since the early observations of Type~III bursts, it was noted \citep[e.g.,][]{1959AuJPh..12..357S,1974SSRv...16..145F} that the source positions are not coincident with the positions in the solar atmosphere and solar wind with densities corresponding to emission at the plasma frequency (or its double); equivalently, the observed burst frequencies required a higher plasma density than that is normally
observed by optical telescopes \cite[e.g.,][]{1974SSRv...16..145F}. However, as Figure~\ref{fig:obs_source_pos} highlights, radio wave scattering substantially shifts the burst emission away from the Sun, so that the density inferred by associating the burst emission with the plasma frequency (or its double) could be 4-5 times higher than the density at the apparent burst position \citep[e.g.,][]{2018ApJ...868...79C}. Figure~\ref{fig:obs_source_pos} shows the variation of observed source positions versus frequency, compared to the predictions of our simulations under the assumptions that the emission is at the plasma frequency (fundamental mode; left panel) or its second harmonic (right panel). We see that the radio-wave scattering reproduces the observed positions over a wide range of distances from the low corona into the solar wind, as shown by the grey band; only a relatively small spread (factor of 4) in the density fluctuation quantity $\overline{q \, \eps ^2} \, R_\odot $ is required to explain the available observations.

These simulations also highlight that, similar to source sizes (but unlike burst decay times), the simulated apparent source positions are relatively independent of the assumption of whether fundamental or harmonic emission is involved. If there were no scattering, at the same observation frequency, sources corresponding to emission at the harmonic should be further outward than sources corresponding to emission at the fundamental. However, in the presence of strong scattering (Figure~\ref{fig:obs_source_pos}), the apparent source positions are significantly further away from the Sun than the locations expected for either emission at the harmonic mode or at the fundamental. This effect is noticeable for all except the highest emission frequencies $f>100$~MHz, where the apparent source position is close to the location expected from emission at the harmonic. This shows clearly that measured source locations are dominated by scattering effects and succinctly explains the observations reported by \citet[][]{1972PASA....2..100S}, showing that the radial distances for fundamental and harmonic components at $80$~MHz are, on average, equal.

\subsection{Remarks}

\begin{figure}[!htb]
  \centering
  \includegraphics[width=0.49\textwidth]{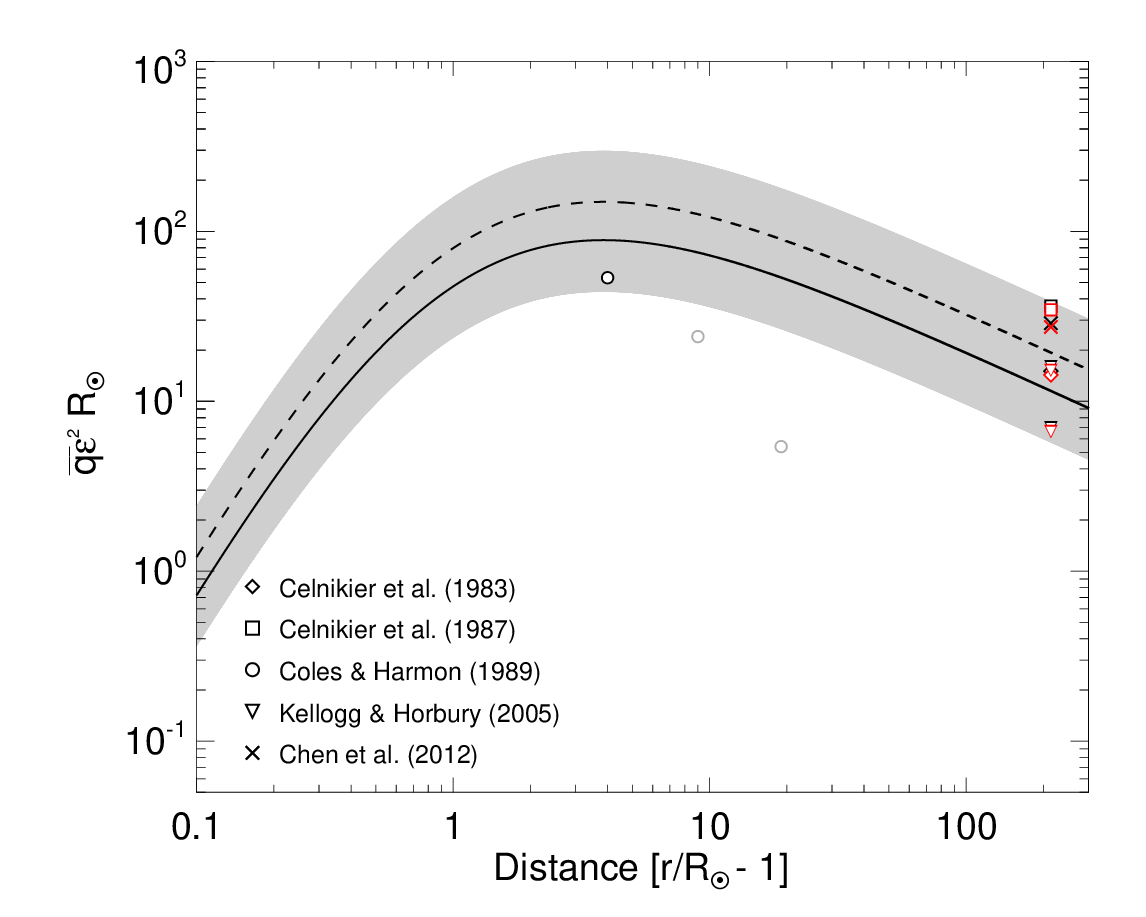}
  \includegraphics[width=0.49\textwidth]{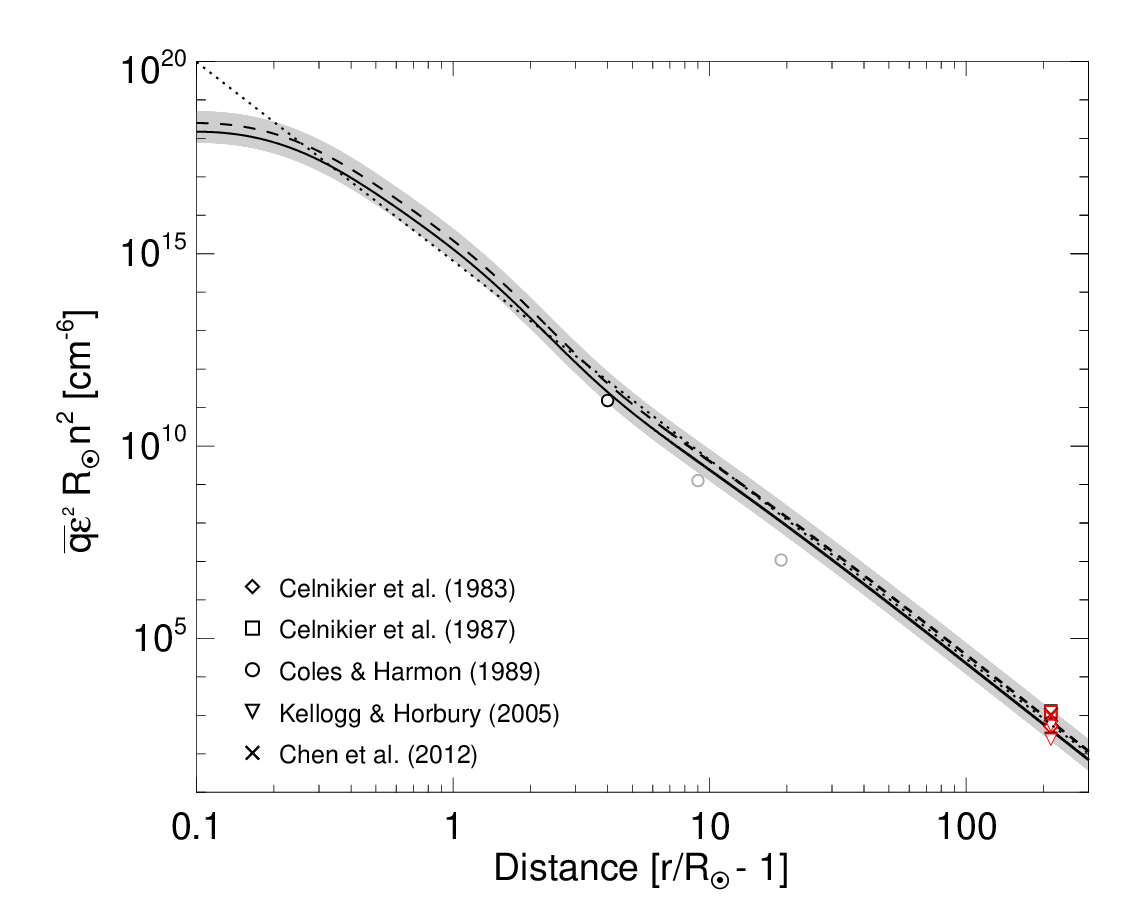}
  \caption{ \textit{Left:} Spectrum-averaged wavenumber $\overline{q \, \eps^2} \, R_\odot$ as a function of heliocentric distance $r$, as given by Equation~\eqref{eq:qbar}, for fundamental emission with an anisotropy factor $\alpha =0.25$ (solid), and for harmonic emission with $\alpha = 0.42$ (dashed). The grey region denotes the range from $[1/2,2] \times \overline{q \, \eps^2} \, R_\odot $. The data from \cite{1989ApJ...337.1023C} are retrieved from the density spectrum in their Figure 4, with a density profile $n(r)$ taken from equation (\ref{eq:density}). The light grey points are not well constrained due to a lack of observations at the (dominant; see Section~\ref{sec:methodology}) inner scale wavenumber $q_i$. The in-situ density fluctuations data at $1$~au \citep{1983A&A...126..293C, 1987A&A...181..138C, 2005AnGeo..23.3765K, 2012PhRvL.109c5001C} are retrieved using Equation~\eqref{fp-int-45}, with the black points corresponding to $\alpha=0.25$ (fundamental) and the red points corresponding to $\alpha=0.42$ (harmonic). The data for \cite{1987A&A...181..138C} are taken from their Figure~7. We use the \cite{2005AnGeo..23.3765K} spectra for densities (5--6)~cm$^{-3}$ and (10--15)~cm$^{-3}$. \textit{Right:}  As in the left panel but multiplied by $n^2$, with $n(r)$ given by Equation~\eqref{eq:density}. The dotted line shows the relation $\overline{q \, \epsilon^2} \, R_\odot  \, n^2 (r) = 6.5 \times10^{14} \, (r/R_\odot - 1)^{-5.17}$.}
  \label{fig:qeps2}
\end{figure}

The observations of Type~III burst decay time, source size and position (Figures~\ref{fig:decay_time} through~\ref{fig:obs_source_pos}) show a noticeable data gap at frequencies between $\sim$5~MHz and $\sim$15~MHz. Observations below the ionospheric cut-off are made from space, while observations above $10-15$~MHz are ground-based. Parker Solar Probe \citep[PSP;][]{2016SSRv..204....7F} has sufficiently high frequency resolution to resolve Type~IIIb burst striae \citep{2020ApJS..246...49P}. Our analysis of $10$~Type~IIIb bursts ($58$~striae from April 2-9, 2019) provides interesting measurements (blue symbols in Figure~\ref{fig:decay_time}), suggesting that Type~IIIb  and Type~III bursts have nearly identical decay times below $1$~MHz, but the fine structures in Type~IIIb bursts decay faster at higher frequencies. This suggests that the Type~III decay is purely due to scattering below $1$~MHz and that electron transport is of relevance at higher frequencies. In other words, the scattering of Type~III radio bursts is so large below $1$~MHz so that it dominates the decay time.

The profiles of $\overline{q \, \eps ^2} \, R_\odot $ (see Figure~\ref{fig:qeps2}) necessary to reproduce the observations of the source size, decay time and source position are consistent with an anisotropic scattering model with $\alpha \simeq 0.25$ (assuming emission at the fundamental) and $\alpha \simeq 0.42$ (assuming purely harmonic emission). We therefore conclude that turbulent density fluctuations, with a magnitude profile consistent with Equation~\eqref{eq:qbar} and an anisotropy parameter $0.2 \lapprox \alpha \lapprox 0.4$, are present in the solar corona and heliosphere.

\section{In-situ observations of density fluctuations}\label{sec:in-situ}

Density fluctuations are often measured in the solar wind
\citep[e.g.,][]{1983A&A...126..293C,1990JGR....9511945M} and can be compared to our density turbulence model by converting the wavenumber spectrum into the frequency power spectrum $P(f)$ (cm$^{-6}$~Hz$^{-1}$) measured by a spacecraft as the solar wind, with velocity ${\vec V}_{\rm SW}$, advects the turbulent fluctuations past the measuring instrumentation. $P(f)$ is related to the wavenumber spectrum $S(\vec q)$ of density fluctuations by \citep[e.g.,][]{1976JGR....81.5591F}:

\begin{equation}\label{pf-general}
  P(f) = n^2 \, \int S ({\vec q}) \, \delta \left ( f- \frac{ {\vec q} \cdot {\vec V}_{\rm SW} }{2 \pi}  \right ) \, \frac{d^3q}{(2 \pi)^3} \,\,\, .
\end{equation}
In Appendix~\ref{appendix:spectral-forms}, we evaluate $P(f)$ for several illustrative forms (both isotropic and anisotropic) of $S({\vec q})$. Here we explore the reverse problem of determining the value of $\overline{q \, \eps^2} \, R_\odot$ from in-situ observations of $P(f)$. We first transform the integration variable from ${\vec q} = (q_{\perp 1}, q_{\perp 2}, q_\parallel)$ to $\widetilde {\vec q} = (q_{\perp 1}, q_{\perp 2}, q_\parallel/\alpha)$ and define $\widetilde{\vec V}_{\rm SW} = (V_{{\rm SW} \, \perp \, 1}, V_{{\rm SW} \, \perp \, 2}, \alpha V_{{\rm SW} \, \parallel})$, so that $\widetilde {\vec q} \cdot \widetilde {\vec V}_{\rm SW} = {\vec q} \cdot {\vec V}_{\rm SW}$ and $d^3 q = \alpha \, d^3 {\widetilde q}$. Then, since $S(\widetilde {\vec q})$ is (by construction\footnote{The analysis of Equations~\eqref{pf-general_2a} through~\eqref{eq_q_bar1} thus applies for the isotropic spectrum $P(f)$ of Appendix~\ref{appendix:isotropic_sq} and even for the anisotropic spectrum of Appendix~\ref{sec:anis_integral} (which can be scaled into an isotropic form), but not for the spectrum of Appendix~\ref{appendix:constant_q}, which cannot be transformed into an isotropic form.}) isotropic,

\begin{equation}\label{pf-general_2a}
  P(f) = \frac{\alpha \, n^2}{(2\pi)^2} \, \int_{{\widetilde q}_\perp=0}^\infty  \, \int_{{\widetilde q}_\parallel=-\infty}^\infty \, S \left ( \sqrt{{\widetilde q}_\perp^2 + {\widetilde q}_\parallel^2} \right ) \, \delta \left ( \frac{ {\widetilde q}_\parallel \, {\widetilde V}_{\rm SW}} {2 \pi} - f \right ) \, {\widetilde q}_\perp \, d{\widetilde q}_\perp\, d{\widetilde q}_\parallel \,\,\, ,
\end{equation}
this can be written as

\begin{eqnarray}\label{pf-general_2b}
  P(f) &=& \frac{\alpha \, n^2}{2\pi \, {\widetilde V}_{\rm SW}} \int_{{\widetilde q}_\perp=0}^\infty  \, \int_{{\widetilde q}_\parallel = -\infty}^\infty \, S \left ( \sqrt{{\widetilde q}_\perp^2 + {\widetilde q}_\parallel^2} \right ) \, \delta \left ( {\widetilde q}_\parallel - \frac{ 2 \pi f }{{\widetilde V}_{\rm SW}} \right ) \, {\widetilde q}_\perp \, d{\widetilde q}_\perp\, d{\widetilde q}_\parallel \cr
  &=& \frac{\alpha \, n^2}{2\pi \, {\widetilde V}_{\rm SW}} \int_{{\widetilde q}_\perp=0}^\infty \, S \left ( \, \left [ {\widetilde q}_\perp^2 + \left ( \frac{2 \pi f}{{\widetilde V}_{\rm SW}} \right )^2 \right ]^{1/2} \, \right ) \, {\widetilde q}_\perp \, d{\widetilde q}_\perp \cr
  &=& \frac{\alpha \, n^2}{2\pi \, {\widetilde V}_{\rm SW}} \int_{{\widetilde q} = 2 \pi f/{\widetilde V}_{\rm SW}}^\infty \, {\widetilde q} \, S ({\widetilde q}) \, d{\widetilde q} \,\,\, ,
\end{eqnarray}
where in the last equality we have used the substitution ${\widetilde q}^2 = {\widetilde q}_\perp^2 +  (2 \pi f/{\widetilde V}_{\rm SW})^2$.

We can now construct the moments of the frequency spectrum of the observed density fluctuations:

\begin{equation}\label{pf-int1}
  \int_0^{\infty} f^m \, P(f) \, df = \frac{\alpha \, n^2}{2\pi \, {\widetilde V}_{\rm SW}} \, \int_{f=0}^\infty f^m \, df \int_{{\widetilde q} = 2 \pi f/{\widetilde V}_{\rm SW}}^\infty  {\widetilde q} \, S ({\widetilde q}) \, d{\widetilde q} \,\,\, ,
\end{equation}
and reversing the order of integration gives

\begin{equation}\label{pf-int2}
  \int_0^{\infty} f^m \, P(f) \, df = \frac{\alpha \, n^2}{2\pi \, {\widetilde V}_{\rm SW}} \, \int_{{\widetilde q}=0}^\infty {\widetilde q} \, S ({\widetilde q}) \, d{\widetilde q} \, \int_0^{{\widetilde q} \, {\widetilde V}_{\rm SW}/2 \pi } f^m \, df \,\,\, .
\end{equation}
Two moments are of particular interest. First, for $m=0$,

\begin{equation}\label{pf-int3}
  \int_0^{\infty}  P(f) \, df = \frac{\alpha \, n^2}{(2\pi)^2} \, \int_{{\widetilde q}=0}^\infty {\widetilde q}^2 \, S ({\widetilde q}) \, d{\widetilde q}  = \alpha \, \frac{\langle \delta n^2 \rangle}{2} \,\,\, ,
\end{equation}
where we have used Equation~\eqref{eq:Sq3d}. The factor of 2 arises because only positive frequencies are considered (e.g., for an isotropic distribution, $\langle \delta n^2 \rangle = 2 \int_o^\infty P(f) \, df = \int_{-\infty}^\infty P(f) \, df  $). Second, for $m=1$,

\begin{equation}\label{pf-int4}
  \int_0^{\infty} f \, P(f) \, df = \frac{\alpha \, n^2 \, {\widetilde V}_{\rm SW}}{2 \, (2\pi)^3} \, \int_{{\widetilde q}=0}^\infty {\widetilde q}^3 \, S ({\widetilde q}) \, d{\widetilde q}  \,\,\, ,
\end{equation}
and comparing this to the expression~\eqref{eq_q_bar} for $\overline{q \, \eps^2} \, R_\odot$, we immediately see that

\begin{equation}\label{eq_q_bar1}
  \overline{q \, \eps ^2} \, R_\odot = 4 \, \frac{2 \pi}{n^2 {\widetilde V}_{\rm SW}} \, R_\odot \, \int_0^{\infty} f \, P(f) \, df \,\,\, .
\end{equation}
Now, ${\widetilde V}_{\rm SW} = V_{\rm SW} \, (\sin^2 \theta_B + \alpha^2 \cos^2 \theta_B)^{1/2}$, where $\theta_B$ is the angle between the solar wind velocity and the (axis of symmetry) magnetic field. Thus, if $\theta_B=0$,

\begin{equation}\label{fp-int}
  \overline{q \, \eps^2} \, R_\odot \vert_{\theta_B = 0} = \frac{4}{\alpha}\, \frac{2\pi}{n^2 \, V_{\rm SW}} \, R_\odot \, \int_{0}^{\infty} f \, P(f) \, df \,\,\, ,
\end{equation}
while, for a typical $1$~au angle $\theta _B \simeq 45^o$,

\begin{equation}\label{fp-int-45}
  \overline{q \, \eps^2} \, R_\odot \vert_{\theta_B = 45^\circ} = 4 \, \sqrt{\frac{2}{1+\alpha^2}} \, \frac{2\pi}{n^2 \, V_{\rm SW}} \, R_\odot \, \int_{0}^{\infty} f \, P(f) \, df \,\,\, .
\end{equation}
The frequency-weighted integral over the measured fluctuation power spectrum $P(f)$ at $1$~au  can be readily evaluated numerically using the data from the spacecraft measurements of \cite{1983A&A...126..293C}, \cite{1987A&A...181..138C}, \cite{2005AnGeo..23.3765K}, and \cite{2012PhRvL.109c5001C}, allowing $\overline{q \, \eps^2} \, R_\odot$ to be determined at $1$~au. The resulting values, for various values of $\alpha$, have been added to Figure~\ref{fig:qeps2}, and are consistent with the general form of $\overline{q \, \eps^2} \, R_\odot $ in Equation~\eqref{eq:qbar}.

\section{Amplitude of density fluctuations}\label{sec:amplitude}

Here we provide further observational constraints on the profile of the spectrum-weighted mean wavenumber of density fluctuations $\overline{q \, \eps^2} \, R_\odot $, using both observations of the (dominant -- Section~\ref{sec:methodology}) inner scale at the boundary between the inertial and dissipative ranges of the turbulence wavenumber spectrum (Equation~\eqref{sq-form-general-delta}), and published results related to scintillations of extra-solar point radio sources.

\subsection{The inner scale of the turbulent wavenumber spectrum}

As noted in Section~\ref{sec:methodology}, $\overline{q \, \eps ^2} \, R_\odot $ is mostly determined by density fluctuations near the inner scale $q_i^{-1}$, where $q_i$ is the boundary between the inertial and dissipative ranges of the wavenumber spectrum $S(q)$ (Equation~\eqref{sq-form-general-delta}). Observations of the inner turbulence scale thus allow to infer the magnitude of the spectrum-weighted mean wavenumber of density fluctuations $\overline{q \, \eps ^2} \, R_\odot $.

\begin{figure}[!htb]
  \centering
  \includegraphics[width=0.49\textwidth]{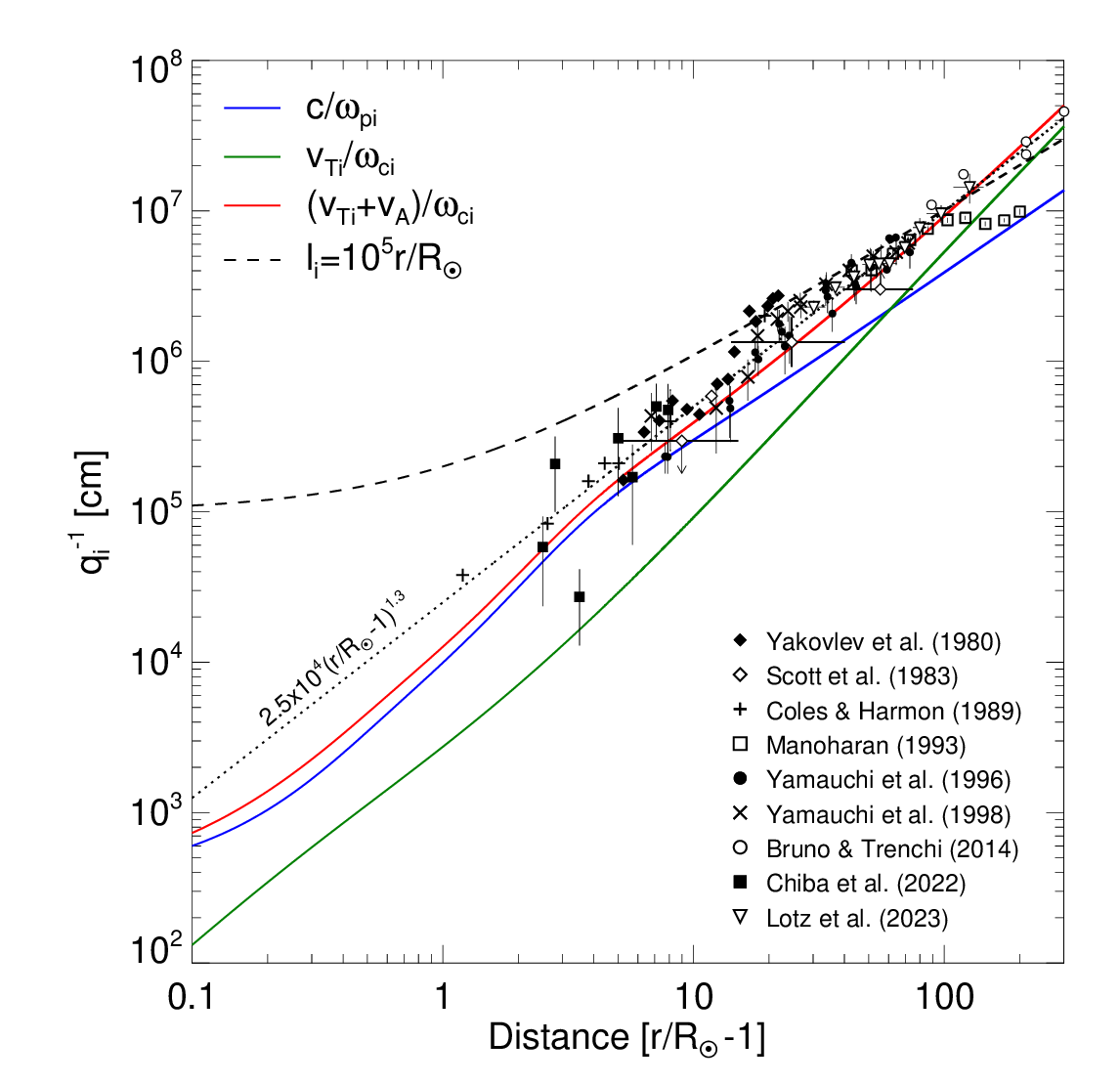}
  \includegraphics[width=0.49\textwidth]{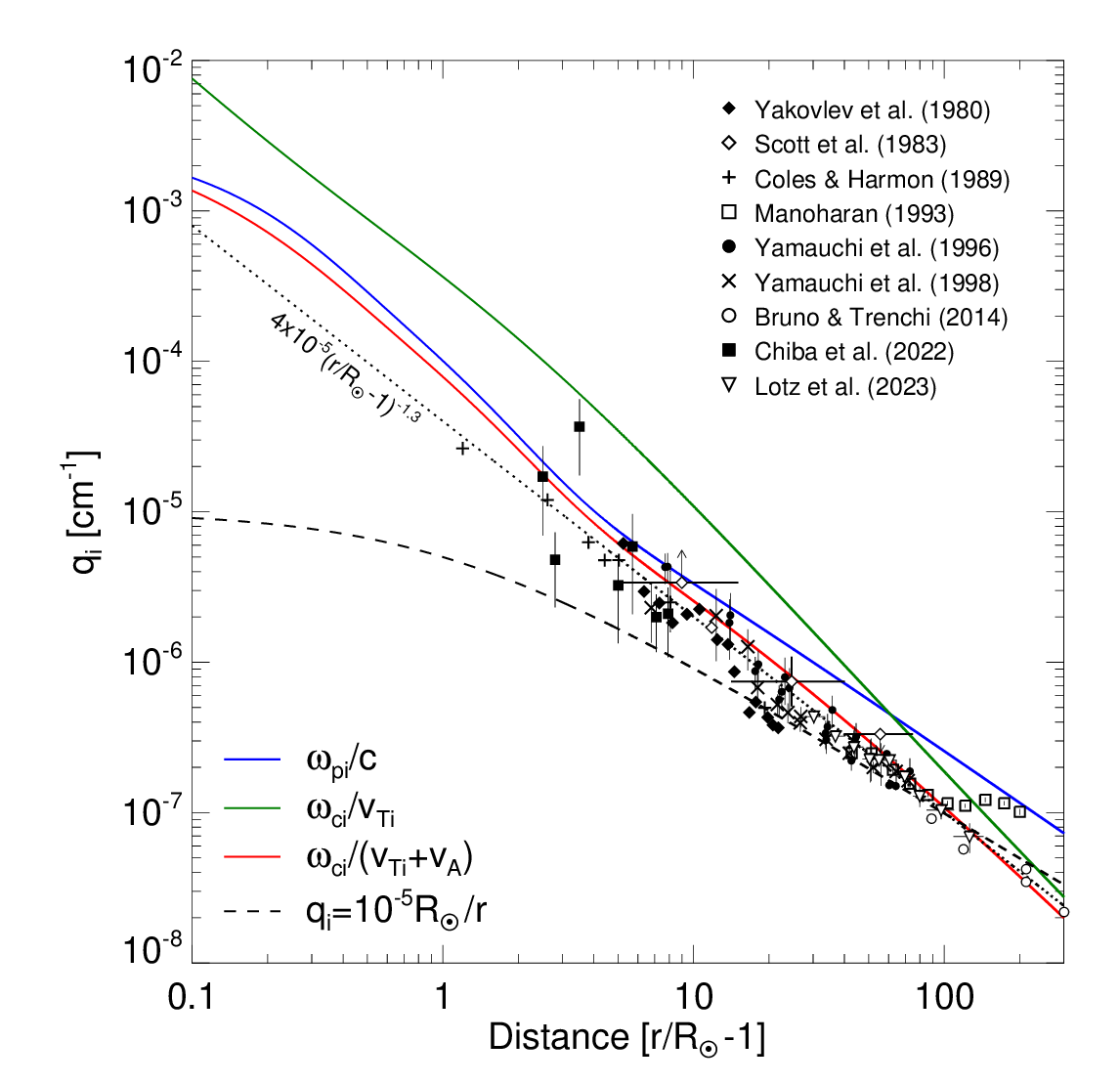}
  \caption{\textit{Left:} Measured turbulence inner scales $q_i^{-1}$ \citep{1980SvA....24..454Y,1983A&A...123..207S,1989ApJ...337.1023C,1993BASI...21..383M,1994JApA...15..387A,1996AIPC..382..366Y,1998JGR...103.6571Y}, and magnetic field fluctuations \citep{2014ApJ...787L..24B,2022SoPh..297...34C,2023ApJ...942...93L}, together with theoretical profiles of key plasma parameters: the proton inertial length $d_i=c/\omega_{\mathrm{pi}}$, the proton gyroradius $\rho_i=v_\mathrm{Ti}/\omega_{\mathrm{ci}}$, the proton resonance distance
  $d_r=(v_\mathrm{Ti} + v_A)/\omega_{\mathrm{ci}}$, calculated using the plasma and magnetic field parameters in Appendix~\ref{sec:plasma_params}. The dotted curve shows the relation $q_i^{-1} =2.5 \times 10^4 \, (r/R_\odot-1)^{1.3}$~cm and the dashed line shows the inner scale model $q_i^{-1} = 10^5 \, r/R_\odot$~cm by \citet{1989ApJ...337.1023C}. \textit{Right:} The same as the left panel, but for the inner scale wavenumber $q_i$.}
  \label{fig:inner_scale}
\end{figure}

Figure~\ref{fig:inner_scale} summarizes the pertinent observations of this scale, together with the behavior of several plasma parameters as a function of distance from the Sun. These parameters include the proton inertial length $d_i=c/\omega_{\mathrm{pi}}$, the proton gyroradius $\rho_i = v_\mathrm{Ti}/\omega_{\mathrm{ci}}$, the  the resonance distance $d_r=(v_\mathrm{Ti} + v_A)/\omega_{\mathrm{ci}}$ for parallel-propagating Alfv\'en waves \citep{1998JGR...103.4775L,2014ApJ...787L..24B}, the proton thermal velocity $v_{\rm {Ti}}=\sqrt{2k_B T_i/m_p}$, the proton plasma frequency $\omega_\mathrm{pi} = \sqrt{4\pi n \, e^2/m_p}$, the proton gyrofrequency $\omega_\mathrm{ci} = e B/m_p c$, and the Alfv\'{e}n speed $v_A = B/\sqrt{4\pi n m_p}$. The density, temperature, and magnetic field profiles used to construct these parameter variations are given in Appendix~\ref{sec:plasma_params}. Over most of the range from the lower corona ($0.1 \, R_\odot$) to $1$~au and beyond, the inner scale $q_i^{-1}$ is more consistent with the Alfv\'en wave resonance distance rather than either the proton gyroradius or the proton inertial length. \citet{1989ApJ...337.1023C} suggest that the inner scale between $2$ and $60$~$R_\odot$ is close to $3c/\omega_{ci}$ and that the data can be approximated by the relation $q_i^{-1} = 10^5 \,  r/R_\odot$~(cm) (see Figure~\ref{fig:inner_scale}). Near $1$~au ($r \simeq 200 \, R_\odot$), the gyroradius $\rho_i$ is smaller than the resonant scattering length $d_r$, so that finite gyroradius effects dominate the scattering process. Observations by \citet{2015ApJ...803..107S} suggest that the location of the density break point between these two domains is controlled by the gyrostructure frequency near 1~au.

\begin{figure}[!htb]
  \centering
  \includegraphics[width=0.49\textwidth]{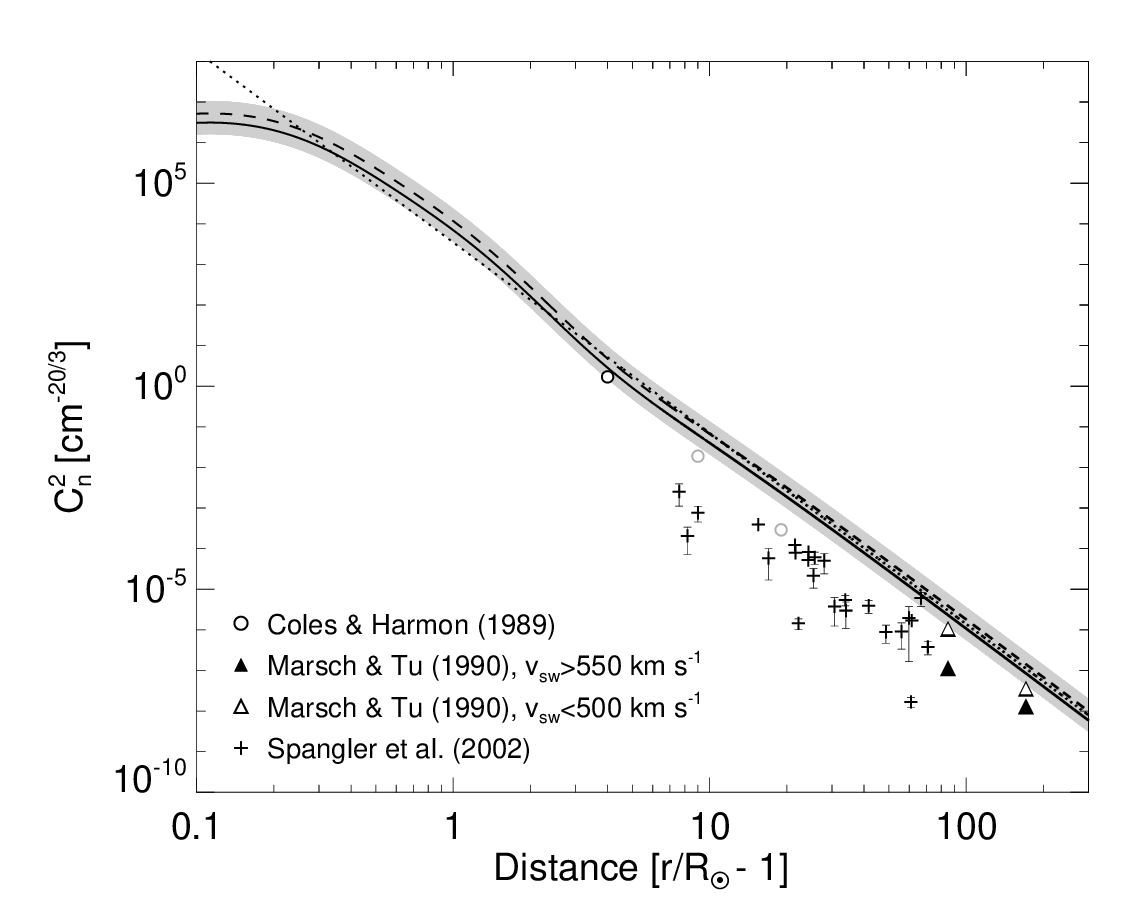}
  \includegraphics[width=0.49\textwidth]{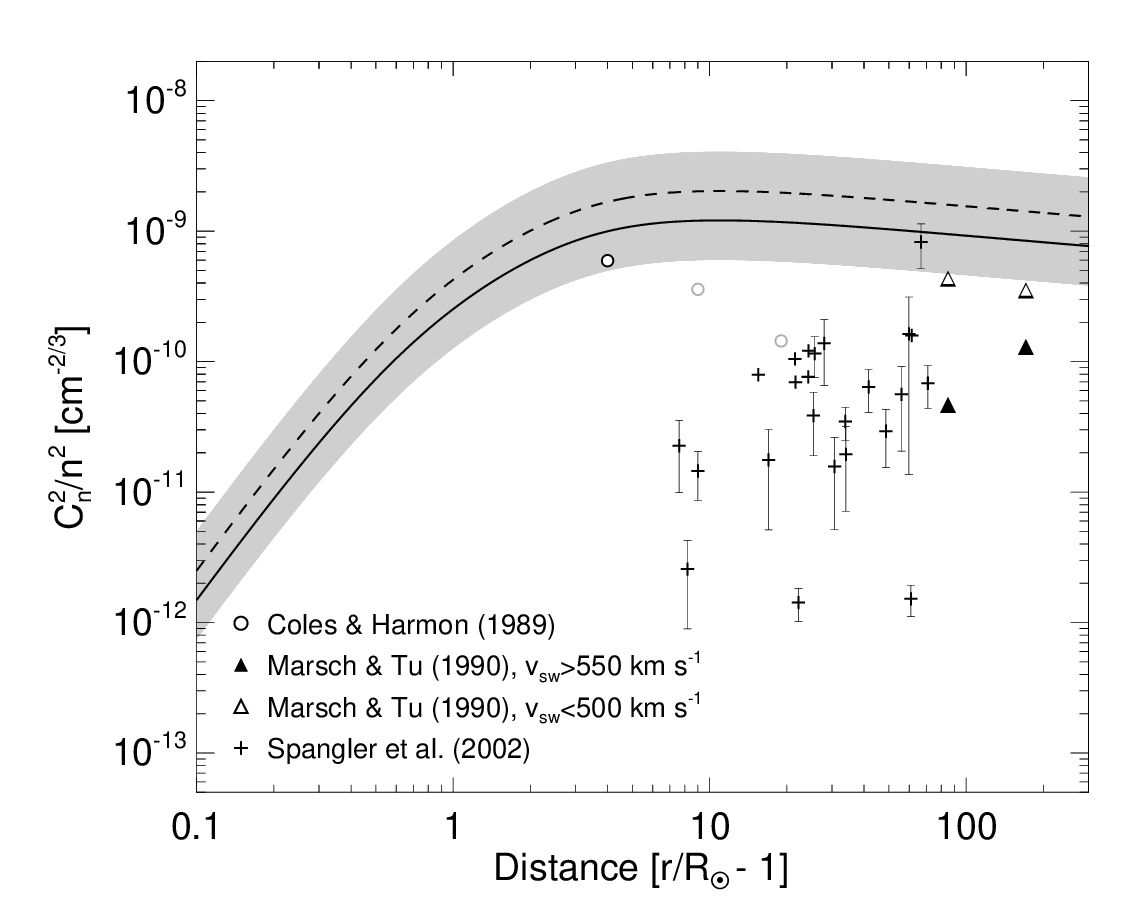}
  \caption{ \textit{Left:} Normalization coefficient $C_n^2(r)$ of the density spectrum given by Equation~\eqref{eq:qepsr2-cn2}, with $q_i=\omega_\mathrm{ci}/(v_\mathrm{Ti} + v_\mathrm{A})$
    (see Figure~\ref{fig:inner_scale}). As in Figure~\ref{fig:qeps2}, the solid/dashed lines are for anisotropy factors $\alpha =0.25$ (fundamental) and $0.42$ (harmonic), respectively, and the grey region denotes the range from $[1/2,2] \times \overline{q \, \eps^2} \, R_\odot$. The data from \cite{1989ApJ...337.1023C} are calculated using the values of $\overline{q \, \eps^2} \, R_\odot$, with $q_i$ inferred from the break at the larger wavenumber of their density spectrum. They grey points show where the inner scale was not well defined in their data. The data from \cite{1990JGR....9511945M} are found via a power law fit of the form $A \, q^{-5/3}$ to the density spectra in their Figure~6, where the slope corresponds to a power law index of $-5/3$, and $C_n^2 (r) = A \, n^2(r)/4\pi$, with $n$ taken from Equation~\eqref{eq:density}. The dotted line shows the relation $C_n^2 (r) = 3.5 \times 10^3 \, (r/R_\odot-1)^{-4.7}$. \textit{Right:} As in the left panel, but divided by $n^2$, with $n$ given by Equation~\eqref{eq:density}.}
  \label{fig:c2_n}
\end{figure}

\subsection{Observations of Extra-Solar Radio Point Sources}

Reports of radio observations of extra-solar radio point sources  \citep{1977ARA&A..15..479R,1989ApJ...337.1023C,1994ApJ...434..773S} typically provide values of the quantity $C^2_n \, (r)$, which is the normalization constant of the Kolmogorov density power spectrum:

\begin{equation}\label{eq:c2_n_spectr}
  \Phi (r ; q) = C^2_n \, (r) \, q^{-11/3}
\end{equation}
between the outer and inner scales $q_o<q<q_i$, so that\footnote{note the $(2\pi)^3$ difference between this normalization and that of Equation~\eqref{eq:Sq3d}} $\delta n^2 (r) = \int \Phi (r; q) \, d^3q$. Noting that $q_o \ll q_i$, one finds by comparing Equations~\eqref{eq:Sq3d}, \eqref{sq-form-general-delta}, \eqref{eq_q_bar}, and~\eqref{eq:c2_n_spectr} that the scaling quantity $C_n^2 \, (r)$ (cm$^{-20/3}$) is related to the $\overline{q \, \eps^2} \, R_\odot $ profile by

\begin{equation}\label{eq:c2_n_qeps2}
  C^2_n \, (r) = \overline{q \, \eps^2} \, R_\odot  \, \frac{q_i^{-1/3}}{12 \, \pi\,  R_\odot} \, n^2 \, (r)\,\,\, ,
\end{equation}
where $n(r)$ is the plasma density. Including the dissipative range above $q_i$ increases $\overline {q \, \eps^2} \, R_\odot$ by a factor of $\sim$5/3 due to power-law in dissipative range (see Equation~\eqref{eq:qeps2_iso}), so that a more appropriate expression, applicable to the wavenumber spectrum in Equation~\eqref{sq-form-general-delta}, is

\begin{equation}\label{eq:qepsr2-cn2}
  C^2_n \, (r) = \overline{q \, \eps^2} \, R_\odot  \, \frac{q_i^{-1/3}}{20 \, \pi\,  R_\odot} \, n^2 \, (r)\,\,\, .
\end{equation}
This $C_n^2 \, (r)$ profile is plotted in the left panel of Figure \ref{fig:c2_n}, with the right panel showing the quantity $C^2_n \, (r) / n^2 \, (r) = \overline{q \, \eps^2} \, R_\odot  \, q_i^{-1/3} / 20 \, \pi\,  R_\odot$. We note that the right side of Equation~\eqref{eq:qepsr2-cn2} depends weakly ($\propto q_i^{-1/3}$) on the inner scale wavenumber, so that even if the inner scale is uncertain to within a factor of two, the value of $C_n^2$ changes by only 26\%.

\begin{figure}[!htb]
  \centering
  \includegraphics[width=0.49\textwidth]{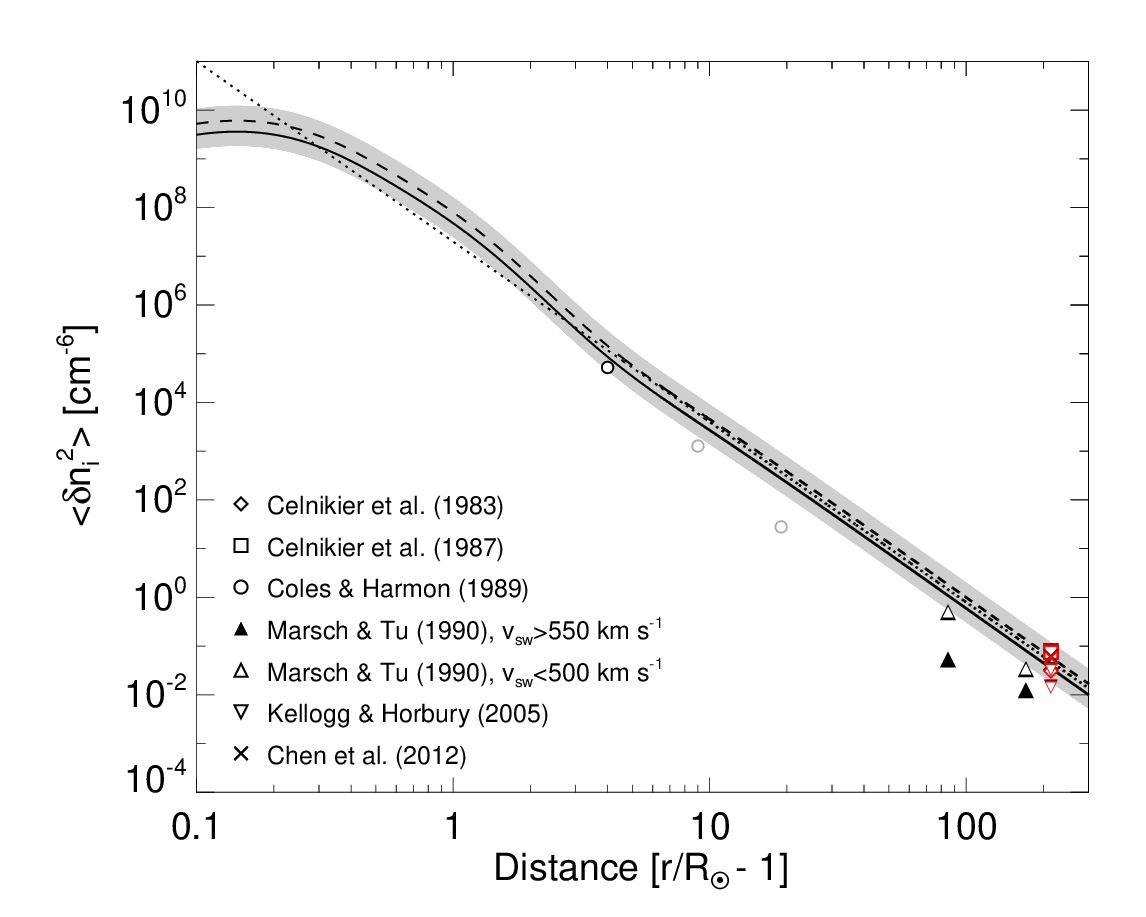}
  \includegraphics[width=0.49\textwidth]{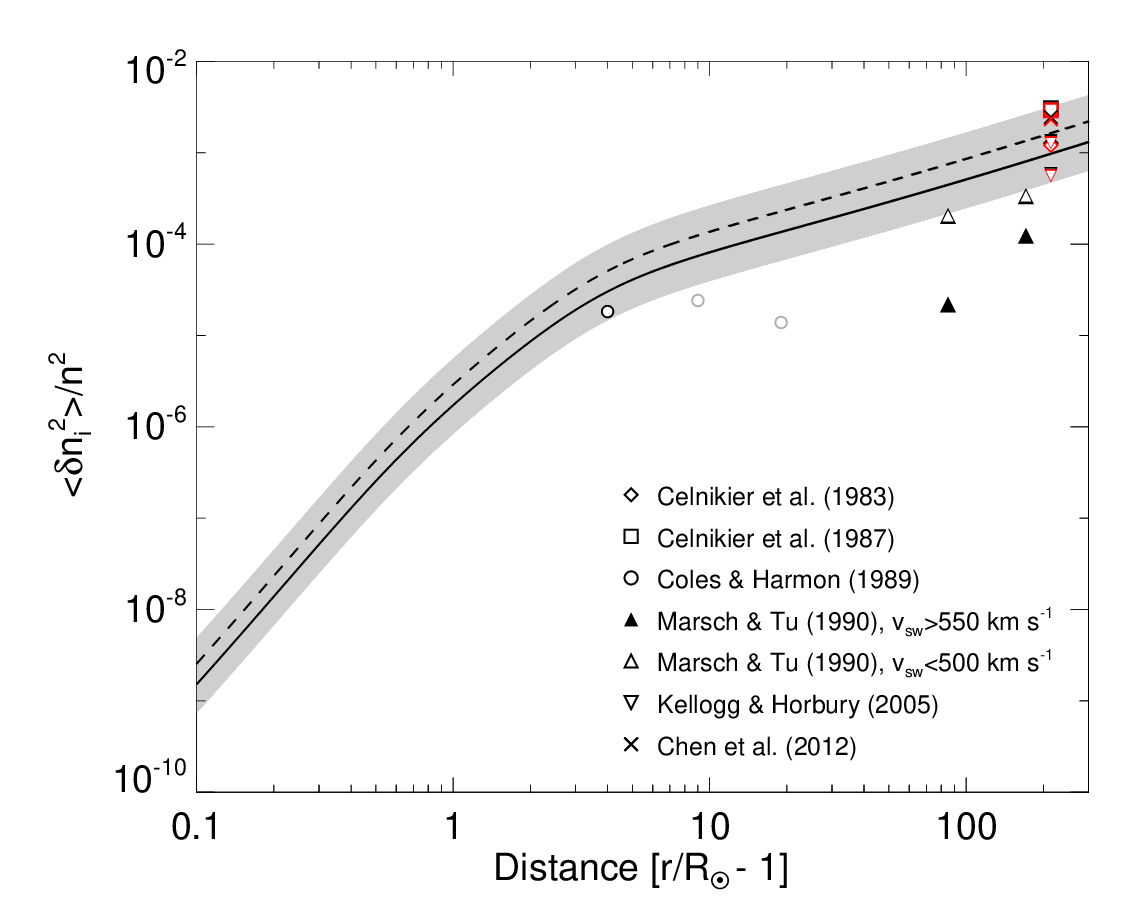}
  \caption{ \textit{Left:} Amplitude of the inner-scale density fluctuations $\langle \delta{n_i}^2 \rangle$ calculated from Equation~\eqref{qeps2-qi-eps2}. The grey regions are the same as in Figure~\ref{fig:qeps2}. The dotted line shows the relation $\langle \delta{n_i}^2 \rangle = 2 \times 10^7 \, (r/R_\odot-1)^{-3.7}$. The inner scale for \cite{1989ApJ...337.1023C} is taken from the start of the dissipation range visible in their model spectra, and the grey points denote where the inner scale was not well observed in their data. The \cite{1990JGR....9511945M} values are calculated using Equation~\eqref{qeps2-approx}; for their results the inner scale is not observed and instead we use the resonant distance $d_r$ as the measure of the inner scale. As in Figure~\ref{fig:qeps2}, the black and red data points correspond to the assumption of fundamental and harmonic emission, respectively. \textit{Right:} As in the left panel, but divided by $n^2$, with $n(r)$ given by Equation~\eqref{eq:density}.}
  \label{fig:dn_i_n}
\end{figure}

In addition to the model profile~\eqref{eq:qepsr2-cn2}, Figure~\ref{fig:c2_n} shows values of $C^2_n$ deduced from observations of scintillations of extra-solar sources \citep{2002A&A...384..654S}. The $C^2_n$ values from \cite{2002A&A...384..654S} appear to be significantly smaller than those in our simulations and in the in-situ observations at $0.4$ and $0.8$~au by \citet{1990JGR....9511945M}. Since Type~III solar radio bursts are generated by electrons from active regions/flares and thus propagate not far from the ecliptic
\citep[e.g.,][]{2021A&A...656A..34M}, our measurements are more consistent with the larger $C^2_n$ values from the slow dense solar wind \citep[see similar discussion and conclusion by][]{2002A&A...384..654S}. \citet{2016JGRA..12111605S} suggest that solar cycle variations of $C^2_n$ might contribute to variations between observations at different times.

The profiles of $\overline {q \, \eps^2} \, R_\odot $ (Equation~\eqref{eq:qbar}) and of the inner scale $q_i^{-1} (r)$ (from Figure~\ref{fig:inner_scale}) also allows us to estimate the amplitude of the density fluctuations at the dominant inner scale. Following, e.g., \cite{2009ApJ...707.1668C} and \cite{2016JGRA..12111605S}, we define the squared fractional density perturbation amplitude at the inner-scale wavenumber $q_i$ by

\begin{equation}\label{qeps2-approx}
  \frac{\langle\delta n_i^2\rangle}{n^2} = 4\pi q_i^3 \, \frac{S(q_i)}{(2\pi)^3} = 4\pi q_i^3 \, \frac{\Phi (q_i)}{n^2 \, (r)}= 4\pi \, q_i^{-2/3} \, \frac{C^2_n \, (r)}{n^2 \, (r)} \,\,\, ,
\end{equation}
which also relates the various wavenumber spectrum normalizations $\langle \delta n_i^2 \rangle/n^2$ (Equation~\eqref{eq:Sq3d}), $S(q_i)$ (Equation~\eqref{sq-form-general-delta}), and $C_n^2$ (Equation~\eqref{eq:c2_n_spectr}). One also finds by using Equation~\eqref{eq:qeps2_iso} that the inner-scale fractional density fluctuation is given by

\begin{equation}\label{qeps2-qi-eps2}
  \frac{\langle\delta n_i^2\rangle \, }{n^2 (r)} = \frac{\overline {q \, \eps^2} \, R_\odot }{\left (  3 + \frac{1}{\delta - 2} \right ) \, q_i (r) \, R_\odot } \simeq \frac{\overline {q \, \eps^2} \, R_\odot }{5 \, q_i (r) \, R_\odot } \,\,\, ,
\end{equation}
which can be evaluated using the turbulence profile of Equation~\eqref{eq:qbar} and the form of $q_i^{-1}(r)$ from Figure~\ref{fig:inner_scale}. This, plus the density profile from Equation~\eqref{eq:density}, gives the profile of the inner-scale squared density fluctuation values $\langle \delta n_i^2 \rangle$ (cm$^{-6}$). This quantity, and its dimensionless fractional value $\langle \delta n_i^2 \rangle/n^2$, are shown as functions of $r$ in Figure~\ref{fig:dn_i_n}.

\section{Discussion and Summary}\label{sec:summary}

We have constructed a density fluctuation model that allows quantitative analysis of radio-wave propagation in a medium that is characterized by an anisotropic density turbulence, symmetric around the direction of magnetic field $\vec{B}$. The density turbulence is characterized by two parameters: the spectrum-weighted mean wavenumber of density fluctuations $\overline{q \, \eps^2}$ (Equation~\eqref{eq_q_bar}) and the anisotropy measure $\alpha \simeq q_\parallel/q_\perp$ (Equation~\eqref{eq:Sq_alpha}). We allow $\overline{q \, \eps^2}$ to be a function of solar distance $r$ from the low corona ($\sim$1.1 $R_\odot$) to $215 \, R_\odot$ ($1$~au), but the anisotropy parameter $\alpha$ is, for simplicity, considered a constant. The inferred profile of the dimensionless quantity $\overline{q \, \eps^2} \, R_\odot $ is found to have a broad maximum with a value of about $100$ located at around $\sim$(4-7) $\, R_\odot$, where the slow solar wind becomes supersonic \citep{1999ApJ...523..812S}. Intriguingly, this is also the range of solar distances where Type~III bursts are observed to have the highest radio spectral flux density \citep[note the broad peak near $1-2$~MHz;][]{2022ApJ...924...58S}. The density turbulence model allows quantitative analysis of radio-wave scattering that in turn could be used to decouple the intrinsic properties of solar radio bursts and the effects of radio-wave propagation.

Analysis of the variation of the inner scale (i.e., the length associated with the smallest eddies in the inertial range of the turbulence spectrum) with solar distance presents a rather coherent picture. Radio measurements of density inner scales \citep[e.g.,][]{1989ApJ...337.1023C} are found to be in good agreement with the inner scales deduced from magnetic fluctuations \citep[e.g.,][]{2023ApJ...942...93L}, supporting a close relation between magnetic fluctuations and density fluctuations. It has been argued that kinetic Alfv\'en waves are a compressive phenomenon that is responsible for both density, magnetic field, and parallel electric field fluctuations near the break between inertial and dissipation ranges \citep[e.g.,][]{2009ApJ...707.1668C,2010A&A...519A.114B}. Over a wide range of distances, from the low solar corona into the solar wind, the turbulence inner scale is comparable (within a factor of two) with the scale of the resonant condition for protons $(v_\mathrm{Ti} + v_A)/\omega_{\mathrm{ci}}$, which is similar to the scale of the break in the spectra of magnetic fluctuations \citep{2014ApJ...787L..24B,2023ApJ...942...93L}. Analysis of the inner scales \citep{1989ApJ...337.1023C} at (10-50)~$R_\odot$ suggests that the proton inertial length $c/\omega_{pi}=v_A/\omega_{ci}$ is a good approximation, while $v_{Ti}/\omega_{pi}$ correlates with the break near $1$~au \citep{2015ApJ...803..107S}. Since $(v_\mathrm{Ti} + v_A)/\omega_{\mathrm{ci}}$ approaches $v_{Ti}/\omega_{pi}$ at $r > 50 \, R_\odot$ and $(v_\mathrm{Ti} + v_A)/\omega_{\mathrm{ci}}$ is dominated by $v_A/\omega_{ci}$ at $r > 50 \, R_\odot$, the resonance condition for protons captures both sets of observations.

Figure~\ref{fig:qeps2} shows the profiles $\overline {q \, \eps^2} \, R_\odot $ inferred from solar, non-solar and in-situ density fluctuation measurements; an acceptable fit to the observations requires the anisotropy parameter to have a value in the range $\alpha \simeq 0.2 - 0.4$. Unfortunately, observations of Type~III burst sizes and decay times do not systematically identify whether the observed emission is at the fundamental frequency or its (second) harmonic, and while this ambiguity does affect the best-fit value of the anisotropy parameter $\alpha$, it has a much smaller effect on the inferred values of $\overline{q \, \eps^2} \, R_\odot $. Assuming that the sources correspond to emission at the fundamental suggests $\alpha \simeq 0.25$; however, if the emission is at the harmonic then the anisotropy parameter is closer to $0.4$. Assuming harmonic emission also requires a factor-of-two larger value of $\overline{q \, \eps^2} \, R_\odot $, due to the $\alpha$-dependence in Equation~\eqref{eq:qbar}; see solid and dashed lines in Figure~\ref{fig:qeps2}. Since the scattering rate parallel to the magnetic field is a factor $\alpha^2 \ll 1$ times the perpendicular scattering rate (Equation~\eqref{eq:dtheta2_perp}), the radio waves escape predominantly parallel to the magnetic field, so that the degree of anisotropy chiefly affects the duration and the decay time of the bursts, while the source size is mostly controlled by the amplitude of the density fluctuations. The high accuracy of the decay times of Type~III bursts allows us to place rather significant constraints on the anisotropy parameter; for example, any value $\alpha > 0.5$ is inconsistent with the available data.

Comparing the results of numerical simulations with observations of source sizes and time profiles, over the range of frequencies $\sim$$0.1-300$~MHz, we find that the turbulence profile $\overline{q \, \eps^2} \, R_\odot $ is, within a factor of two, well represented by the empirical form Equation~\eqref{eq:qbar}. The spatial variation of the fractional density fluctuation amplitude at the inner scale increases monotonically with heliocentric distance, with a noticeable ``knee'' near the location corresponding to the maximum in $\overline{q \, \eps^2} \, R_\odot$. The amplitude of the \cite{1941DoSSR..30..301K} density turbulence spectrum varies with distance $r$ from the Sun as $C_n^2 \, (r) \simeq 3.5 \times 10^3 \, (r/R_\odot -1)^{-4.7}$~cm$^{-20/3}$. The quantity $C_n^2$ matches well with in-situ density fluctuation measurements at $0.4$ and $0.8$~au for the slow solar wind and is generally above the measurements of the fast solar wind \citep{1990JGR....9511945M}. However, only the largest values of $C_n^2$ inferred from the slow solar wind scintillation measurements of \citet{2002A&A...384..654S} are in agreement with our model, possibly because Type~III bursts propagate mostly in the ecliptic and the scintillation measurements are mostly taken away from the ecliptic. The normalized quantity $C_n^2/n^2$ demonstrates two distinct domains: it is a constant (or close to a constant) for distances $>10R_\odot$, supporting the conclusions of \citet{2002ApJ...576..997S}, but exhibits a growth of three orders of magnitude close to the Sun, reaching a maximum value near $(5-7) \, R_\odot$.

      The variance of density fluctuations decreases with solar distance approximately as $\langle\delta{n_i}^2 \rangle \, (r) = 2 \times 10^7 \, (r/R_\odot-1)^{-3.7}$~cm$^{-6}$. The exponent in this expression is close to $4$ and therefore is not inconsistent with the $\langle \delta{n_i}^2 \rangle \propto r^{-4}$ expectation \citep{2017ApJ...835..147Z,2022ApJ...928..125T} associated with nearly-incompressible magnetohydrodynamic turbulence. Comparison of the model with observations does, however, suggest that a photosphere-rooted scaling (with a dependence on the quantity $(r/R_\odot-1)$) could be a better rough symmetry than a simple heliocentric symmetry (i.e., $\propto r^{-4}$), probably due to the turbulence evolution along flux tubes rooted into the photosphere.

      \begin{figure}[!htb]
        \centering
        \includegraphics[width=0.49\textwidth]{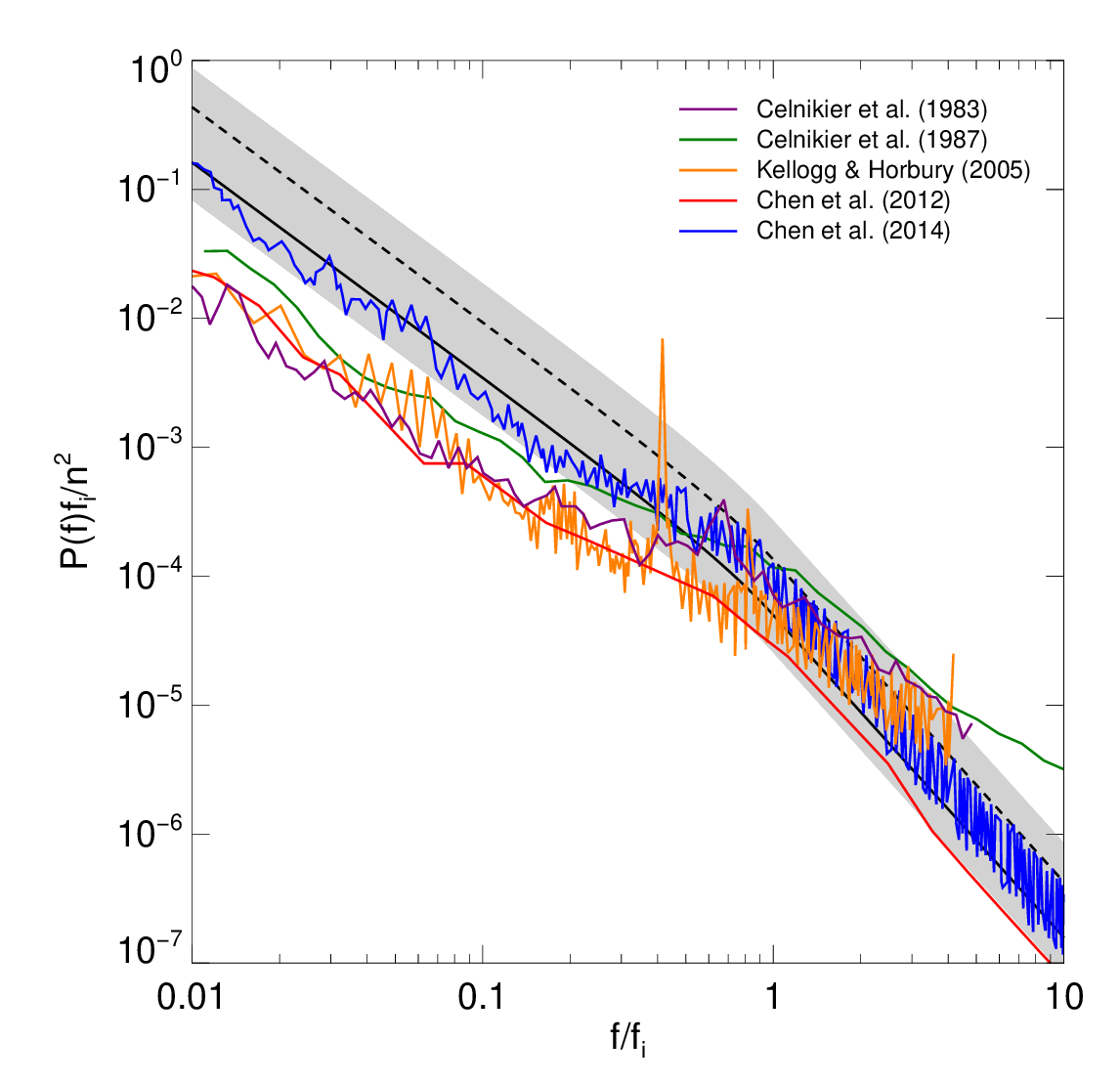}
        \caption{Frequency spectrum of density fluctuations (fundamental is solid, harmonic is dashed line) in accordance with the density fluctuation model, using Equation~\eqref{pf-anisotropic} with the values of $\langle\delta n_i^2\rangle/n^2$ at $1$~au taken from Figure~\ref{fig:dn_i_n}. The grey area shows the [1/2,2] multiplication-factor interval used in the previous figures. The in-situ fluctuations data at $1$~au \citep{1983A&A...126..293C, 1987A&A...181..138C, 2005AnGeo..23.3765K, 2012PhRvL.109c5001C,2014ApJ...789L...8C} are shown for comparison.}
        \label{fig:P_f}
      \end{figure}

      The density fluctuations model presented in Equation~\eqref{eq:qbar} provides a unifying picture for interpretation of solar radio bursts, extra-solar source broadening, and in-situ measurements of density fluctuations in the solar wind. As the model extends to $1$~au, we can predict (Appendix~\ref{sec:anis_integral}; Equation~\eqref{pf-anisotropic}) the frequency spectrum and compare with those measured by spacecraft in the slow solar wind. For a $1$~au value $\theta_B \simeq 45^\circ$, the ratio of the spectral break frequency $\widetilde {f}_i$ to its value $f_i$ for an isotropic ($\alpha = 1$) spectrum is $\sqrt{0.5 (1+\alpha^2)}$ (see remark after Equation~\eqref{pf-anisotropic}), which is weakly sensitive to $\alpha$ in the range $0.25 - 0.4$, so that the spectral break point is about the same for all $\alpha$ values in this range. Figure~\ref{fig:P_f} demonstrates good agreement between the model and observations near the break of the spectrum at $f \simeq f_i$. The discrepancy below the break frequency is likely to be associated with flattening of the density fluctuation spectra due to kinetic Alfv\'en waves \citep{2009ApJ...707.1668C} and has been observed  close to the Sun before \citep[e.g.,][]{1989ApJ...337.1023C}. As discussed, this flattening is not included into our model, but could be used for further improvement of the model.

      \bigskip

      An important feature of the model is that it provides a quantitative description for solar burst source sizes, apparent positions, and decay times, with the results summarized below:

      \bigskip

      \textit{Source sizes:} Type~III burst source sizes are predominantly determined by radio-wave scattering over the entire range of frequencies and follow a $1/f$ trend (both in the simulations and observations) from $300$~MHz down to $0.1$~MHz. The source sizes do not depend on burst intensity \citep{2013ApJ...762...60S} and the source sizes for fine structures in Type~III bursts and Type~III spikes are rather similar \citep[e.g.,][]{2017NatCo...8.1515K,2023ApJ...946...33C}. At frequencies below $0.1$~MHz, the source sizes exceed $120$~degrees (see Figure~\ref{fig:point_source_size}), comparable to half the sky, so that the source effectively surrounds the observer.

      \bigskip

      \textit{Source positions:} From Figure~\ref{fig:obs_source_pos} we see that for high frequency sources between $100$ and~$300$~MHz the apparent source positions for both fundamental ($\omega = \omega_{pe}$) and harmonic ($\omega = 2 \, \omega_{pe}$) emission are at a location corresponding to emission at double the plasma frequency. However, for emission frequencies from $\sim$$100$~MHz down to $\sim$$0.1$~MHz, the apparent source location is further away from the Sun than the location corresponding to emission at the plasma frequency or even to the location corresponding to emission at double the plasma frequency. This result elegantly resolves the long-standing conundrum that source positions observed at the fundamental and the harmonic are virtually co-spatial \citep{1985srph.book..289S}: the apparent source position is moved away from the Sun by scattering effects, which are largely independent of whether fundamental or harmonic emission is involved. At low frequencies $\lapprox 0.2$~MHz, this effect is even more pronounced: the source positions are shifted by 0.5~au, so that an observer at $1$~au is actually embedded in the apparent source.

      \bigskip

      \textit{Type~III burst decay times:} The decay time of Type~III solar radio bursts is well approximated by an inverse-frequency ($1/f$) dependence between $0.1$ and~$300$~MHz. Our scattering simulations reveal a rather intricate picture, in which scattering serves to provide a fundamental ``floor'' on the observed duration of radio bursts emitted via plasma emission. Below $1$~MHz, the average Type~III burst decay time is consistent with the scattering simulations, but at frequencies above a few~MHz, the typical Type~III burst decay time is longer than the simulated value, suggesting that the decay time for such bursts is determined by processes that operate on timescales longer than scattering, such as electron propagation, electron injection, and/or intrinsic emission time scales. The shortest features in the dynamic spectrum, such as Type~IIIb striae \citep[e.g.,][]{2017NatCo...8.1515K,2018SoPh..293..115S}, drift pairs \citep{2019A&A...631L...7K}, spikes \citep{2021ApJ...917L..32C}, and fine structures in type II solar radio bursts \citep{2018ApJ...868...79C} are all consistent with the scattering model. Statistical analysis of $\sim1$$100$ fine structures \citep{2023ApJ...946...33C} near~$30$~MHz, with $0.02$~s resolution, indeed shows that there are virtually no bursts shorter than $0.2$~s and that the typical burst decay time is about $0.3$~s, consistent with the simulations in Figure~\ref{fig:decay_time}.

    \bigskip

    In summary, solar burst shortest time profiles, source sizes, and positions are determined mainly by propagation effects (mostly anisotropic scattering) and not by intrinsic properties of the radio emission source. A detailed knowledge of the scattering process paves the way to disentangling scattering effects from observations and so better constraining the intrinsic properties of solar radio burst sources. Since individual source sizes and decay times are typically measured more accurately than the spread in measurements of multiple sources, it is therefore likely that the spread in observational properties is due to varying levels of turbulence and plasma density in different events. Varying the magnitude of the $\overline{q \, \eps^2} \, R_\odot $ profile (Equation~\eqref{eq:qbar}) by a factor in the range $0.5$ to~$2$ covers the majority of the observations, while extending this multiplicative factor by a further factor of two (to a range between $0.25$ to $4$) covers virtually all observed data points (except for some extreme outliers). We thus conclude that the $\overline{q \, \eps^2} \, R_\odot $ profile of Equation~\eqref{eq:qbar} is variable within a factor of about two. The broadening of extra-solar point sources by the turbulent solar atmosphere and solar radio burst measurements are complementary data sets. We note the considerable data gap between ground-based and space-based solar burst observations in the range $3-20$~MHz (where extra-solar observations appear essential), and encourage the development of observations to fill this gap and hence further constrain the level of turbulence in the inner heliosphere. \href{https://www.jpl.nasa.gov/missions/sun-radio-interferometer-space-experiment}{The Sun Radio Interferometer Space Experiment (SunRISE)} is likely to provide much needed data.

    \begin{acknowledgments}

We thank P. Massa, Prasad Subramanian, Tim Bastian, Arnaud Zaslavsky, and Yingjie Luo for helpful discussions. EPK, DLC and XC acknowledge financial support from the STFC Consolidated Grant ST/T000422/1. AGE was supported by NASA Kentucky under NASA award number 80NSSC21M0362.
NC acknowledges funding support from the Initiative Physique
des Infinis (IPI), a research training program of the Idex SUPER at Sorbonne University.
We also acknowledge support from the International Space Science Institute for the LOFAR \url{http://www.issibern.ch/teams/lofar/}
and solar flare \url{http://www.issibern.ch/teams/solflareconnectsolenerg/} teams.

    \end{acknowledgments}

\appendix

    \section{Plasma density, magnetic field and temperature profiles}\label{sec:plasma_params}

\begin{figure}[htb!]
      \centering
      \includegraphics[width=0.49\textwidth]{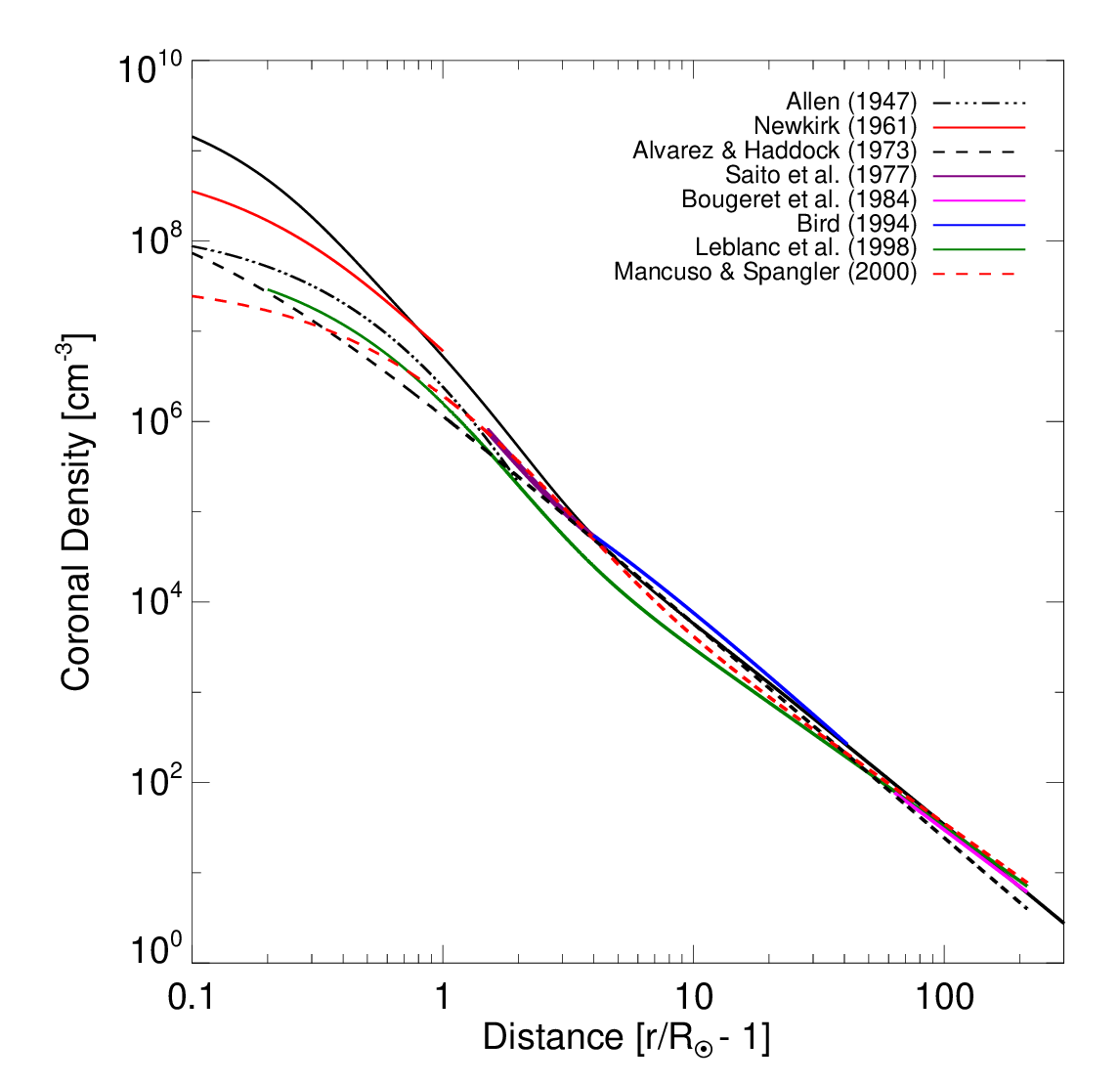}
      \includegraphics[width=0.49\textwidth]{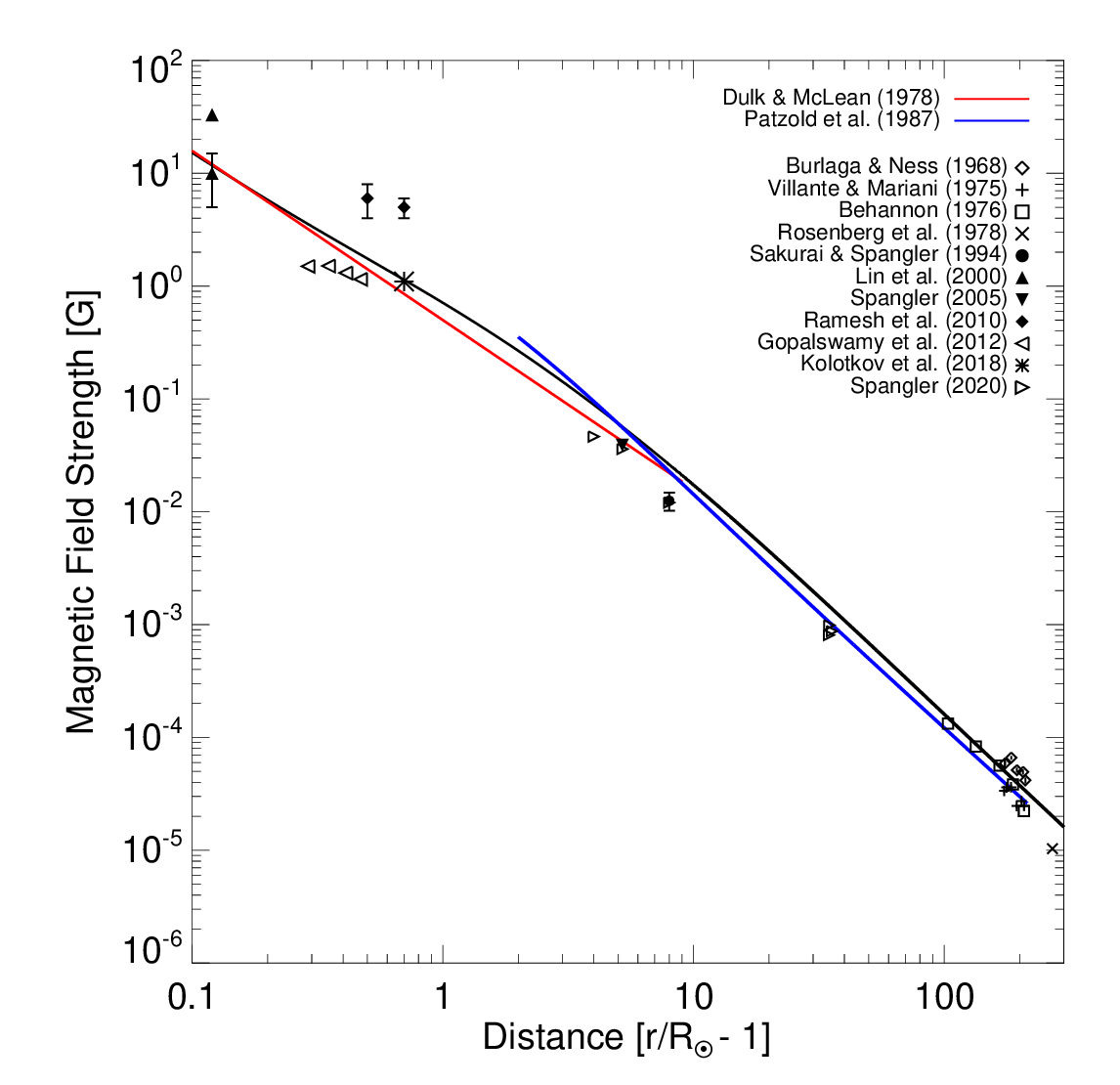}
      \caption{\textit{Left panel}: Coronal density model (black solid line) given by Equation~\eqref{eq:density}. Other coronal density models \citep{1947MNRAS.107..426A, 1961ApJ...133..983N, 1973SoPh...29..197A, 1977SoPh...55..121S, 1984SoPh...90..401B, 1994ApJ...426..373B, 1998SoPh..183..165L, 2000ApJ...539..480M} are also shown, within the ranges of their respective validity. The models from \cite{1994ApJ...426..373B} and \cite{2000ApJ...539..480M} are as used in \cite{2002ApJ...576..997S}. \textit{Right panel}: Magnetic field model (black solid line) given by Equation~\eqref{eq:b_r}, together with observations \citep{1968CaJPS..46..962B, 1975GeoRL...2...73V, 1976pspe.proc..332B, 1978JGR....83.4165R, 1978SoPh...57..279D, 1987SoPh..109...91P, 1994ApJ...434..773S, 2000ApJ...541L..83L, 2005SSRv..121..189S, 2010ApJ...711.1029R, 2012ApJ...744...72G, 2018ApJ...861...33K, 2020RNAAS...4..147S}. The red line shows the model of \citet{1978SoPh...57..279D}}
      \label{fig:density-magnetic-field}
    \end{figure}

    The density profile $n(r)$ (cm$^{-3}$) used for the solar burst scattering simulations by \cite{2019ApJ...884..122K}, \cite{2020ApJ...898...94K}, and \cite{2020ApJ...905...43C, 2023MNRAS.520.3117C} is

    \begin{equation}\label{eq:density}
      n(r)= 4.8\times 10^{9}\left(\frac{R_{\odot}}{r}\right)^{14}
      + 3\times 10^8\left(\frac{R_{\odot}}{r}\right)^6
      +1.4\times 10^6\left(\frac{R_{\odot}}{r}\right)^{2.3} \,\,\, .
    \end{equation}
    This density profile (i) closely follows the \cite{1958ApJ...128..664P} density model, (ii) has an analytical form that can be easily differentiated to find the derivatives that are useful in solving the ray-tracing equations, (iii) is consistent at low heights with the higher densities observed in flaring active regions that produce Type~III bursts \citep[e.g.,][]{2011SSRv..159..107H,2014A&A...567A..85R}, and (iv) is suitable for Type~III burst modelling in the solar corona and heliosphere \citep{1999A&A...348..614M, 2001SoPh..202..131K,2021NatAs...5..796R}. The left panel of Figure~\ref{fig:density-magnetic-field} shows the model profile compared with different density models frequently used in the literature.

    The magnetic field $B(r)$ (G) model, shown in the right panel of Figure \ref{fig:density-magnetic-field}, is given by

    \begin{equation}\label{eq:b_r}
      B(r)= 0.5 \, \frac{\left ( \frac{r}{R_\odot}- 1 \right )^{-1.5}}{ \, \left ( \frac{r} {10 \, R_\odot}+1 \right ) } + 1.18 \, \left ( \frac{R_\odot}{r} \right )^2 \,\,\, ,
    \end{equation}
    where the first term is introduced to accommodate the results of \cite{1978SoPh...57..279D}, applicable for near-solar distances $r \lapprox 10 \, R_\odot$, into the form of the interplanetary magnetic field for $r \gg 10 \, R_\odot$ \citep[see discussion by][]{1987SoPh..109...91P}.

    The ion temperature $T_i$ (K) is given by

    \begin{equation}\label{eq:T_i_better}
      T_i(r) = 2.4 \times 10^6 \, \left ( \frac{r} {10 \, R_\odot} + 1 \right )^{-0.74} \, \mbox{ K} \,\,\, ,
    \end{equation}
    where the interplanetary dependency $r^{-0.74}$ (with $T_i = 2.5 \times 10^5$~K at $1$~au $= 215 \, R_\odot$) is taken from \cite{2011JGRA..116.9105H} and has been adjusted to accommodate the range $r \lapprox 10 \, R_\odot$ in accordance with the results of \cite{1978SoPh...57..279D}.

    \section{Angular broadening of extra-solar point sources}\label{appendix:angular}

    The overall angular broadening of a point source at infinity can be calculated by integrating the angular broadening rate along the path ($z$-direction) traversed by the radio waves. For the most significant broadening, i.e., that along the $\perp_2$ direction perpendicular to the (radial) magnetic field in the plane of the sky, the rate of diffusive broadening is given by Equation~\eqref{eq:dtheta2_perp}, repeated here:

    \begin{equation}\label{eq:theta_perp-rate}\nonumber
      \frac{d \, \langle \theta_\perp ^2\rangle}{dz}
      =\frac{\pi}{8} \, \frac{\omega _{pe}^4}{\omega^4} \,\, \overline{q \, \eps^2} \,  \,\,\, .
    \end{equation}
    If $r$ is the distance of closest approach to the Sun of a ray from a distance point source (Figure~\ref{fig:point_source_size}), then the apparent source, observed at $1$~au, will be broadened by an amount

    \begin{equation}\label{eq:theta_perp}
      \langle \theta_\perp ^2\rangle
      = \frac{\pi}{8 \, \omega^4} \, \int_{-\infty}^{1~{\rm au}} \omega_{pe}^4 \left ( \sqrt{r^2+z^2} \right ) \,\, \overline{q \, \eps^2} \left ( \sqrt{r^2+z^2} \right ) \, dz
      = \frac{2 \, \pi^3 \, e^4}{m_e^2 \, \omega^4} \, \int_{-\infty}^{1~{\rm au}} n^2 \left ( \sqrt{r^2+z^2}  \right ) \,\, \overline{q \, \eps^2} \left ( \sqrt{r^2+z^2}  \right ) \, dz \,\,\, ,
    \end{equation}
    where we have used $\omega_{pe} = \sqrt{4 \pi n e^2/m_e}$. With the substitution $z = r \tan \chi$, this can be written

    \begin{equation}\label{eq:theta_perp-1}
      \langle \theta_\perp ^2\rangle
      =\alpha  \frac{2 \, \pi^3 \, e^4}{m_e^2 \, \omega^4} \, \left ( \frac{r}{R_\odot} \right ) \, \int_{-\pi/2}^{\tan^{-1}(215 \, R_\odot/r)} n^2 ( r \sec \chi ) \,\, \overline{q \, \eps^2} \, R_\odot  ( r \sec \chi ) \, \sec^2 \chi \, d\chi \,\,\, .
    \end{equation}
    Equation~\eqref{eq:theta_perp-1} provides the formal expression corresponding to the approximate proportionalities in Equation~\eqref{eq:perp_diffusion}. Using the form of $\overline{q \, \eps^2} \, R_\odot$ from our nominal model~\eqref{eq:qbar} and the density model from Equation~\eqref{eq:density}, we obtain

    \begin{eqnarray}\label{eq:theta_perp-2}
      \langle \theta_\perp ^2\rangle
      &=& 2 \times 10^3 \, \alpha \, \times \, \left ( \frac{e^4}{8 \pi \, m_e^2 \, f^4} \right ) \, \left ( \frac{r}{R_\odot} \right )^{0.3} \, \times \cr
      & \, & \int_{-\pi/2}^{\tan^{-1}(215 \, R_\odot/r)} \, \left [ 4.8\times 10^{9} \left ( \frac{R_{\odot}}{r} \right)^{14} \, \cos^{14} \chi + 3 \times 10^8 \left ( \frac{R_{\odot}}{r} \right )^6 \, \cos^6 \chi + 1.4 \times 10^6 \left ( \frac{R_{\odot}}{r} \right )^{2.3} \, \cos^{2.3} \chi \right ]^2 \, \times \cr
      && \qquad \qquad \times \, \left( 1 - \frac{R_\odot \, \cos \chi}{r} \right)^{2.7} \, \frac{d\chi}{\cos^{1.3} \chi} \,\,\, .
    \end{eqnarray}
    At large closest-approach distances $r \gg R_\odot$, the density profile~\eqref{eq:density} is well approximated by the last term in the square brackets, i.e., by the simple power-law $n(r) \propto r^{-2.3}$, and the term $(1 - R_\odot \cos \chi/r)^{2.7} \simeq 1$. In this situation, the FWHM size assumes the relatively simple form

    \begin{equation}\label{eq:theta_perp-3}
      {\rm FWHM}_\perp = 2.35 \, \sqrt{\alpha} \times \sqrt{2 \times 10^3} \times (1.4 \times 10^6) \times \, \left [ \frac{{\rm B} \, (1; 2.15, 0.5) + {\rm B} \, (\psi; 2.15, 0.5)}{16 \pi} \,  \right ]^{1/2} \, \left ( \frac{e^2}{m_e} \right ) \, \left ( \frac{r} {R_{\odot}}\right )^{-2.15} \,  \times \frac{1}{f^2} \,\,\, ,
    \end{equation}
    where ${\rm B} (\psi; u,v)$ are the (incomplete if $\psi < 1$) beta functions corresponding to the integrals $2 \, \int_{-\pi/2}^{0} \cos^{3.3} \chi \, d\chi$ and $ 2 \, \int_{0}^{\tan^{-1}(215 R_\odot/r)} \cos^{3.3} \chi \, d\chi$, respectively, and $\psi= [1 + (r/215 \, R_\odot)^2]^{-1}$. Scaling to the nominal frequency $f = 1.5$~GHz (Figure~\ref{fig:point_source_size}), this evaluates to

    \begin{equation}\label{eq:FWHM_perp}
      \text{FWHM}_\perp \simeq 4 \times 10^{-3} \, \sqrt{\alpha} \, \left ( \frac{r}{R_\odot} \right )^{-2.15} \left(\frac{1.5 \, \text{GHz}}{f}\right)^2\, {\rm radians} \, \simeq \, 800 \, \sqrt{\alpha} \, \left ( \frac{r}{R_\odot} \right )^{-2.15} \left(\frac{1.5 \, \text{GHz}}{f}\right)^2 \, {\rm arc seconds}  \,\,\, .
    \end{equation}
    Equation~\eqref{eq:FWHM_perp} provides a simple, but accurate, analytical approximation for FWHM$_\perp$, valid for $r \gg R_\odot$.

    \section{Forms of the in situ density fluctuation frequency spectra}\label{appendix:spectral-forms}

    The in situ frequency power spectrum $P(f)$ of density fluctuations measured by a spacecraft is related to the wavenumber spectrum $S({\vec q})$ at a single location in the solar wind frame through the relation~\eqref{pf-general}, repeated here:

    \begin{equation}\label{pf-general_3}\nonumber
      P(f) = \frac{n^2}{(2\pi)^3} \, \int S ({\vec q}) \, \delta \left ( \frac{ {\vec q} \cdot {\vec V}_{\rm SW} }{2 \pi} - f \right ) \, d^3q \,\,\, .
    \end{equation}
    In this Appendix we explore the forms of $P(f)$ for different forms of the wavenumber spectrum $S(\vec q)$.

    \subsection{Isotropic Wavenumber Spectrum}\label{appendix:isotropic_sq}

    Our first example is the simplest case of an isotropic spectrum: $S({\vec q}) = S(q)$. For such a case, Equation~\eqref{pf-general} may be written

    \begin{eqnarray}\label{pf-isotropic-1}
      P(f) \, &=& \, \frac{n^2}{(2\pi)^2} \, \int_{q_\perp = 0 }^\infty \, \int_{q_\parallel = -\infty}^\infty S \left ( \sqrt{q_\parallel^2 + q_\perp^2} \right ) \, \delta \left ( \frac{ q_\parallel V_{\rm SW} }{2 \pi} - f \right ) \, q_\perp \, dq_\perp \, dq_\parallel \cr
      &=& \, \frac{n^2}{2\pi \, V_{\rm SW}} \, \int_{q_\perp = 0 }^\infty S \left ( \sqrt{q_\perp^2 + \left ( \frac{2 \pi f}{V_{\rm SW}} \right )^2 } \right ) \, q_\perp \, dq_\perp \,\,\, ,
    \end{eqnarray}
    Here, we evaluate this for a specific form of $S(q)$, viz. the broken power-law around $q_i$ (cf. Equation~\eqref{sq-form-general-delta}), with an outer scale $q_o^{-1} \gg q_i^{-1}$, so that the contribution to $\overline{q \, \eps ^2} \, R_\odot$ from the large scales $> q_o^{-1}$ is negligible:

    \begin{equation}\label{sq-form}
      S(q) = S(q_i) \, \times \, \begin{cases}
        \left ( \frac{q}{q_i} \right )^{-11/3} ;  \qquad q \le q_i \cr \left ( \frac{q}{q_i} \right )^{-2-\delta} ; \qquad q > q_i \,\,\, .
      \end{cases}
    \end{equation}
    Evaluation of the fluctuation frequency spectrum $P(f)$ requires consideration of two domains:

    \bigskip
    (i) $f \le f_i = \frac{q_i V_{\rm SW}}{2 \pi}$
    \bigskip

    Here the resonant $q_\parallel = 2 \pi f/V_{\rm SW} < q_i$, so that for values of $q_\perp^2 < q_i^2 - q_\parallel^2$ the lower (inertial) branch of the wavenumber spectrum applies, while for values of $q_\perp^2 > q_i^2 - q_\parallel^2$ the upper (dissipative) branch of the wavenumber spectrum is relevant. The integral is thus composed of two parts:

    \begin{eqnarray}\label{eq:Pf_iso_small_f}
      P(f) \!\!\!\! &= \frac{n^2}{2\pi \, V_{\rm SW}} \, S(q_i) & \!\!\!\! \left [ q_i^{11/3} \, \int_{q_\perp =0}^{\sqrt{q_i^2 - (2 \pi f/V_{\rm SW})^2}} \left ( q_\perp^2 + \left ( \frac{2 \pi f}{V_{\rm SW}} \right )^2 \right )^{-11/6} \, q_\perp \, dq_\perp  \, \right . + \cr
      & & \qquad + \left . \, q_i^{\delta + 2} \, \int_{q_\perp ={\sqrt{q_i^2 - (2 \pi f/V_{\rm SW})^2}}}^\infty \left ( q_\perp^2 + \left ( \frac{2 \pi f}{V_{\rm SW}} \right )^2 \right )^{-1 - \frac{\delta}{2}} \, q_\perp \, dq_\perp  \right ] \cr
      & \,\,\,\, = \frac{n^2}{4\pi^2 \, f_i} \, q_i^3 \, S(q_i) &\, \left [ \frac{3}{5} \left ( \left ( \frac{2 \pi f}{q_i \, V_{\rm SW}} \right )^{-5/3} - 1 \right ) + \frac{1}{\delta} \right ] \cr
      & \!\!\!\!\!\!\!\!\!\!\!\!\!\!\!\!\!\!\!\! = \frac{\langle \delta n_i^2\rangle}{2 \, f_i}  & \!\!\!\!\!\!\!\!\!\!\!\!\!\!\!\!\!\!\!\!\!\! \left [ \frac{3}{5} \left ( \left ( \frac{f}{f_i } \right )^{-5/3} - 1 \right ) + \frac{1}{\delta} \right ]  \,\,\, ,
    \end{eqnarray}
    where in the last equality we have used Equation~\eqref{qeps2-approx}.

    \bigskip
    (ii) $f > f_i = \frac{q_i V_{\rm SW}}{2 \pi}$

    \bigskip

    Here the resonant $q_\parallel = 2 \pi f/V_{\rm SW} > q_i$, so that for all values of $q_\perp$ only the upper branch of the wavenumber spectrum is relevant. Thus

    \begin{eqnarray}\label{eq:Pf_iso_large_f}
      P(f) \, &=& \, \frac{n^2}{2\pi \, V_{\rm SW}} \, S(q_i) \, q_i^{\delta + 2} \, \int_{q_\perp =0}^\infty \left ( q_\perp^2 + \left ( \frac{2 \pi f}{V_{\rm SW}} \right )^2 \right )^{-1 - \frac{\delta}{2}} \, q_\perp \, dq_\perp \cr
      &=& \, \frac{n^2}{4\pi^2 \, f_i} \, q_i^3 \, S(q_i) \,  \frac{1}{\delta} \, \left ( \frac{2 \pi f}{q_i \, V_{\rm SW}} \right )^{-\delta} \cr
      &=& \, \frac{\langle\delta n_i^2\rangle}{2 \, f_i} \frac{1}{\delta} \, \left ( \frac{f}{f_i } \right )^{-\delta}
    \end{eqnarray}
    Combining Equations~\eqref{eq:Pf_iso_small_f} and~\eqref{eq:Pf_iso_large_f},

    \begin{equation}\label{pf-isotropic}
      P(f) =  \frac{\langle \delta n_i^2 \rangle}{2 \,  f_i }  \, \times \, \begin{cases}
        \frac{3}{5} \, \left  [ \left ( \frac{f}{f_i} \right )^{-5/3} \, - \, \left ( 1 - \frac{5}{3 \delta} \right ) \, \right ] \,\,\,\,\,\,\,\,\,   ; \qquad f \le f_i \cr
        \frac{1}{\delta} \, \left ( \frac{f}{f_i} \right )^{-\delta} \qquad \qquad \qquad \qquad \,\, ; \qquad
        f > f_i  \,\,\, .
      \end{cases}
    \end{equation}

    As a check on this result, we can evaluate (using algebra) the quantity

    \begin{equation}\label{fp-int-check}
      \int_0^\infty f \, P(f) \, df = \frac{\langle \delta n_i^2 \rangle}{2 \,  f_i }  \, \left (3 + \frac{1}{\delta - 2}  \right ) \, \frac{f_i^2}{2} = \frac{1}{4} \,  \left ( 3 + \frac{1}{\delta - 2} \right) \, \langle\delta n_i^2\rangle \, f_i \,\,\, .
    \end{equation}
    so that, using the general result~\eqref{eq_q_bar1}, with $\alpha = 1$ for an isotropic spectrum, the value of $\overline{q \, \eps^2} \, R_\odot$ is

    \begin{equation}\label{qeps2-check}
      \overline{q \, \eps^2} \, R_\odot =
      \frac{8 \pi}{n^2 \, V_{\rm SW}} \, R_\odot \, \int_0^\infty f \, P(f) \, df = \left ( 3 + \frac{1}{\delta - 2} \right) \, (q_i \, R_\odot) \, \frac{\langle\delta n_i^2\rangle}{n^2}  \,\,\, ,
    \end{equation}
    in agreement with Equation~\eqref{qeps2-qi-eps2}.

    The local power-law index of the frequency power spectrum is the negative of its logarithmic derivative, viz.

    \begin{equation}\label{pf-isotropic-slope}
      \delta_{\rm loc} = \, - \, \frac{d \ln P(f)}{d \ln f}  \equiv - \, \frac{f}{P} \, \frac{d P}{df} \, = \begin{cases}
        \frac{5/3}{1 - \left ( 1 - \frac{5}{3 \delta} \right ) \left (  \frac{f}{f_i} \right )^{5/3} } \,\,\,\,\,\,\,  ; \qquad f \le f_i \cr
        \delta \,\,\, \qquad \qquad \qquad \,\, ; \qquad
        f > f_i  \,\,\, .
      \end{cases}
    \end{equation}
    At low frequencies $f \ll f_i$, the frequency spectrum has a power-law index $\delta_{\rm loc} = 5/3$; as $f$ increases, the spectrum gradually steepens to $\delta_{\rm loc} = \delta$ as $f$ approaches $f_i$, and continues with that value thereafter. Thus a double-power law spectrum in wavenumber does not correspond to a double power-law spectrum in frequency, neither does a double-power law in frequency correspond to a double power-law in wavenumber space. Interestingly, the variation in the spectrum of magnetic fluctuations reported by \citet[][their Fig. 2]{2023ApJ...943L...8S} near the break frequency $f_i$, is broadly consistent with Equation~\eqref{pf-isotropic-slope}.

    \subsection{Anisotropic density turbulence Spectrum with anisotropic spectral break points}\label{sec:anis_integral}

\begin{figure}[htb!]
      \centering
      \includegraphics[width=0.45\textwidth]{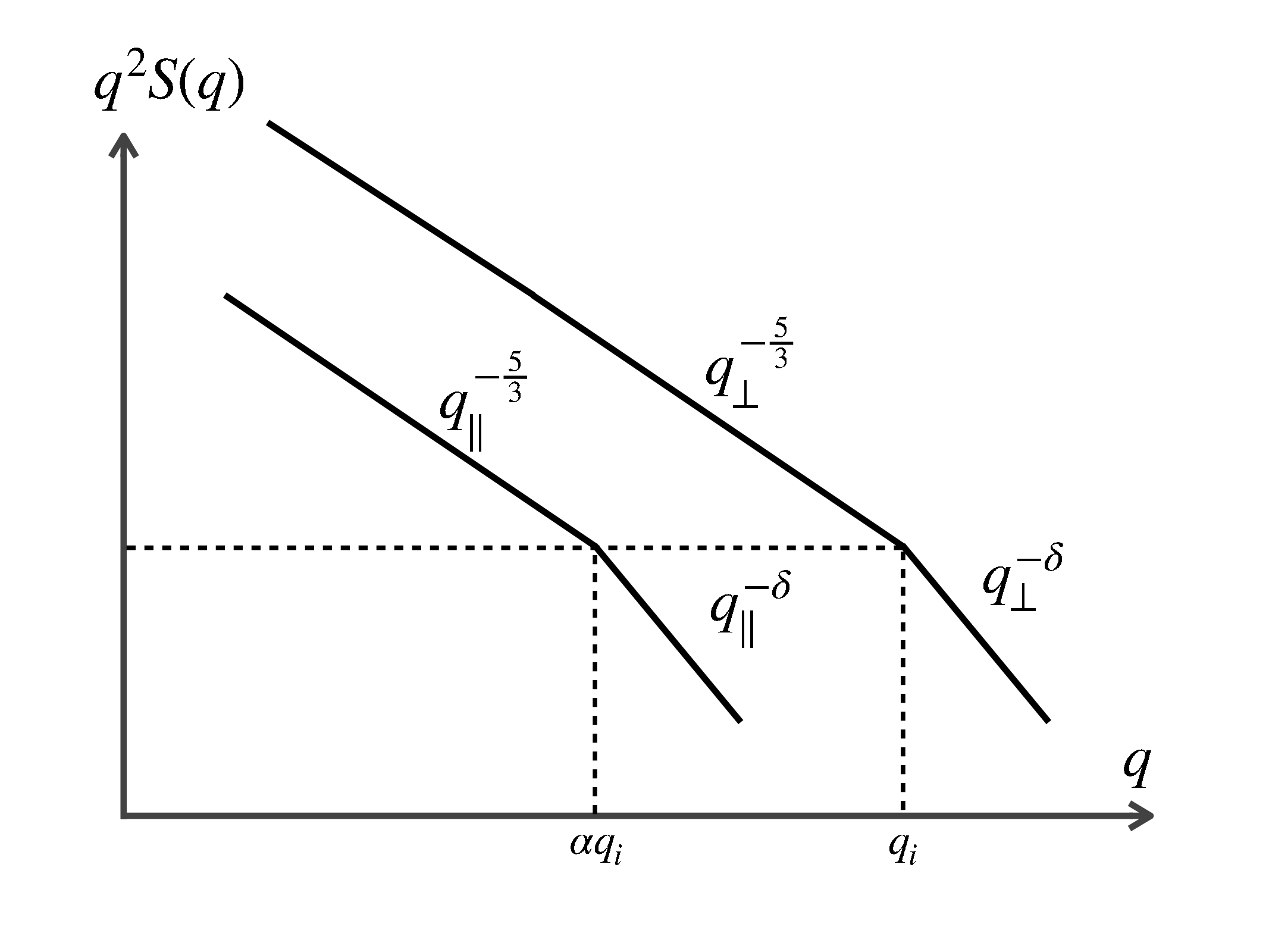}
      \includegraphics[width=0.45\textwidth]{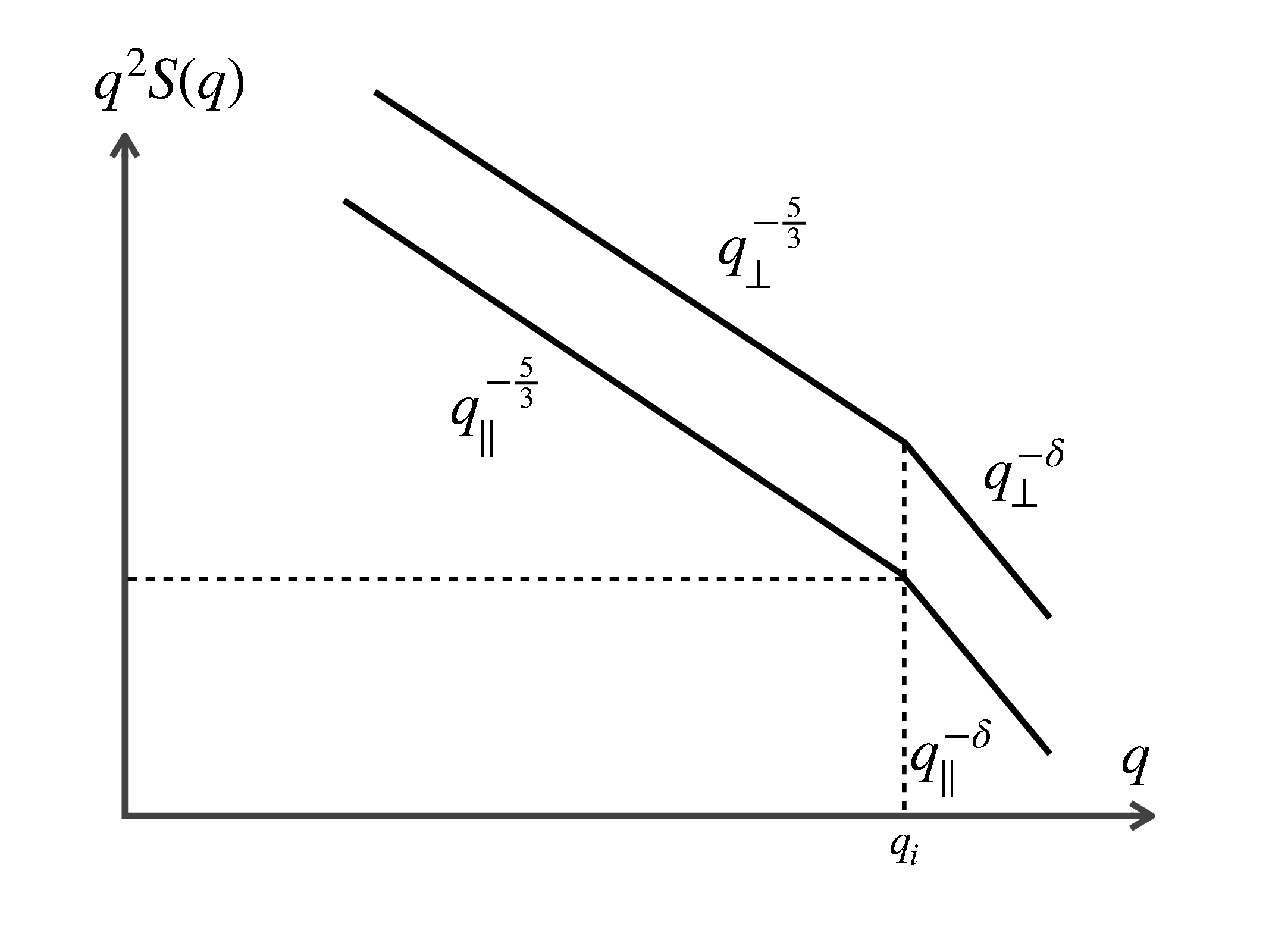}
      \caption{Forms of the parallel and perpendicular wavenumber spectra corresponding to the two cases described in Appendices~\ref{sec:anis_integral} (Equation~\eqref{sq-ani}; left) and~\ref{appendix:constant_q} (Equation~\eqref{sq-aniB}; right). In the former case, the spectral break points are different for parallel and perpendicular directions, while in the latter case the spectral break points for both directions are the same. In both plots, both axes are on logarithmic scales.}
      \label{fig:sq-case-AB}
    \end{figure}

    For anisotropic turbulence with a spectral break point that is also anisotropic (see left panel of Figure~\ref{fig:sq-case-AB}),

    \begin{equation}\label{sq-ani}
      S(q) = S(q_i) \, \times \, \begin{cases}
        \left ( \frac{\sqrt{q_\perp^2 + \frac{q_\parallel^2}{\alpha^2}}}{q_i} \right )^{-11/3} ;  \qquad \sqrt{q_\perp^2 + \frac{q_\parallel^2}{\alpha^2}} \le q_i \cr
        \left ( \frac{\sqrt{q_\perp^2 + \frac{q_\parallel^2}{\alpha^2}}}{q_i} \right )^{-2-\delta} ; \qquad \sqrt{q_\perp^2 + \frac{q_\parallel^2}{\alpha^2}} > q_i \,\,\, .
      \end{cases}
    \end{equation}
    The basic result~\eqref{pf-general} can be transformed from the wavevector ${\vec q}$ to the variable $\widetilde{\vec q}$ (cf. Equation~\eqref{pf-general_2a}), giving

    \begin{equation}\label{pf-anisotropic-scaled-isotropic}
      P(f) =\alpha \frac{n^2}{(2\pi)^3} \, \int S (\widetilde{\vec q}) \, \delta \left ( \frac{ \widetilde{\vec q} \cdot \widetilde{\vec V}_{\rm SW} }{2 \pi} - f \right ) \, d^3\widetilde{q} \,\,\, ,
    \end{equation}
    where $\widetilde{\vec V}_{\rm SW}=(V_\perp, \alpha V_\parallel)$. Since $S({\widetilde {\vec q}})$ is isotropic, a similar analysis to that of Appendix~\ref{appendix:isotropic_sq} yields the spectrum

    \begin{equation}\label{pf-anisotropic}
      P(f) = \alpha \, \frac{\langle \delta n_i^2 \rangle}{2 \,  \widetilde{f}_i }  \, \times \, \begin{cases}
        \frac{3}{5} \, \left  [ \left ( \frac{f}{\widetilde{f}_i} \right )^{-5/3} \, - \, \left ( 1 - \frac{5}{3 \delta} \right ) \, \right ] \,\,\,\,\,\,\,\,\,   ; \qquad f \le \widetilde{f}_i \cr
        \frac{1}{\delta} \, \left ( \frac{f}{\widetilde{f}_i} \right )^{-\delta} \qquad \qquad \qquad \qquad \,\, ; \qquad
        f > \widetilde{f}_i  \,\,\, ,
      \end{cases}
    \end{equation}
    where $\widetilde{f}_i= q_i{\widetilde { V}}_{\rm SW}/2\pi$, with $\widetilde{V}_{\rm SW} = V_{\rm SW} \left( \sin^2\theta_{B} +\alpha^2 \cos^2 \theta_{B} \right)^{1/2}$, $\theta_{B}$ being the angle between the solar wind velocity ${\vec V}_{\rm SW}$ and the magnetic field ${\vec B}$. The form of the local power-law spectral index $\delta_{\rm loc}$ is similar to Equation~\eqref{pf-isotropic-slope}, but with $f_i$ replaced by $\widetilde{f}_i$. In the case where the solar wind velocity ${\vec V}_{\rm SW}$ is parallel to the magnetic field ${\vec B}$, ${\widetilde V}_{\rm SW} = \alpha V_{\rm SW}$, ${\widetilde f}_i = \alpha \, f_i$, and so

    \begin{equation}\label{pf-sw-parallel-to-B}
      P(f) = \frac{\langle \delta n_i^2 \rangle}{2 \,  f_i }  \, \times \, \begin{cases}
        \frac{3}{5} \, \left  [ \left ( \frac{f}{\alpha \, f_i} \right )^{-5/3} \, - \, \left ( 1 - \frac{5}{3 \delta} \right ) \, \right ] \,\,\,\,\,\,\,\,\,   ; \qquad f \le \alpha \, f_i \cr
        \frac{1}{\delta} \, \left ( \frac{f}{\alpha \, f_i} \right )^{-\delta} \qquad \qquad \qquad \qquad \,\, ; \qquad
        f > \alpha \, f_i  \,\,\, ,
      \end{cases}
    \end{equation}

    \subsection{Anisotropic spectrum with isotropic break point}\label{appendix:constant_q}

    It has been argued by \cite{1990ApJ...358..685A} that a wavenumber spectrum that breaks at a prescribed value of $q = \sqrt{q_\parallel^2 + q_\perp^2}$ in all directions (despite the anisotropy in the wavenumber distribution) is a better fit to the observations. This corresponds to the following density turbulence spectrum (see right panel of Figure~\ref{fig:sq-case-AB}):

    \begin{equation}\label{sq-aniB}
      S(\vec{q}) = S(q_i)  \, \times \, \begin{cases}
        \left ( \frac{\sqrt{q_\perp^2 + \frac{q_\parallel^2}{\alpha^2}}}{q_i} \right )^{-11/3} ;  \qquad \sqrt{q_\perp^2 + q_\parallel^2} \le q_i \cr
        \left ( \frac{\sqrt{q_\perp^2 + \frac{q_\parallel^2}{\alpha^2}}}{q_i} \right )^{-2-\delta} ; \qquad \sqrt{q_\perp^2 + q_\parallel^2} > q_i \,\,\, .
      \end{cases}
    \end{equation}
    Note that the argument of $S$ reflects the anisotropy in $S({\vec q})$ and so is different than the isotropic case of of Appendix~\ref{appendix:isotropic_sq} (Equation~\eqref{sq-form}). However, the spectral break point is the same as in Appendix~\ref{appendix:isotropic_sq}, and so differs from Appendix~\ref{sec:anis_integral}, where the spectral break point occurs at different values of $q$ for different directions (Equation~\eqref{sq-ani}).

    As before, the resonance occurs when $q_\parallel = 2 \pi f /V_{\rm SW}$, and so we have, similar to Equation~\eqref{pf-isotropic-1}, but with an additional factor of $\alpha$ appearing in the argument of $S$,

    \begin{equation}\label{pf-anisotropic-constant-break}
      P(f) = \frac{n^2}{2\pi \, V_{\rm SW}} \, \int_{q_\perp = 0 }^\infty S \left ( \sqrt{q_\perp^2 + \left ( \frac{2 \pi f}{\alpha \, V_{\rm SW}} \right )^2 } \right ) \, q_\perp \, dq_\perp \,\,\, .
    \end{equation}
    The derivation of $P(f)$ proceeds similarly to Appendix~\ref{appendix:isotropic_sq}, but recognizes the appearance of the factor $\alpha$ in the second term in the argument of $S$:

    \bigskip
    (i) $f \le f_i = \frac{q_i V_{\rm SW}}{2 \pi}$

    \begin{eqnarray}\label{eq:Pf_constant-break_small_f}
      P(f) &= \frac{n^2}{2\pi \, V_{\rm SW}} \, S(q_i) & \,\left [ q_i^{11/3} \, \int_{q_\perp =0}^{\sqrt{q_i^2 - (2 \pi f/V_{\rm SW})^2}} \left ( q_\perp^2 + \left ( \frac{2 \pi f}{\alpha \, V_{\rm SW}} \right )^2 \right )^{-11/6} \, q_\perp \, dq_\perp  \, \right . + \cr
      & & \qquad + \left . \, q_i^{\delta + 2} \, \int_{q_\perp ={\sqrt{q_i^2 - (2 \pi f/V_{\rm SW})^2}}}^\infty \left ( q_\perp^2 + \left ( \frac{2 \pi f}{\alpha \, V_{\rm SW}} \right )^2 \right )^{-1 - \frac{\delta}{2}} \, q_\perp \, dq_\perp  \right ] \cr
      &= \frac{n^2}{4\pi^2 \, f_i} \, q_i^3 \, S(q_i) & \, \left [ \frac{3}{5} \left \{ \left ( \frac{2 \pi f}{\alpha \, q_i \, V_{\rm SW}} \right )^{-5/3} - \left ( 1 - \left ( \frac{2 \pi f}{q_i \, V_{\rm SW}} \right )^2 \, \left ( 1 - \frac{1}{\alpha^2} \right ) \right )^{-5/6}  \right \} \right . + \cr
        & & \qquad + \left . \frac{1}{\delta} \, \left ( 1 - \left ( \frac{2 \pi f}{q_i \, V_{\rm SW}} \right )^2 \, \left ( 1 - \frac{1}{\alpha^2} \right ) \right )^{-\delta/2} \right ] \cr
      &\!\!\!\!\!\!\!\!\!\!\!\!\!\!\!\!\!\!\!\!\!\!\!\!\! = \frac{\langle\delta n_i^2\rangle}{2 \, f_i} \,  &\!\!\!\!\!\!\!\!\!\!\!\!\!\!\!\!\!\!\!\!\!\!\! \left [ \frac{3}{5} \left \{ \left ( \frac{f}{\alpha f_i } \right )^{-5/3} - \left ( 1 - \left ( \frac{f}{f_i} \right )^2 \, \left ( 1 - \frac{1}{\alpha^2} \right ) \right )^{-5/6}  \right \} + \frac{1}{\delta} \, \left ( 1 - \left ( \frac{f}{f_i} \right )^2 \, \left ( 1 - \frac{1}{\alpha^2} \right ) \right )^{-\delta/2} \right ]  \,\,\, .
    \end{eqnarray}

    \bigskip

    (ii) $f > f_i = \frac{q_i V_{\rm SW}}{2 \pi}$

    \begin{eqnarray}\label{eq:Pf_constant-break_large_f}
      P(f) \, &=& \, \frac{n^2}{2\pi \, V_{\rm SW}} \, S(q_i) \, q_i^{\delta + 2} \, \int_{q_\perp =0}^\infty \left ( q_\perp^2 + \left ( \frac{2 \pi f}{\alpha \, V_{\rm SW}} \right )^2 \right )^{-1 - \frac{\delta}{2}} \, q_\perp \, dq_\perp \cr
      &=& \, \frac{n^2}{4\pi \, V_{\rm SW}} \, S(q_i) \,  \frac{2}{\delta} \, q_i^{\delta+2} \left ( \frac{2 \pi f}{\alpha \, V_{\rm SW}} \right )^{-\delta} \cr
      &=& \, \frac{\langle\delta n_i^2\rangle}{2 \, f_i} \frac{1}{\delta} \, \left ( \frac{f}{\alpha \, f_i } \right )^{-\delta}
    \end{eqnarray}
    Results~\eqref{eq:Pf_constant-break_small_f} and~\eqref{eq:Pf_constant-break_large_f} match at $f=f_i$ and also reduce to the isotropic result~\eqref{pf-isotropic} when $\alpha =1$.  However, the results~\eqref{pf-general_2a} through~\eqref{eq_q_bar1} relating $\overline {q \, \eps^2} \, R_\odot$ and $\int_0^\infty f \, P(f) \, df$ do not hold in the general case $\alpha \ne 1$, because the wavenumber spectrum $S(\widetilde q)$ in Equation~\eqref{sq-aniB} is not, and cannot be reduced to, an isotropic form.

\bibliography{refs}

\begin{thebibliography}{}
\expandafter\ifx\csname natexlab\endcsname\relax\def\natexlab#1{#1}\fi
\providecommand{\url}[1]{\href{#1}{#1}}
\providecommand{\dodoi}[1]{doi:~\href{http://doi.org/#1}{\nolinkurl{#1}}}
\providecommand{\doeprint}[1]{\href{http://ascl.net/#1}{\nolinkurl{http://ascl.net/#1}}}
\providecommand{\doarXiv}[1]{\href{https://arxiv.org/abs/#1}{\nolinkurl{https://arxiv.org/abs/#1}}}

\bibitem[{{Abranin} {et~al.}(1976){Abranin}, {Bazelian}, {Goncharov},
  {Zaitsev}, {Zinichev}, {Rapoport}, \& {Tsybko}}]{1976SvA....19..602A}
{Abranin}, E.~P., {Bazelian}, L.~L., {Goncharov}, N.~I., {et~al.} 1976,
  \sovast, 19, 602

\bibitem[{{Abranin} {et~al.}(1978){Abranin}, {Bazelian}, {Goncharov},
  {Zaitsev}, {Zinichev}, {Rapoport}, \& {Tsybko}}]{1978SoPh...57..229A}
---. 1978, \solphys, 57, 229, \dodoi{10.1007/BF00152056}

\bibitem[{{Alexander} {et~al.}(1969){Alexander}, {Malitson}, \&
  {Stone}}]{1969SoPh....8..388A}
{Alexander}, J.~K., {Malitson}, H.~H., \& {Stone}, R.~G. 1969, \solphys, 8,
  388, \dodoi{10.1007/BF00155385}

\bibitem[{{Alexandrova} {et~al.}(2013){Alexandrova}, {Chen}, {Sorriso-Valvo},
  {Horbury}, \& {Bale}}]{2013SSRv..178..101A}
{Alexandrova}, O., {Chen}, C.~H.~K., {Sorriso-Valvo}, L., {Horbury}, T.~S., \&
  {Bale}, S.~D. 2013, \ssr, 178, 101, \dodoi{10.1007/s11214-013-0004-8}

\bibitem[{{Allen}(1947)}]{1947MNRAS.107..426A}
{Allen}, C.~W. 1947, \mnras, 107, 426, \dodoi{10.1093/mnras/107.5-6.426}

\bibitem[{{Alvarez}(1976)}]{1976SoPh...46..483A}
{Alvarez}, H. 1976, \solphys, 46, 483, \dodoi{10.1007/BF00149873}

\bibitem[{{Alvarez} \& {Haddock}(1973{\natexlab{a}})}]{1973SoPh...30..175A}
{Alvarez}, H., \& {Haddock}, F.~T. 1973{\natexlab{a}}, \solphys, 30, 175,
  \dodoi{10.1007/BF00156186}

\bibitem[{{Alvarez} \& {Haddock}(1973{\natexlab{b}})}]{1973SoPh...29..197A}
---. 1973{\natexlab{b}}, \solphys, 29, 197, \dodoi{10.1007/BF00153449}

\bibitem[{{Anantharamaiah} {et~al.}(1994){Anantharamaiah}, {Gothoskar}, \&
  {Cornwell}}]{1994JApA...15..387A}
{Anantharamaiah}, K.~R., {Gothoskar}, P., \& {Cornwell}, T.~J. 1994, Journal of
  Astrophysics and Astronomy, 15, 387, \dodoi{10.1007/BF02714823}

\bibitem[{{Armand} {et~al.}(1987){Armand}, {Efimov}, \&
  {Iakovlev}}]{1987A&A...183..135A}
{Armand}, N.~A., {Efimov}, A.~I., \& {Iakovlev}, O.~I. 1987, \aap, 183, 135

\bibitem[{{Armstrong} {et~al.}(1990){Armstrong}, {Coles}, {Kojima}, \&
  {Rickett}}]{1990ApJ...358..685A}
{Armstrong}, J.~W., {Coles}, W.~A., {Kojima}, M., \& {Rickett}, B.~J. 1990,
  \apj, 358, 685, \dodoi{10.1086/169022}

\bibitem[{{Arzner} \& {Magun}(1999)}]{1999A&A...351.1165A}
{Arzner}, K., \& {Magun}, A. 1999, \aap, 351, 1165

\bibitem[{{Barrow} \& {Achong}(1975)}]{1975SoPh...45..459B}
{Barrow}, C.~H., \& {Achong}, A. 1975, \solphys, 45, 459,
  \dodoi{10.1007/BF00158462}

\bibitem[{{Barrow} {et~al.}(1994){Barrow}, {Zarka}, \&
  {Aubier}}]{1994A&A...286..597B}
{Barrow}, C.~H., {Zarka}, P., \& {Aubier}, M.~G. 1994, \aap, 286, 597

\bibitem[{{Bastian}(1994)}]{1994ApJ...426..774B}
{Bastian}, T.~S. 1994, \apj, 426, 774, \dodoi{10.1086/174114}

\bibitem[{{Behannon}(1976)}]{1976pspe.proc..332B}
{Behannon}, K.~W. 1976, in Physics of Solar Planetary Environments, ed. D.~J.
  {Williams}, Vol.~1, 332--345

\bibitem[{{Bian} {et~al.}(2019){Bian}, {Emslie}, \&
  {Kontar}}]{2019ApJ...873...33B}
{Bian}, N.~H., {Emslie}, A.~G., \& {Kontar}, E.~P. 2019, \apj, 873, 33,
  \dodoi{10.3847/1538-4357/ab0411}

\bibitem[{{Bian} {et~al.}(2010){Bian}, {Kontar}, \&
  {Brown}}]{2010A&A...519A.114B}
{Bian}, N.~H., {Kontar}, E.~P., \& {Brown}, J.~C. 2010, \aap, 519, A114+,
  \dodoi{10.1051/0004-6361/201014048}

\bibitem[{{Bird} {et~al.}(2002){Bird}, {Efimov}, {Andreev}, {Samoznaev},
  {Chashei}, {Edenhofer}, {Plettemeier}, \& {Wohlmuth}}]{2002AdSpR..30..447B}
{Bird}, M.~K., {Efimov}, A.~I., {Andreev}, V.~E., {et~al.} 2002, Advances in
  Space Research, 30, 447, \dodoi{10.1016/S0273-1177(02)00334-4}

\bibitem[{{Bird} {et~al.}(1994){Bird}, {Volland}, {Paetzold}, {Edenhofer},
  {Asmar}, \& {Brenkle}}]{1994ApJ...426..373B}
{Bird}, M.~K., {Volland}, H., {Paetzold}, M., {et~al.} 1994, \apj, 426, 373,
  \dodoi{10.1086/174073}

\bibitem[{{Blesing} \& {Dennison}(1972)}]{1972PASA....2...84B}
{Blesing}, R.~G., \& {Dennison}, P.~A. 1972, \pasa, 2, 84,
  \dodoi{10.1017/S1323358000012947}

\bibitem[{{Bougeret} {et~al.}(1970){Bougeret}, {Caroubalos}, {Mercier}, \&
  {Pick}}]{1970A&A.....6..406B}
{Bougeret}, J.-L., {Caroubalos}, C., {Mercier}, C., \& {Pick}, M. 1970, \aap,
  6, 406

\bibitem[{{Bougeret} {et~al.}(1984{\natexlab{a}}){Bougeret}, {Fainberg}, \&
  {Stone}}]{1984A&A...141...17B}
{Bougeret}, J.~L., {Fainberg}, J., \& {Stone}, R.~G. 1984{\natexlab{a}}, \aap,
  141, 17

\bibitem[{{Bougeret} {et~al.}(1984{\natexlab{b}}){Bougeret}, {King}, \&
  {Schwenn}}]{1984SoPh...90..401B}
{Bougeret}, J.~L., {King}, J.~H., \& {Schwenn}, R. 1984{\natexlab{b}},
  \solphys, 90, 401, \dodoi{10.1007/BF00173965}

\bibitem[{{Bougeret} {et~al.}(2008){Bougeret}, {Goetz}, {Kaiser}, {Bale},
  {Kellogg}, {Maksimovic}, {Monge}, {Monson}, {Astier}, {Davy}, {Dekkali},
  {Hinze}, {Manning}, {Aguilar-Rodriguez}, {Bonnin}, {Briand}, {Cairns},
  {Cattell}, {Cecconi}, {Eastwood}, {Ergun}, {Fainberg}, {Hoang}, {Huttunen},
  {Krucker}, {Lecacheux}, {MacDowall}, {Macher}, {Mangeney}, {Meetre},
  {Moussas}, {Nguyen}, {Oswald}, {Pulupa}, {Reiner}, {Robinson}, {Rucker},
  {Salem}, {Santolik}, {Silvis}, {Ullrich}, {Zarka}, \&
  {Zouganelis}}]{2008SSRv..136..487B}
{Bougeret}, J.~L., {Goetz}, K., {Kaiser}, M.~L., {et~al.} 2008, \ssr, 136, 487,
  \dodoi{10.1007/s11214-007-9298-8}

\bibitem[{{Bradford} \& {Routledge}(1980)}]{1980MNRAS.190P..73B}
{Bradford}, H.~M., \& {Routledge}, D. 1980, \mnras, 190, 73P,
  \dodoi{10.1093/mnras/190.1.73P}

\bibitem[{{Breen} {et~al.}(1999){Breen}, {Mikic}, {Linker}, {Lazarus},
  {Thompson}, {Biesecker}, {Moran}, {Varley}, {Williams}, \&
  {Lecinski}}]{1999JGR...104.9847B}
{Breen}, A.~R., {Mikic}, Z., {Linker}, J.~A., {et~al.} 1999, \jgr, 104, 9847,
  \dodoi{10.1029/1998JA900091}

\bibitem[{{Bruno} \& {Trenchi}(2014)}]{2014ApJ...787L..24B}
{Bruno}, R., \& {Trenchi}, L. 2014, \apjl, 787, L24,
  \dodoi{10.1088/2041-8205/787/2/L24}

\bibitem[{{Burlaga} \& {Ness}(1968)}]{1968CaJPS..46..962B}
{Burlaga}, L.~F., \& {Ness}, N.~F. 1968, Canadian Journal of Physics
  Supplement, 46, 962, \dodoi{10.1139/p68-394}

\bibitem[{{Cecconi}(2014)}]{2014CRPhy..15..441C}
{Cecconi}, B. 2014, Comptes Rendus Physique, 15, 441,
  \dodoi{10.1016/j.crhy.2014.02.005}

\bibitem[{{Celnikier} {et~al.}(1983){Celnikier}, {Harvey}, {Jegou}, {Moricet},
  \& {Kemp}}]{1983A&A...126..293C}
{Celnikier}, L.~M., {Harvey}, C.~C., {Jegou}, R., {Moricet}, P., \& {Kemp}, M.
  1983, \aap, 126, 293

\bibitem[{{Celnikier} {et~al.}(1987){Celnikier}, {Muschietti}, \&
  {Goldman}}]{1987A&A...181..138C}
{Celnikier}, L.~M., {Muschietti}, L., \& {Goldman}, M.~V. 1987, \aap, 181, 138

\bibitem[{{Chandran} {et~al.}(2009){Chandran}, {Quataert}, {Howes}, {Xia}, \&
  {Pongkitiwanichakul}}]{2009ApJ...707.1668C}
{Chandran}, B. D.~G., {Quataert}, E., {Howes}, G.~G., {Xia}, Q., \&
  {Pongkitiwanichakul}, P. 2009, \apj, 707, 1668,
  \dodoi{10.1088/0004-637X/707/2/1668}

\bibitem[{{Chandrasekhar}(1952)}]{1952MNRAS.112..475C}
{Chandrasekhar}, S. 1952, \mnras, 112, 475, \dodoi{10.1093/mnras/112.5.475}

\bibitem[{{Chen} {et~al.}(2012){Chen}, {Salem}, {Bonnell}, {Mozer}, \&
  {Bale}}]{2012PhRvL.109c5001C}
{Chen}, C.~H.~K., {Salem}, C.~S., {Bonnell}, J.~W., {Mozer}, F.~S., \& {Bale},
  S.~D. 2012, \prl, 109, 035001, \dodoi{10.1103/PhysRevLett.109.035001}

\bibitem[{{Chen} {et~al.}(2014){Chen}, {Sorriso-Valvo},
  {{\v{S}}afr{\'a}nkov{\'a}}, \& {N{\v{e}}me{\v{c}}ek}}]{2014ApJ...789L...8C}
{Chen}, C.~H.~K., {Sorriso-Valvo}, L., {{\v{S}}afr{\'a}nkov{\'a}}, J., \&
  {N{\v{e}}me{\v{c}}ek}, Z. 2014, \apjl, 789, L8,
  \dodoi{10.1088/2041-8205/789/1/L8}

\bibitem[{{Chen} \& {Shawhan}(1978)}]{1978SoPh...57..205C}
{Chen}, H.~S.-L., \& {Shawhan}, S.~D. 1978, \solphys, 57, 205,
  \dodoi{10.1007/BF00152055}

\bibitem[{{Chen} {et~al.}(2020){Chen}, {Kontar}, {Chrysaphi}, {Jeffrey},
  {Gordovskyy}, {Yan}, \& {Tan}}]{2020ApJ...905...43C}
{Chen}, X., {Kontar}, E.~P., {Chrysaphi}, N., {et~al.} 2020, \apj, 905, 43,
  \dodoi{10.3847/1538-4357/abc24e}

\bibitem[{{Chen} {et~al.}(2023){Chen}, {Kontar}, {Clarkson}, \&
  {Chrysaphi}}]{2023MNRAS.520.3117C}
{Chen}, X., {Kontar}, E.~P., {Clarkson}, D.~L., \& {Chrysaphi}, N. 2023,
  \mnras, 520, 3117, \dodoi{10.1093/mnras/stad325}

\bibitem[{{Chhetri} {et~al.}(2018){Chhetri}, {Morgan}, {Ekers}, {Macquart},
  {Sadler}, {Giroletti}, {Callingham}, \& {Tingay}}]{2018MNRAS.474.4937C}
{Chhetri}, R., {Morgan}, J., {Ekers}, R.~D., {et~al.} 2018, \mnras, 474, 4937,
  \dodoi{10.1093/mnras/stx2864}

\bibitem[{{Chiba} {et~al.}(2022){Chiba}, {Imamura}, {Tokumaru}, {Shiota},
  {Matsumoto}, {Ando}, {Takeuchi}, {Murata}, {Yamazaki}, {H{\"a}usler}, \&
  {P{\"a}tzold}}]{2022SoPh..297...34C}
{Chiba}, S., {Imamura}, T., {Tokumaru}, M., {et~al.} 2022, \solphys, 297, 34,
  \dodoi{10.1007/s11207-022-01968-9}

\bibitem[{{Chrysaphi} {et~al.}(2018){Chrysaphi}, {Kontar}, {Holman}, \&
  {Temmer}}]{2018ApJ...868...79C}
{Chrysaphi}, N., {Kontar}, E.~P., {Holman}, G.~D., \& {Temmer}, M. 2018, \apj,
  868, 79, \dodoi{10.3847/1538-4357/aae9e5}

\bibitem[{{Chrysaphi} {et~al.}(2020){Chrysaphi}, {Reid}, \&
  {Kontar}}]{2020ApJ...893..115C}
{Chrysaphi}, N., {Reid}, H. A.~S., \& {Kontar}, E.~P. 2020, \apj, 893, 115,
  \dodoi{10.3847/1538-4357/ab80c1}

\bibitem[{{Clarkson} {et~al.}(2021){Clarkson}, {Kontar}, {Gordovskyy},
  {Chrysaphi}, \& {Vilmer}}]{2021ApJ...917L..32C}
{Clarkson}, D.~L., {Kontar}, E.~P., {Gordovskyy}, M., {Chrysaphi}, N., \&
  {Vilmer}, N. 2021, \apjl, 917, L32, \dodoi{10.3847/2041-8213/ac1a7d}

\bibitem[{{Clarkson} {et~al.}(2023){Clarkson}, {Kontar}, {Vilmer},
  {Gordovskyy}, {Chen}, \& {Chrysaphi}}]{2023ApJ...946...33C}
{Clarkson}, D.~L., {Kontar}, E.~P., {Vilmer}, N., {et~al.} 2023, \apj, 946, 33,
  \dodoi{10.3847/1538-4357/acbd3f}

\bibitem[{{Cohen} {et~al.}(1967){Cohen}, {Gundermann}, {Hardebeck}, \&
  {Sharp}}]{1967ApJ...147..449C}
{Cohen}, M.~H., {Gundermann}, E.~J., {Hardebeck}, H.~E., \& {Sharp}, L.~E.
  1967, \apj, 147, 449, \dodoi{10.1086/149028}

\bibitem[{{Coles}(1978)}]{1978SSRv...21..411C}
{Coles}, W.~A. 1978, \ssr, 21, 411, \dodoi{10.1007/BF00173067}

\bibitem[{{Coles} \& {Harmon}(1989)}]{1989ApJ...337.1023C}
{Coles}, W.~A., \& {Harmon}, J.~K. 1989, \apj, 337, 1023,
  \dodoi{10.1086/167173}

\bibitem[{{DeForest} {et~al.}(2018){DeForest}, {Howard}, {Velli}, {Viall}, \&
  {Vourlidas}}]{2018ApJ...862...18D}
{DeForest}, C.~E., {Howard}, R.~A., {Velli}, M., {Viall}, N., \& {Vourlidas},
  A. 2018, \apj, 862, 18, \dodoi{10.3847/1538-4357/aac8e3}

\bibitem[{{Dennison} \& {Blesing}(1972)}]{1972PASAu...2...86D}
{Dennison}, P.~A., \& {Blesing}, R.~G. 1972, Proceedings of the Astronomical
  Society of Australia, 2, 86, \dodoi{10.1017/S1323358000012959}

\bibitem[{{Dulk} \& {McLean}(1978)}]{1978SoPh...57..279D}
{Dulk}, G.~A., \& {McLean}, D.~J. 1978, \solphys, 57, 279,
  \dodoi{10.1007/BF00160102}

\bibitem[{{Dulk} \& {Suzuki}(1980)}]{1980A&A....88..203D}
{Dulk}, G.~A., \& {Suzuki}, S. 1980, \aap, 88, 203

\bibitem[{{Elgaroy} \& {Lyngstad}(1972)}]{1972A&A....16....1E}
{Elgaroy}, O., \& {Lyngstad}, E. 1972, \aap, 16, 1

\bibitem[{{Fainberg} \& {Stone}(1974)}]{1974SSRv...16..145F}
{Fainberg}, J., \& {Stone}, R.~G. 1974, \ssr, 16, 145,
  \dodoi{10.1007/BF00240885}

\bibitem[{{Fokker}(1965)}]{1965BAN....18..111F}
{Fokker}, A.~D. 1965, \bain, 18, 111

\bibitem[{{Fox} {et~al.}(2016){Fox}, {Velli}, {Bale}, {Decker}, {Driesman},
  {Howard}, {Kasper}, {Kinnison}, {Kusterer}, {Lario}, {Lockwood}, {McComas},
  {Raouafi}, \& {Szabo}}]{2016SSRv..204....7F}
{Fox}, N.~J., {Velli}, M.~C., {Bale}, S.~D., {et~al.} 2016, \ssr, 204, 7,
  \dodoi{10.1007/s11214-015-0211-6}

\bibitem[{{Fredricks} \& {Coroniti}(1976)}]{1976JGR....81.5591F}
{Fredricks}, R.~W., \& {Coroniti}, F.~V. 1976, \jgr, 81, 5591,
  \dodoi{10.1029/JA081i031p05591}

\bibitem[{{Ginzburg} \& {Zhelezniakov}(1958)}]{1958SvA.....2..653G}
{Ginzburg}, V.~L., \& {Zhelezniakov}, V.~V. 1958, \sovast, 2, 653

\bibitem[{{Gopalswamy} {et~al.}(2012){Gopalswamy}, {Nitta}, {Akiyama},
  {M{\"a}kel{\"a}}, \& {Yashiro}}]{2012ApJ...744...72G}
{Gopalswamy}, N., {Nitta}, N., {Akiyama}, S., {M{\"a}kel{\"a}}, P., \&
  {Yashiro}, S. 2012, \apj, 744, 72, \dodoi{10.1088/0004-637X/744/1/72}

\bibitem[{{Gordovskyy} {et~al.}(2019){Gordovskyy}, {Kontar}, {Browning}, \&
  {Kuznetsov}}]{2019ApJ...873...48G}
{Gordovskyy}, M., {Kontar}, E., {Browning}, P., \& {Kuznetsov}, A. 2019, \apj,
  873, 48, \dodoi{10.3847/1538-4357/ab03d8}

\bibitem[{{Gorgolewski} {et~al.}(1962){Gorgolewski}, {Hanasz}, {Iwaniszewski},
  \& {Tur{\l}o}}]{1962AcA....12..251G}
{Gorgolewski}, S., {Hanasz}, J., {Iwaniszewski}, H., \& {Tur{\l}o}, Z. 1962,
  \actaa, 12, 251

\bibitem[{{Gurnett} {et~al.}(1978){Gurnett}, {Baumback}, \&
  {Rosenbauer}}]{1978JGR....83..616G}
{Gurnett}, D.~A., {Baumback}, M.~M., \& {Rosenbauer}, H. 1978, \jgr, 83, 616,
  \dodoi{10.1029/JA083iA02p00616}

\bibitem[{{Harries} {et~al.}(1970){Harries}, {Blesing}, \&
  {Dennison}}]{1970PASA....1..319H}
{Harries}, J.~R., {Blesing}, R.~G., \& {Dennison}, P.~A. 1970, \pasa, 1, 319,
  \dodoi{10.1017/S1323358000012091}

\bibitem[{{Hellinger} {et~al.}(2011){Hellinger}, {Matteini},
  {{\v{S}}tver{\'a}k}, {Tr{\'a}vn{\'\i}{\v{c}}ek}, \&
  {Marsch}}]{2011JGRA..116.9105H}
{Hellinger}, P., {Matteini}, L., {{\v{S}}tver{\'a}k}, {\v{S}}.,
  {Tr{\'a}vn{\'\i}{\v{c}}ek}, P.~M., \& {Marsch}, E. 2011, Journal of
  Geophysical Research (Space Physics), 116, A09105,
  \dodoi{10.1029/2011JA016674}

\bibitem[{{Hewish}(1958)}]{1958MNRAS.118..534H}
{Hewish}, A. 1958, \mnras, 118, 534, \dodoi{10.1093/mnras/118.6.534}

\bibitem[{{Hewish} {et~al.}(1964){Hewish}, {Scott}, \&
  {Wills}}]{1964Natur.203.1214H}
{Hewish}, A., {Scott}, P.~F., \& {Wills}, D. 1964, \nat, 203, 1214,
  \dodoi{10.1038/2031214a0}

\bibitem[{{Hewish} \& {Wyndham}(1963)}]{1963MNRAS.126..469H}
{Hewish}, A., \& {Wyndham}, J.~D. 1963, \mnras, 126, 469,
  \dodoi{10.1093/mnras/126.5.469}

\bibitem[{{H{\"o}gbom}(1960)}]{1960MNRAS.120..530H}
{H{\"o}gbom}, J.~A. 1960, \mnras, 120, 530, \dodoi{10.1093/mnras/120.6.530}

\bibitem[{{Holman} {et~al.}(2011){Holman}, {Aschwanden}, {Aurass}, {Battaglia},
  {Grigis}, {Kontar}, {Liu}, {Saint-Hilaire}, \&
  {Zharkova}}]{2011SSRv..159..107H}
{Holman}, G.~D., {Aschwanden}, M.~J., {Aurass}, H., {et~al.} 2011, \ssr, 159,
  107, \dodoi{10.1007/s11214-010-9680-9}

\bibitem[{{Kellogg} \& {Horbury}(2005)}]{2005AnGeo..23.3765K}
{Kellogg}, P.~J., \& {Horbury}, T.~S. 2005, Annales Geophysicae, 23, 3765,
  \dodoi{10.5194/angeo-23-3765-2005}

\bibitem[{{Kolmogorov}(1941)}]{1941DoSSR..30..301K}
{Kolmogorov}, A. 1941, Akademiia Nauk SSSR Doklady, 30, 301

\bibitem[{{Kolotkov} {et~al.}(2018){Kolotkov}, {Nakariakov}, \&
  {Kontar}}]{2018ApJ...861...33K}
{Kolotkov}, D.~Y., {Nakariakov}, V.~M., \& {Kontar}, E.~P. 2018, \apj, 861, 33,
  \dodoi{10.3847/1538-4357/aac77e}

\bibitem[{{Kontar}(2001)}]{2001SoPh..202..131K}
{Kontar}, E.~P. 2001, \solphys, 202, 131, \dodoi{10.1023/A:1011894830942}

\bibitem[{{Kontar} {et~al.}(2017){Kontar}, {Yu}, {Kuznetsov}, {Emslie},
  {Alcock}, {Jeffrey}, {Melnik}, {Bian}, \&
  {Subramanian}}]{2017NatCo...8.1515K}
{Kontar}, E.~P., {Yu}, S., {Kuznetsov}, A.~A., {et~al.} 2017, Nature
  Communications, 8, 1515, \dodoi{10.1038/s41467-017-01307-8}

\bibitem[{{Kontar} {et~al.}(2019){Kontar}, {Chen}, {Chrysaphi}, {Jeffrey},
  {Emslie}, {Krupar}, {Maksimovic}, {Gordovskyy}, \&
  {Browning}}]{2019ApJ...884..122K}
{Kontar}, E.~P., {Chen}, X., {Chrysaphi}, N., {et~al.} 2019, \apj, 884, 122,
  \dodoi{10.3847/1538-4357/ab40bb}

\bibitem[{{Krupar} {et~al.}(2014){Krupar}, {Maksimovic}, {Santolik}, {Cecconi},
  \& {Kruparova}}]{2014SoPh..289.4633K}
{Krupar}, V., {Maksimovic}, M., {Santolik}, O., {Cecconi}, B., \& {Kruparova},
  O. 2014, \solphys, 289, 4633, \dodoi{10.1007/s11207-014-0601-z}

\bibitem[{{Krupar} {et~al.}(2018){Krupar}, {Maksimovic}, {Kontar}, {Zaslavsky},
  {Santolik}, {Soucek}, {Kruparova}, {Eastwood}, \&
  {Szabo}}]{2018ApJ...857...82K}
{Krupar}, V., {Maksimovic}, M., {Kontar}, E.~P., {et~al.} 2018, \apj, 857, 82,
  \dodoi{10.3847/1538-4357/aab60f}

\bibitem[{{Krupar} {et~al.}(2020){Krupar}, {Szabo}, {Maksimovic}, {Kruparova},
  {Kontar}, {Balmaceda}, {Bonnin}, {Bale}, {Pulupa}, {Malaspina}, {Bonnell},
  {Harvey}, {Goetz}, {Dudok de Wit}, {MacDowall}, {Kasper}, {Case}, {Korreck},
  {Larson}, {Livi}, {Stevens}, {Whittlesey}, \&
  {Hegedus}}]{2020ApJS..246...57K}
{Krupar}, V., {Szabo}, A., {Maksimovic}, M., {et~al.} 2020, \apjs, 246, 57,
  \dodoi{10.3847/1538-4365/ab65bd}

\bibitem[{{Kuznetsov} {et~al.}(2020){Kuznetsov}, {Chrysaphi}, {Kontar}, \&
  {Motorina}}]{2020ApJ...898...94K}
{Kuznetsov}, A.~A., {Chrysaphi}, N., {Kontar}, E.~P., \& {Motorina}, G. 2020,
  \apj, 898, 94, \dodoi{10.3847/1538-4357/aba04a}

\bibitem[{{Kuznetsov} \& {Kontar}(2019)}]{2019A&A...631L...7K}
{Kuznetsov}, A.~A., \& {Kontar}, E.~P. 2019, \aap, 631, L7,
  \dodoi{10.1051/0004-6361/201936447}

\bibitem[{{Leamon} {et~al.}(1998){Leamon}, {Smith}, {Ness}, {Matthaeus}, \&
  {Wong}}]{1998JGR...103.4775L}
{Leamon}, R.~J., {Smith}, C.~W., {Ness}, N.~F., {Matthaeus}, W.~H., \& {Wong},
  H.~K. 1998, \jgr, 103, 4775, \dodoi{10.1029/97JA03394}

\bibitem[{{Leblanc} {et~al.}(1998){Leblanc}, {Dulk}, \&
  {Bougeret}}]{1998SoPh..183..165L}
{Leblanc}, Y., {Dulk}, G.~A., \& {Bougeret}, J.-L. 1998, \solphys, 183, 165,
  \dodoi{10.1023/A:1005049730506}

\bibitem[{{Lee} \& {Jokipii}(1975)}]{1975ApJ...196..695L}
{Lee}, L.~C., \& {Jokipii}, J.~R. 1975, \apj, 196, 695, \dodoi{10.1086/153458}

\bibitem[{{Lin} {et~al.}(2000){Lin}, {Penn}, \&
  {Tomczyk}}]{2000ApJ...541L..83L}
{Lin}, H., {Penn}, M.~J., \& {Tomczyk}, S. 2000, \apjl, 541, L83,
  \dodoi{10.1086/312900}

\bibitem[{{Lotz} {et~al.}(2023){Lotz}, {Nel}, {Wicks}, {Roberts},
  {Engelbrecht}, {Strauss}, {Botha}, {Kontar}, {Pit{\v{n}}a}, \&
  {Bale}}]{2023ApJ...942...93L}
{Lotz}, S., {Nel}, A.~E., {Wicks}, R.~T., {et~al.} 2023, \apj, 942, 93,
  \dodoi{10.3847/1538-4357/aca903}

\bibitem[{{Machin} \& {Smith}(1952)}]{1952Natur.170..319M}
{Machin}, K.~E., \& {Smith}, F.~G. 1952, \nat, 170, 319,
  \dodoi{10.1038/170319b0}

\bibitem[{{Maguire} {et~al.}(2021){Maguire}, {Carley}, {Zucca}, {Vilmer}, \&
  {Gallagher}}]{2021ApJ...909....2M}
{Maguire}, C.~A., {Carley}, E.~P., {Zucca}, P., {Vilmer}, N., \& {Gallagher},
  P.~T. 2021, \apj, 909, 2, \dodoi{10.3847/1538-4357/abda51}

\bibitem[{{Mancuso} \& {Spangler}(2000)}]{2000ApJ...539..480M}
{Mancuso}, S., \& {Spangler}, S.~R. 2000, \apj, 539, 480,
  \dodoi{10.1086/309205}

\bibitem[{{Mann} {et~al.}(1999){Mann}, {Jansen}, {MacDowall}, {Kaiser}, \&
  {Stone}}]{1999A&A...348..614M}
{Mann}, G., {Jansen}, F., {MacDowall}, R.~J., {Kaiser}, M.~L., \& {Stone},
  R.~G. 1999, \aap, 348, 614

\bibitem[{{Manoharan}(1993{\natexlab{a}})}]{1993SoPh..148..153M}
{Manoharan}, P.~K. 1993{\natexlab{a}}, \solphys, 148, 153,
  \dodoi{10.1007/BF00675541}

\bibitem[{{Manoharan}(1993{\natexlab{b}})}]{1993BASI...21..383M}
---. 1993{\natexlab{b}}, Bulletin of the Astronomical Society of India, 21, 383

\bibitem[{{Marsch} \& {Tu}(1990)}]{1990JGR....9511945M}
{Marsch}, E., \& {Tu}, C.~Y. 1990, \jgr, 95, 11945,
  \dodoi{10.1029/JA095iA08p11945}

\bibitem[{{McCauley} {et~al.}(2018){McCauley}, {Cairns}, \&
  {Morgan}}]{2018SoPh..293..132M}
{McCauley}, P.~I., {Cairns}, I.~H., \& {Morgan}, J. 2018, \solphys, 293, 132,
  \dodoi{10.1007/s11207-018-1353-y}

\bibitem[{{McKim Malville} {et~al.}(1967){McKim Malville}, {Aller}, \&
  {Jensen}}]{1967ApJ...147..711M}
{McKim Malville}, J., {Aller}, H.~D., \& {Jensen}, C.~J. 1967, \apj, 147, 711,
  \dodoi{10.1086/149048}

\bibitem[{{Miyamoto} {et~al.}(2014){Miyamoto}, {Imamura}, {Tokumaru}, {Ando},
  {Isobe}, {Asai}, {Shiota}, {Toda}, {H{\"a}usler}, {P{\"a}tzold}, {Nabatov},
  \& {Nakamura}}]{2014ApJ...797...51M}
{Miyamoto}, M., {Imamura}, T., {Tokumaru}, M., {et~al.} 2014, \apj, 797, 51,
  \dodoi{10.1088/0004-637X/797/1/51}

\bibitem[{{Mohan}(2021)}]{2021A&A...655A..77M}
{Mohan}, A. 2021, \aap, 655, A77, \dodoi{10.1051/0004-6361/202142029}

\bibitem[{{Morimoto} \& {Kai}(1961)}]{1961PASJ...13..294M}
{Morimoto}, M., \& {Kai}, K. 1961, \pasj, 13, 294

\bibitem[{{Murphy} {et~al.}(2021){Murphy}, {Carley}, {Ryan}, {Zucca}, \&
  {Gallagher}}]{2021A&A...645A..11M}
{Murphy}, P.~C., {Carley}, E.~P., {Ryan}, A.~M., {Zucca}, P., \& {Gallagher},
  P.~T. 2021, \aap, 645, A11, \dodoi{10.1051/0004-6361/202038518}

\bibitem[{{Musset} {et~al.}(2021){Musset}, {Maksimovic}, {Kontar}, {Krupar},
  {Chrysaphi}, {Bonnin}, {Vecchio}, {Cecconi}, {Zaslavsky}, {Issautier},
  {Bale}, \& {Pulupa}}]{2021A&A...656A..34M}
{Musset}, S., {Maksimovic}, M., {Kontar}, E., {et~al.} 2021, \aap, 656, A34,
  \dodoi{10.1051/0004-6361/202140998}

\bibitem[{{Narayan} {et~al.}(1989){Narayan}, {Anantharamaiah}, \&
  {Cornwell}}]{1989MNRAS.241..403N}
{Narayan}, R., {Anantharamaiah}, K.~R., \& {Cornwell}, T.~J. 1989, \mnras, 241,
  403, \dodoi{10.1093/mnras/241.3.403}

\bibitem[{{Newkirk}(1961)}]{1961ApJ...133..983N}
{Newkirk}, Jr., G. 1961, \apj, 133, 983, \dodoi{10.1086/147104}

\bibitem[{{Parker}(1958)}]{1958ApJ...128..664P}
{Parker}, E.~N. 1958, \apj, 128, 664, \dodoi{10.1086/146579}

\bibitem[{{Patzold} {et~al.}(1987){Patzold}, {Bird}, {Volland}, {Levy},
  {Seidel}, \& {Stelzried}}]{1987SoPh..109...91P}
{Patzold}, M., {Bird}, M.~K., {Volland}, H., {et~al.} 1987, \solphys, 109, 91,
  \dodoi{10.1007/BF00167401}

\bibitem[{{Pulupa} {et~al.}(2020){Pulupa}, {Bale}, {Badman}, {Bonnell}, {Case},
  {de Wit}, {Goetz}, {Harvey}, {Hegedus}, {Kasper}, {Korreck},
  {Krasnoselskikh}, {Larson}, {Lecacheux}, {Livi}, {MacDowall}, {Maksimovic},
  {Malaspina}, {Mart{\'\i}nez Oliveros}, {Meyer-Vernet}, {Moncuquet},
  {Stevens}, \& {Whittlesey}}]{2020ApJS..246...49P}
{Pulupa}, M., {Bale}, S.~D., {Badman}, S.~T., {et~al.} 2020, \apjs, 246, 49,
  \dodoi{10.3847/1538-4365/ab5dc0}

\bibitem[{{Ramesh} {et~al.}(2010){Ramesh}, {Kathiravan}, \&
  {Sastry}}]{2010ApJ...711.1029R}
{Ramesh}, R., {Kathiravan}, C., \& {Sastry}, C.~V. 2010, \apj, 711, 1029,
  \dodoi{10.1088/0004-637X/711/2/1029}

\bibitem[{{Reid} \& {Kontar}(2018)}]{2018A&A...614A..69R}
{Reid}, H. A.~S., \& {Kontar}, E.~P. 2018, \aap, 614, A69,
  \dodoi{10.1051/0004-6361/201732298}

\bibitem[{{Reid} \& {Kontar}(2021)}]{2021NatAs...5..796R}
---. 2021, Nature Astronomy, 5, 796, \dodoi{10.1038/s41550-021-01370-8}

\bibitem[{{Reid} {et~al.}(2014){Reid}, {Vilmer}, \&
  {Kontar}}]{2014A&A...567A..85R}
{Reid}, H. A.~S., {Vilmer}, N., \& {Kontar}, E.~P. 2014, \aap, 567, A85,
  \dodoi{10.1051/0004-6361/201321973}

\bibitem[{{Reiner} {et~al.}(1998){Reiner}, {Fainberg}, {Kaiser}, \&
  {Stone}}]{1998JGR...103.1923R}
{Reiner}, M.~J., {Fainberg}, J., {Kaiser}, M.~L., \& {Stone}, R.~G. 1998, \jgr,
  103, 1923, \dodoi{10.1029/97JA02646}

\bibitem[{{Reiner} {et~al.}(2009){Reiner}, {Goetz}, {Fainberg}, {Kaiser},
  {Maksimovic}, {Cecconi}, {Hoang}, {Bale}, \&
  {Bougeret}}]{2009SoPh..259..255R}
{Reiner}, M.~J., {Goetz}, K., {Fainberg}, J., {et~al.} 2009, \solphys, 259,
  255, \dodoi{10.1007/s11207-009-9404-z}

\bibitem[{{Rickett}(1977)}]{1977ARA&A..15..479R}
{Rickett}, B.~J. 1977, \araa, 15, 479,
  \dodoi{10.1146/annurev.aa.15.090177.002403}

\bibitem[{{Riddle}(1974)}]{1974SoPh...35..153R}
{Riddle}, A.~C. 1974, \solphys, 35, 153, \dodoi{10.1007/BF00156964}

\bibitem[{{Roberts} {et~al.}(2018){Roberts}, {Narita}, \&
  {Escoubet}}]{2018AnGeo..36..527R}
{Roberts}, O.~W., {Narita}, Y., \& {Escoubet}, C.~P. 2018, Annales Geophysicae,
  36, 527, \dodoi{10.5194/angeo-36-527-2018}

\bibitem[{{Rosenberg} {et~al.}(1978){Rosenberg}, {Kivelson}, {Coleman}, \&
  {Smith}}]{1978JGR....83.4165R}
{Rosenberg}, R.~L., {Kivelson}, M.~G., {Coleman}, P.~J., J., \& {Smith}, E.~J.
  1978, \jgr, 83, 4165, \dodoi{10.1029/JA083iA09p04165}

\bibitem[{{Saint-Hilaire} {et~al.}(2013){Saint-Hilaire}, {Vilmer}, \&
  {Kerdraon}}]{2013ApJ...762...60S}
{Saint-Hilaire}, P., {Vilmer}, N., \& {Kerdraon}, A. 2013, \apj, 762, 60,
  \dodoi{10.1088/0004-637X/762/1/60}

\bibitem[{{Saito} {et~al.}(1977){Saito}, {Poland}, \&
  {Munro}}]{1977SoPh...55..121S}
{Saito}, K., {Poland}, A.~I., \& {Munro}, R.~H. 1977, \solphys, 55, 121,
  \dodoi{10.1007/BF00150879}

\bibitem[{{Sakurai} \& {Spangler}(1994)}]{1994ApJ...434..773S}
{Sakurai}, T., \& {Spangler}, S.~R. 1994, \apj, 434, 773,
  \dodoi{10.1086/174780}

\bibitem[{{Sasikumar Raja} {et~al.}(2016){Sasikumar Raja}, {Ingale}, {Ramesh},
  {Subramanian}, {Manoharan}, \& {Janardhan}}]{2016JGRA..12111605S}
{Sasikumar Raja}, K., {Ingale}, M., {Ramesh}, R., {et~al.} 2016, Journal of
  Geophysical Research (Space Physics), 121, 11, \dodoi{10.1002/2016JA023254}

\bibitem[{{Sasikumar Raja} {et~al.}(2017){Sasikumar Raja}, {Subramanian},
  {Ramesh}, {Vourlidas}, \& {Ingale}}]{2017ApJ...850..129S}
{Sasikumar Raja}, K., {Subramanian}, P., {Ramesh}, R., {Vourlidas}, A., \&
  {Ingale}, M. 2017, \apj, 850, 129, \dodoi{10.3847/1538-4357/aa94cd}

\bibitem[{{Sasikumar Raja} {et~al.}(2022){Sasikumar Raja}, {Maksimovic},
  {Kontar}, {Bonnin}, {Zarka}, {Lamy}, {Reid}, {Vilmer}, {Lecacheux}, {Krupar},
  {Cecconi}, {Nora}, \& {Denis}}]{2022ApJ...924...58S}
{Sasikumar Raja}, K., {Maksimovic}, M., {Kontar}, E.~P., {et~al.} 2022, \apj,
  924, 58, \dodoi{10.3847/1538-4357/ac34ed}

\bibitem[{{Scott} {et~al.}(1983){Scott}, {Coles}, \&
  {Bourgois}}]{1983A&A...123..207S}
{Scott}, S.~L., {Coles}, W.~A., \& {Bourgois}, G. 1983, \aap, 123, 207

\bibitem[{{Shain} \& {Higgins}(1959)}]{1959AuJPh..12..357S}
{Shain}, C.~A., \& {Higgins}, C.~S. 1959, Australian Journal of Physics, 12,
  357, \dodoi{10.1071/PH590357}

\bibitem[{{Sharma} \& {Oberoi}(2021)}]{2021ApJ...913..153S}
{Sharma}, R., \& {Oberoi}, D. 2021, \apj, 913, 153,
  \dodoi{10.3847/1538-4357/ac01df}

\bibitem[{{Sharykin} {et~al.}(2018){Sharykin}, {Kontar}, \&
  {Kuznetsov}}]{2018SoPh..293..115S}
{Sharykin}, I.~N., {Kontar}, E.~P., \& {Kuznetsov}, A.~A. 2018, \solphys, 293,
  115, \dodoi{10.1007/s11207-018-1333-2}

\bibitem[{{Shevchuk} {et~al.}(2016){Shevchuk}, {Melnik}, {Poedts}, {Dorovskyy},
  {Magdalenic}, {Konovalenko}, {Brazhenko}, {Briand}, {Frantsuzenko}, {Rucker},
  \& {Zarka}}]{2016SoPh..291..211S}
{Shevchuk}, N.~V., {Melnik}, V.~N., {Poedts}, S., {et~al.} 2016, \solphys, 291,
  211, \dodoi{10.1007/s11207-015-0799-4}

\bibitem[{{Sioulas} {et~al.}(2023){Sioulas}, {Huang}, {Shi}, {Velli},
  {Tenerani}, {Bowen}, {Bale}, {Huang}, {Vlahos}, {Woodham}, {Horbury}, {de
  Wit}, {Larson}, {Kasper}, {Owen}, {Stevens}, {Case}, {Pulupa}, {Malaspina},
  {Bonnell}, {Livi}, {Goetz}, {Harvey}, {MacDowall}, {Maksimovi{\'c}},
  {Louarn}, \& {Fedorov}}]{2023ApJ...943L...8S}
{Sioulas}, N., {Huang}, Z., {Shi}, C., {et~al.} 2023, \apjl, 943, L8,
  \dodoi{10.3847/2041-8213/acaeff}

\bibitem[{{Sittler} \& {Guhathakurta}(1999)}]{1999ApJ...523..812S}
{Sittler}, Edward~C., J., \& {Guhathakurta}, M. 1999, \apj, 523, 812,
  \dodoi{10.1086/307742}

\bibitem[{{Slee}(1959)}]{1959AuJPh..12..134S}
{Slee}, O.~B. 1959, Australian Journal of Physics, 12, 134,
  \dodoi{10.1071/PH590134}

\bibitem[{{Slee}(1966)}]{1966P&SS...14..255S}
---. 1966, \planss, 14, 255, \dodoi{10.1016/0032-0633(66)90125-5}

\bibitem[{{Spangler}(2002)}]{2002ApJ...576..997S}
{Spangler}, S.~R. 2002, \apj, 576, 997, \dodoi{10.1086/341889}

\bibitem[{{Spangler}(2005)}]{2005SSRv..121..189S}
---. 2005, \ssr, 121, 189, \dodoi{10.1007/s11214-006-4719-7}

\bibitem[{{Spangler}(2020)}]{2020RNAAS...4..147S}
---. 2020, Research Notes of the American Astronomical Society, 4, 147,
  \dodoi{10.3847/2515-5172/abb29a}

\bibitem[{{Spangler} {et~al.}(2002){Spangler}, {Kavars}, {Kortenkamp}, {Bondi},
  {Mantovani}, \& {Alef}}]{2002A&A...384..654S}
{Spangler}, S.~R., {Kavars}, D.~W., {Kortenkamp}, P.~S., {et~al.} 2002, \aap,
  384, 654, \dodoi{10.1051/0004-6361:20020028}

\bibitem[{{Steinberg}(1972)}]{1972A&A....18..382S}
{Steinberg}, J.~L. 1972, \aap, 18, 382

\bibitem[{{Steinberg} {et~al.}(1971){Steinberg}, {Aubier-Giraud}, {Leblanc}, \&
  {Boischot}}]{1971A&A....10..362S}
{Steinberg}, J.~L., {Aubier-Giraud}, M., {Leblanc}, Y., \& {Boischot}, A. 1971,
  \aap, 10, 362

\bibitem[{{Steinberg} {et~al.}(1985){Steinberg}, {Hoang}, \&
  {Dulk}}]{1985A&A...150..205S}
{Steinberg}, J.~L., {Hoang}, S., \& {Dulk}, G.~A. 1985, \aap, 150, 205

\bibitem[{{Stewart}(1972)}]{1972PASA....2..100S}
{Stewart}, R.~T. 1972, \pasa, 2, 100, \dodoi{10.1017/S1323358000013059}

\bibitem[{{Suzuki} \& {Dulk}(1985)}]{1985srph.book..289S}
{Suzuki}, S., \& {Dulk}, G.~A. 1985, in Solar Radiophysics: Studies of Emission
  from the Sun at Metre Wavelengths, ed. D.~J. {McLean} \& N.~R. {Labrum}
  (Cambridge University Press), 289--332

\bibitem[{{Tasnim} {et~al.}(2022){Tasnim}, {Zank}, {Cairns}, \&
  {Adhikari}}]{2022ApJ...928..125T}
{Tasnim}, S., {Zank}, G.~P., {Cairns}, I.~H., \& {Adhikari}, L. 2022, \apj,
  928, 125, \dodoi{10.3847/1538-4357/ac5031}

\bibitem[{{Thejappa} \& {MacDowall}(2008)}]{2008ApJ...676.1338T}
{Thejappa}, G., \& {MacDowall}, R.~J. 2008, \apj, 676, 1338,
  \dodoi{10.1086/528835}

\bibitem[{{Tyul'bashev} {et~al.}(2023){Tyul'bashev}, {Chashei}, \&
  {Kitaeva}}]{2023MNRAS.tmp.1421T}
{Tyul'bashev}, S.~A., {Chashei}, I.~V., \& {Kitaeva}, M.~A. 2023, \mnras,
  \dodoi{10.1093/mnras/stad1461}

\bibitem[{{Villante} \& {Mariani}(1975)}]{1975GeoRL...2...73V}
{Villante}, U., \& {Mariani}, F. 1975, \grl, 2, 73,
  \dodoi{10.1029/GL002i003p00073}

\bibitem[{{{\v{S}}afr{\'a}nkov{\'a}} {et~al.}(2015){{\v{S}}afr{\'a}nkov{\'a}},
  {N{\v{e}}me{\v{c}}ek}, {N{\v{e}}mec}, {P{\v{r}}ech}, {Pit{\v{n}}a}, {Chen},
  \& {Zastenker}}]{2015ApJ...803..107S}
{{\v{S}}afr{\'a}nkov{\'a}}, J., {N{\v{e}}me{\v{c}}ek}, Z., {N{\v{e}}mec}, F.,
  {et~al.} 2015, \apj, 803, 107, \dodoi{10.1088/0004-637X/803/2/107}

\bibitem[{{Wild} {et~al.}(1959){Wild}, {Sheridan}, \&
  {Trent}}]{1959IAUS....9..176W}
{Wild}, J.~P., {Sheridan}, K.~V., \& {Trent}, G.~H. 1959, in IAU Symposium,
  Vol.~9, URSI Symp. 1: Paris Symposium on Radio Astronomy, ed. R.~N.
  {Bracewell}, 176

\bibitem[{{Woo}(1977)}]{1977Natur.266..514W}
{Woo}, R. 1977, \nat, 266, 514, \dodoi{10.1038/266514a0}

\bibitem[{{Woo}(1978)}]{1978ApJ...219..727W}
---. 1978, \apj, 219, 727, \dodoi{10.1086/155831}

\bibitem[{{Woo} {et~al.}(1977){Woo}, {Yang}, \&
  {Ishimaru}}]{1977ApJ...218..557W}
{Woo}, R., {Yang}, F.~C., \& {Ishimaru}, A. 1977, \apj, 218, 557,
  \dodoi{10.1086/155710}

\bibitem[{{Yakovlev} {et~al.}(1980){Yakovlev}, {Efimov}, {Razmanov}, \&
  {Shtrykov}}]{1980SvA....24..454Y}
{Yakovlev}, O.~I., {Efimov}, A.~I., {Razmanov}, V.~M., \& {Shtrykov}, V.~K.
  1980, \sovast, 24, 454

\bibitem[{{Yamauchi} {et~al.}(1998){Yamauchi}, {Tokumaru}, {Kojima},
  {Manoharan}, \& {Esser}}]{1998JGR...103.6571Y}
{Yamauchi}, Y., {Tokumaru}, M., {Kojima}, M., {Manoharan}, P.~K., \& {Esser},
  R. 1998, \jgr, 103, 6571, \dodoi{10.1029/97JA03598}

\bibitem[{{Yamauchi} {et~al.}(1996){Yamauchi}, {Tokumaru}, {Kojima}, {Misawa},
  {Mori}, {Takaba}, {Kondo}, {Tanaka}, {Manoharan}, \&
  {Esser}}]{1996AIPC..382..366Y}
{Yamauchi}, Y., {Tokumaru}, M., {Kojima}, M., {et~al.} 1996, in American
  Institute of Physics Conference Series, Vol. 382, Proceedings of the eigth
  International solar wind Conference: Solar wind eight, ed. D.~{Winterhalter},
  J.~T. {Gosling}, S.~R. {Habbal}, W.~S. {Kurth}, \& M.~{Neugebauer}, 366--366,
  \dodoi{10.1063/1.51472}

\bibitem[{{Zank} {et~al.}(2017){Zank}, {Adhikari}, {Hunana}, {Shiota}, {Bruno},
  \& {Telloni}}]{2017ApJ...835..147Z}
{Zank}, G.~P., {Adhikari}, L., {Hunana}, P., {et~al.} 2017, \apj, 835, 147,
  \dodoi{10.3847/1538-4357/835/2/147}

\bibitem[{{Zhang} {et~al.}(2019){Zhang}, {Yu}, {Kontar}, \&
  {Wang}}]{2019ApJ...885..140Z}
{Zhang}, P., {Yu}, S., {Kontar}, E.~P., \& {Wang}, C. 2019, \apj, 885, 140,
  \dodoi{10.3847/1538-4357/ab458f}

\end{thebibliography}

\end{document}